%% file: PaperI_Main.tex
\documentclass[fleqn,usenatbib]{mnras}
\usepackage[T1]{fontenc}
\usepackage{newtxtext,newtxmath}
\DeclareRobustCommand{\VAN}[3]{#2}
\let\VANthebibliography\thebibliography
\def\thebibliography{\DeclareRobustCommand{\VAN}[3]{##3}\VANthebibliography}
\usepackage{subcaption}
\usepackage{graphicx}	
\usepackage{amsmath}	
\usepackage{breqn}
\usepackage[section]{placeins}
\newcommand{\Msun}{{M$_{\odot}$}}
\newcommand{\Rsun}{{R$_{\odot}$}}
\newcommand{\kmps}{{km\,s$^{-1}$}}
\newcommand{\degree}{$^{\circ}$}

\newcommand{\vsys}{\upsilon_{\rm sys}}
\newcommand{\vpec}{\upsilon_{\rm pec}}

\newcommand{\vorb}{\upsilon_{\rm rel}}

\newcommand{\vnk}{\upsilon_{\rm NK}}
\newcommand{\vmlk}{\upsilon_{\rm MLK}}
\newcommand{\thetank}{\theta_{\rm NK}}
\newcommand{\phink}{\phi_{\rm NK}}
\newcommand{\manomaly}{$\mathcal{M}$}
\newcommand{\vnkconf}{\upsilon_{\rm NK, 68\%}}

\newcommand{\sphi}{\phi_{\mathrm{s}}}
\newcommand{\cphi}{\phi_{\mathrm{c}}}
\newcommand{\stheta}{\theta_{\mathrm{s}}}
\newcommand{\ctheta}{\theta_{\mathrm{c}}}
\newcommand{\calpha}{\Omega_\mathrm{c}}
\newcommand{\salpha}{\Omega_\mathrm{s}}
\newcommand{\G}{G}
\newcommand{\Mc}{M_{\mathrm{CO}}}

\newcommand{\mone}{M_{\rm 1}}
\newcommand{\mtwo}{M_{\rm 2}}
\newcommand{\porb}{P_{\rm orb}}
\newcommand{\monei}{$M_{\rm 1, i}$}
\newcommand{\mtwoi}{$M_{\rm 2, i}$}
\newcommand{\porbi}{$P_{\rm orb, i}$}
\newcommand{\sepi}{$r_{\rm orb, i}$}
\newcommand{\ei}{$e_{\rm i}$}
\newcommand{\monepre}{$M_{\rm 1, SN}$}
\newcommand{\mtwopre}{$M_{\rm 2, SN}$}
\newcommand{\porbpre}{$P_{\rm orb, SN}$}
\newcommand{\seppre}{$r_{\rm orb, SN}$}
\newcommand{\epre}{$e_{\rm SN}$}
\newcommand{\monepost}{$M_{\rm 1, post-SN}$}
\newcommand{\mtwopost}{$M_{\rm 2, post-SN}$}
\newcommand{\porbpost}{$P_{\rm orb, post-SN}$}
\newcommand{\seppost}{$r_{\rm orb, post-SN}$}
\newcommand{\epost}{$e_{\rm post-SN}$}
\newcommand{\moneobs}{$M_{\rm 1, obs}$}
\newcommand{\mtwoobs}{$M_{\rm 2, obs}$}
\newcommand{\porbobs}{$P_{\rm orb, obs}$}
\newcommand{\eobs}{$e_{\rm obs}$}
\newcommand{\mdot}{$\dot{M}$}

\newcommand{\fouru}{{4U\,1700--37}}
\newcommand{\cyg}{{Cyg\,X--1}}
\newcommand{\xtej}{{XTE\,J1118+480}}
\newcommand{\aqlx}{{Aql\,X--1}}
\newcommand{\btwelve}{{PSR\,B1259--63}}
\newcommand{\gxthree}{{GX\,301--2}}
\newcommand{\jtwo}{{PSR\,J2339--0522}}
\newcommand{\gaia}{{\em Gaia}}
\newcommand{\cosmic}{{\tt COSMIC}}
\newcommand{\mesa}{{\tt MESA}}
\newcommand{\StarTrack}{{\tt StarTrack}}
\defcitealias{Zhao2023}{{Z23}}

\title[Inferring Natal Kicks]{Natal Kicks of Compact Binaries:\\ I. Population Synthesis Framework and Illustrative Examples}

\author[C. Dashwood Brown et al.]{Cordelia Dashwood Brown,$^{1}$\thanks{E-mail: c.dashwood.brown@soton.ac.uk},
Poshak Gandhi,$^{1}$
Christian Knigge,$^{1}$
Yue Zhao,$^{1}$
Phil Charles,$^{1,2,3}$
\newauthor
Pornisara Nuchvanichakul,$^{1}$
Katelyn Breivik$^4$, Akarsh Sahay$^{5,1}$, and Siddharth Malu$^6$
\\
$^{1}$School of Physics \& Astronomy, University of Southampton, Highfield, Southampton SO17 1BJ, UK\\
$^{2}$Astrophysics, Department of Physics, University of Oxford, Keble Road, Oxford OX1 3RH, UK\\
$^{3}$Department of Physics, University of the Free State, PO Box 339, Bloemfontein 9300, South Africa\\
$^{4}$McWilliams Center for Cosmology \& Astrophysics, Department of Physics Carnegie Mellon University, Pittsburgh, PA 15213, USA\\
$^{5}$Department of Physics, University of Oxford, Keble Road, Oxford OX1 3RH, UK\\
$^{6}$AI Center of Excellence, Shiv Nadar Institution of Eminence, NCR Delhi, India}

\date{Accepted 2026 September 8. Received 2026 September 4; in original form 2026 July 1.}
\pubyear{2026}

\begin{document}
\label{firstpage}
\pagerange{\pageref{firstpage}--\pageref{lastpage}}
\maketitle

\begin{abstract}
The natal kicks imparted to neutron stars and black holes at birth are critical for understanding massive-star evolution, compact-object formation, and compact binary merger rates. However, constraining these kicks is challenging, since they cannot be directly observed. Here, we outline a framework for inferring natal kick probability distributions for compact binaries. Our method embeds the rapid binary-population code COSMIC in a Monte-Carlo workflow that evolves ensembles of simulated binaries. From those which host compact objects, we can select subsets that match the observational characteristics of any given system (e.g. component masses, orbital period, mass-transfer status, systemic velocity). As a proof-of-concept, we apply this framework to six compact binaries spanning a wide range of companion masses and compact-object types. For each system, we identify the evolutionary channel(s) that could have led to the current observational characteristics, and isolate regions of allowed natal kicks and progenitor parameter space. We reconfirm that at least one black hole system, XTE J1118+480, must have received a natal kick $\vnk>100$\,\kmps. The neutron-star system, GX 301--2, is also inferred to have received an impulsive kick exceeding 280\,\kmps. We discuss the strengths and limitations associated with population-synthesis-based inference methods and explore the effects of adopting different prescriptions for critical binary evolution phases. We provide a detailed demonstration of the non-linear relationship between $\vnk$ and $\vsys$, highlighting the shortcomings of relying on observations alone to measure natal kicks. The inferred kick velocity distributions across our benchmark sample are generally consistent with previous constraints.

\end{abstract}




\section{Introduction}\label{Sect:Intro}

Peculiar velocity ($\vpec$) in the context of stars refers to the motion of a stellar system beyond that predicted by freefall within an assumed Galactic gravitational potential. The peculiar velocities of Galactic objects (including stars, stellar clusters, and compact objects) have been measured for decades \citep[e.g.,][ for very early examples]{Williams1948,Shatsova1967}. A variety of gravitational and impulsive mechanisms are thought to drive $\vpec$, including environmental density fluctuations, mass-loss from the system, and violent momentum driving in stellar explosions. The deaths of massive stars, in particular, can impart strong acceleration to the compact remnant at the instant of a supernova (SN) explosion - an impulse referred to as a `natal kick.' These kick velocities have a profound effect on the remnant's subsequent dynamical evolution \citep{Kalogera1996,Fryer1997,vandenheuvel97, Nelemans1999,Hobbs2005}; compact object spin \& misalignment with the orbit, its spin \citep{Callister2021,Coleman2022,Fragione2023,Burrows2024a,Baibhav2024}; the rate of gravitational wave merger events in binary systems \citep{Belczynski2020}; the rate of lensing events they may give rise to \citep{Koshimoto2024}; compact remnant retention fraction in star clusters \citep{Pfahl2002a, Strader2012, Giesler2018}; and the dynamical evolution and characteristics of these clusters \citep{Sippel2012,Rostami2024}. Natal kicks can also disrupt the binary, resulting in isolated black holes (BHs) (and neutron stars, NSs) and a population of high-velocity `run-away' ($\vpec\geq$30\,\kmps) or, perhaps, `walk-away' ($\vpec\leq$10\,\kmps) stars \citep{Renzo2019}. These isolated black holes are difficult to find except through microlensing \citep{Sahu2022,Lam2022}; constraining their prevalence will be the next frontier of compact object population studies. 

Conversely, natal kicks may be responsible for binary components merging. In the case of surviving binaries, the natal kicks impact not only the velocity but also the eccentricity and orbital separation \citep[e.g.,][]{Hills1983,Kalogera1996,Fragione2021,Magana,VignaGomez2025}, which in turn govern whether there will be periods of accretion and mass transfer, and have a hand in determining how likely we are to observe these systems. 

These natal kicks could originate from a number of different physical processes. In a binary system, explosive decoupling and ejection of mass when one star undergoes supernova results in a recoil of the binary's centre-of-mass. This was first investigated for massive runaway stars \citep{Blaauw1961} and is now colloquially referred to as a `Blaauw' kick or mass-loss kick. This mechanism operates irrespective of the geometrical distribution of ejected matter. By contrast, other processes invoke inherent inhomogeneities as a necessary condition for kicked compact objects (recalling Newton's ubiquitous Third Law); these asymmetries may be intrinsic to the ejected matter (baryonic) or the neutrino distributions \citep{Harrison1975,Janka1994,Burrows1996,Fryer2004,Scheck2004,Scheck2006,Kochanek2008}. Inhomogeneities likely exist in both the density of the stellar progenitor \citep{Shklovskii1970,Burrows1996}, the neutrino-heated region behind the shock wave \citep{Burrows1996,Lai2000b}, and the shock wave itself \citep{Scheck2004}; Rayleigh-Taylor instabilities, for instance, are a well-established feature of supernova hydrodynamics \citep{Burrows1995,Herant1994,Janka1995}. These non-negligible perturbations likely grow with time -- it is for this reason that a slower onset of explosion is conducive to larger asymmetries and therefore larger recoils imparted to the newly formed compact object \citep{Fryer2004,Scheck2004}. 

In the absence of an agreed-upon first-principles theory of natal kicks for compact objects, observations provide concrete constraints that advance our understanding. Pulsar kinematics have been the subject of intensive studies over the past decades, owing to precision astrometry from radio interferometers and high-quality timing \citep[e.g.][]{Lyne1994,tauris96,Hansen1997,Cordes1998,Toscano1999,Hobbs2005}. These studies have inferred a relatively high mean $\vpec$ of several hundred km\,s$^{-1}$ for isolated pulsars \citep{Hobbs2005, Igoshev2021,Disberg2024,Valli2025}. By contrast, studies of black hole kinematics are handicapped by the difficulty of detecting them. Constraints on BH kinematics are available for those in binaries, whose motion can be tracked by the luminous companion star. In addition, dedicated radio follow up with Very-long-baseline interferometry (VLBI) has put constraints on the kinematics of a couple of accreting black holes \citep{Mirabel2001, Reid2014, Atri2020, MillerJones2021}.

Improvements in optical imaging technologies have resulted in progressive gains in optical astrometric precision over the years, delivering proper motions and astrometric parallaxes for accreting binaries from which their peculiar motions have been inferred \citep[e.g.,][]{mirabel02, Mirabel2017}, with the most recent gains from the \gaia\ satellite and its all-sky survey \citep{Gandhi2019, Arnason2021, Fortin2023, Nuchvanichakul25, Zhao2023, ODoherty2023, Nagarajan2024}. 

These studies identify the necessity for high $\vpec$ values (up to several hundred km\,s$^{-1}$ in at least some binaries). Previous literature has made a distinction between kicks associated with neutron stars \citep[e.g.,][]{Gunn1970,Trimble1971,Lyne1982,Lyne1994,Brandt1995b,Hobbs2005,Verbunt2017} versus black holes \citep[e.g.,][]{Jonker2004,Gualandris2005,Repetto2012,Janka2013,Mandel2015}, with the former being more extensively studied. More recently comprehensive sample studies encompassing neutron star as well as black hole systems show statistically similar $\vpec$ distributions irrespective of compact object type \citep{Zhao2023}; instead, the key differentiator appears to be the total {\em system mass} including the donor star, with heavier systems exhibiting slower peculiar motions, on average. This is broadly aligned with expectations from momentum conservation, irrespective of the kind of natal kick imparted to the remnant \citep{Brandt1995a, Kalogera1996, tauris98}.\footnote{In asymmetric kicks, the ejecta can carry away linear momentum anisotropically, so conservation of momentum cannot be applied to the stellar/compact remnants alone.}

In practice, $\vpec$ is not necessarily a good proxy for natal kick. Any impulsive perturbation will be moderated by the inertia of the donor star as well as any change in orbital angular momentum at supernova. As a result, the final systemic velocity could show a wide scatter \citep{Brandt1995a, tauris98, Kalogera1996, Mandel2015}. Estimating the true natal kick then requires determination of the likely pre-supernova system characteristics followed by an evaluation of the range of natal kicks that could have been survived, resulting ultimately in the binary as observed today \citep{Belczynski2016}. 

Recent years have witnessed much progress in stellar evolutionary modelling, with several computational codes now capable of simulating populations of binaries through the compact object phase with varying degrees of sophistication (e.g. \cosmic, \citealt{cosmic}; \StarTrack, \citealt{Belczynski2008}; \mesa, \citealt{Paxton2011}; {\tt POSYDON}, \citealt{posydon}). Detailed simulations of individual binaries have placed constraints on natal kicks in a handful of systems \citep[e.g., ][]{Willems2005,Fragos2009,Wong2012,Kimball2023}. These studies have revealed the need for high natal kicks in some cases, including for black holes \citep[e.g., ][]{DashwoodBrown2023, Shah2026, li26}. This is further supported by the spatial distribution of black hole systems in the Milky Way --- specifically, an anti-correlation between binary orbital period, and height above the Galactic plane, could originate from kicks ejecting sources from their natal sites in the Galactic disc, with only tight binaries surviving substantial kicks \citep{Gandhi2020, Jonker2021}.

However, strong natal kicks in black holes remain controversial. One major factor that counteracts strong natal kicks is the fallback of material onto a massive compact remnant during the supernova. This fallback can reduce the recoil associated with anisotropic ejecta, resulting in weaker natal kicks. Canonical work embeds fallback using mass-dependent scaling laws \citep[e.g., ][]{Fryer2012}, and significant kicks in black holes would require a reassessment of the impact of fallback and potentially, compact object formation channels.

In a series of works, we are attempting to understand the natal kick distribution of X-ray binaries (XRBs). The novelty that motivates this work is the increasing availability of state-of-the-art astrometric observational results from \gaia. These observations have provided new inferences on systemic peculiar motions, and are collated in \citet[][building upon some of our other works on smaller subsamples; \citealt{Gandhi2019, Rao2020, Nuchvanichakul25}]{Zhao2023}. We then conducted proof-of-concept modelling on one system, H\,1705--250, a known black hole low-mass X-ray binary which had prior indications of a putative natal kick \citep{Repetto2015}. Our work \citep{DashwoodBrown2023} confirmed and refined the need for a substantial natal kick and refined its magnitude. 

In this paper, we present a limited expansion to a subset of other XRBs in order to explore in detail the applicability and limitations of our methodology for constraining the plausible natal kicks any given system may have been subject to. We outline our theoretical framework based on the binary population code \cosmic, along with 
the data that serve as boundary conditions for our modelling. Six systems are chosen for reference here, with discussions on the limitations of the observational constraints and evolutionary modelling in each case, with a specific focus on our ability (or inability, in some cases) to constrain natal kicks.

In a follow-up work (Dashwood Brown et al., in prep., Paper II), we will extend this method to the full population of astrometrically constrained XRBs. Whilst previous works have conducted binary population synthesis on individual systems, our efforts are the first to apply joint constraints on systemic velocities together with binary characterisation on a population-wide scale across black-hole as well as neutron-star systems, in order to infer natal kick distributions. 

\section{Sample of X-ray Binaries}\label{Sect:Sample}

Our base sample for selecting binaries to model is drawn from \citet[][hereafter, Z23]{Zhao2023}, who compiled a catalogue of 89 compact object binaries\footnote{here, `compact object binaries' refers to a black hole or a neutron star in a binary with a luminous star.} with available five-parameter astrometric solutions from \gaia\ data release 3 \citep[DR3,][]{Gaia2016, Gaia2023} or dedicated VLBI observations, as well as systemic radial velocities from the literature. There are 20 BH and 60 NS interacting binaries (XRBs and binary pulsars), all of which are undergoing accretion onto the compact object or experiencing significant interaction with their companion. The catalogue also included 9 non-interacting systems, which are not the focus of our work.

These parameters (position, proper motion, distance, and systemic radial velocity) are combined to estimate the $\vpec$ for each system following the formulation of \citet{Reid2009}. The \citetalias{Zhao2023} sample shows a bimodal distribution in $\vpec$, with low-mass binaries moving at higher $\vpec$ values than high-mass systems. Consistently, the sample also exhibits an anti-correlation between $\vpec$ and the total system mass ($\vpec\,\propto\,M_{\rm tot}^{-0.5}$).

\subsection{Testbed targets}\label{SubSect:Targets}

For illustration and verification of our framework, we focus here on the six systems listed in Table\,\ref{tab:sample}. These are chosen to span a representative range of parameter space in terms of compact object types: BH (\xtej, \cyg), NS (\aqlx, \gxthree), and donor mass: low-mass (\xtej, \aqlx, \jtwo), high-mass (\cyg, \gxthree, \btwelve). These binaries are all interacting through Roche lobe overflow, stellar wind, or some combination thereof. Two binary pulsars (\jtwo, \btwelve) are also included. 

We include one additional source, \fouru, as an instance of the limitations of the current framework. \fouru\ is a high-mass X-ray binary (HMXB) with an extreme mass ratio of 0.04 (amongst the most extreme in high-mass systems) and the only eclipsing system amongst our testbed targets. The nature of the compact object in this system remains controversial allowing a wider exploration of parameter space. As discussed in Section \ref{SubSect:4U170037}, we are unable to reproduce the joint constraints of binary properties and kinematics for this system, and we discuss lessons learnt from this. 

Further details of individual systems will be discussed in dedicated sub-sections further down. Our full sample study, to be presented in Paper II, will include the remaining compact object XRBs from \citetalias{Zhao2023}.

\begin{table}
{\fontsize{7}{8}\selectfont 
    \caption{Observed system characteristics of the six testbed targets}
    \begin{tabular}{lccccr}
    \hline
    Target &  \moneobs  & \mtwoobs  & \porbobs & $\vpec$ & Type \\
           &  (\Msun)  & (\Msun)  &  (days)  & (\kmps) & \\
    \hline
    \xtej\   & 7.5$^a$ & 0.18$^a$ & 0.17$^b$ & 143\,$\pm$\,11 & BH-LMXB\\
    \cyg\    & 21.2$^c$ & 40$^c$ & 5.6$^d$ & 22\,$\pm$\,\,3& BH-HMXB \\
    \jtwo\   & 1.64$^e$ & 0.35$^f$ & 0.19$^e$ & 64\,$\pm$\,13 & Binary Pulsar \\
    \btwelve\ & 1.4$^*$ & 23$^g$ & 1236.7$^g$ & 25\,$\pm$\,8 & Binary Pulsar \\
    \aqlx\   & 1.4$^*$ & 0.6$^h$ & 0.79$^h$ & 42\,$\pm$\,12 & NS-LMXB\\
    \gxthree\ & 1.85$^i$ & 43$^i$ & 41.5$^i$ &58\,$\pm$\,4 & NS-HMXB  \\
    \fouru\ & 1.96$^j$ & 46$^j$ & 3.4$^k$ &71\,$\pm$\,11 & NS-HMXB  \\
    \hline
    \end{tabular}
    ~\\
    Observed values for: compact object mass (\moneobs); donor mass (\mtwoobs); orbital period (\porbobs); and peculiar velocity ($\vpec$).\\
    References: (a) \citet{Khargharia2013}, (b) \citet{GonzalezHernandez2014}, (c) \citet{MillerJones2021}, (d) \citet{LaSala1998}, (e) \citet{Pletsch2015}, (f) \citet{Strader2019}, (g) \citet{MillerJones2018}, (h) \citet{MataSanchez2017}, (i) \citet{Kaper2006}, (j) \citet{Falanga2015}, (k) \citet{Islam2016}. \\
    $^*$: No literature value.\\ 
    $\vpec$ values come from \citetalias{Zhao2023}.
    \label{tab:sample}
}
\end{table}

\section{Methodology}\label{Sect:Methods}

\begin{figure*}
    \centering
    \includegraphics[width=0.9\textwidth]{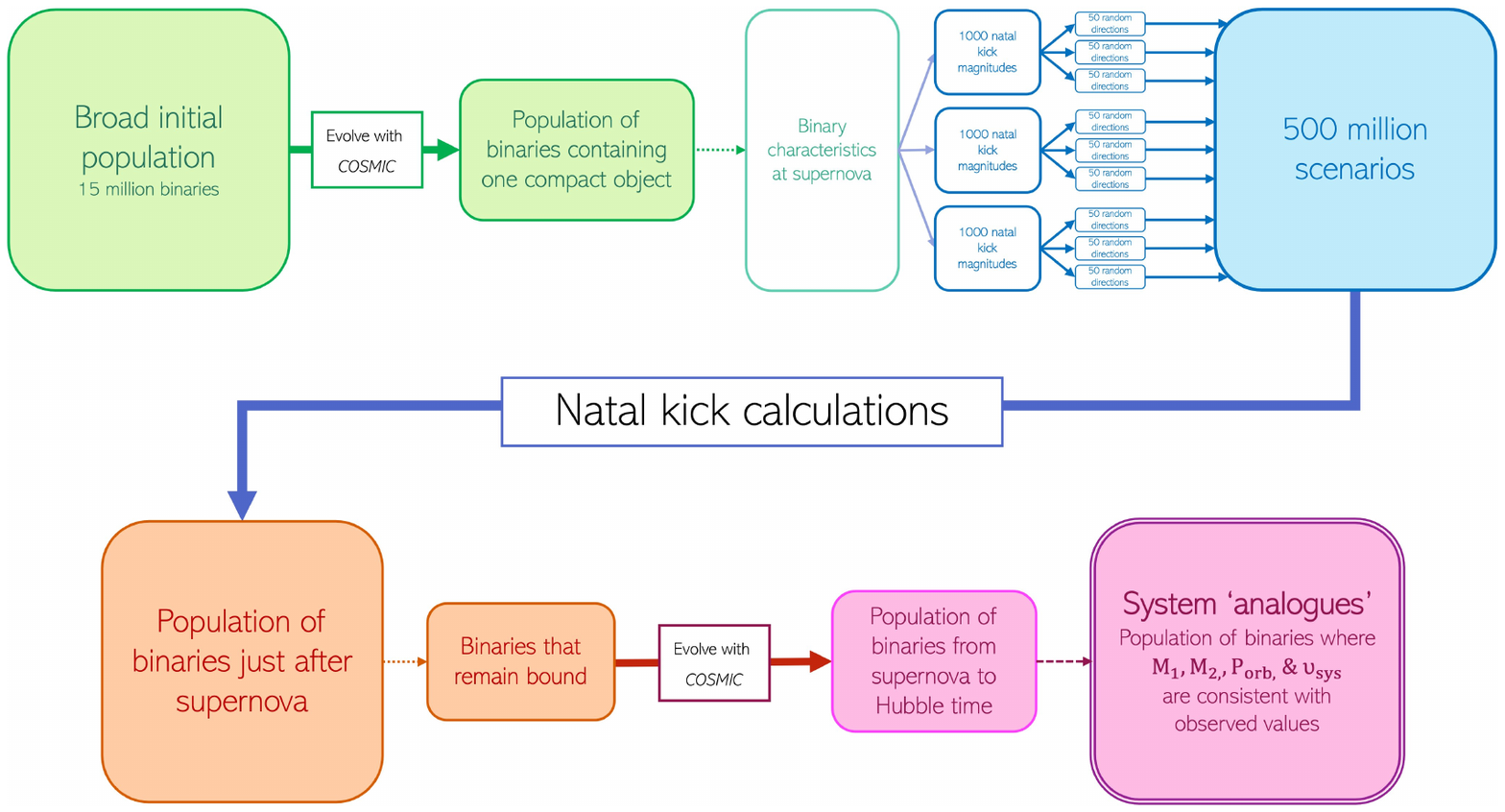}
    \caption{Flowchart outlining our method, as described in Section \ref{Sect:Methods}. We generate an initial population with the priors described in Table \ref{tab:priors}, and evolve using \cosmic\ (for further discussion of binary population synthesis and metaparameters, see Section \ref{SubSect:BSE}). We take the binary parameters (component masses, orbital period, and eccentricity) for each binary (that survives until the first supernova) and subject each one to 50,000 distinct natal kicks, spanning 0--1000\,\kmps, directed uniformly in solid angle. We calculate the resultant orbital separation, eccentricity, and systemic velocity of each binary, according to Equations \ref{eq:a_psn},\ref{eq:e_psn}, and \ref{eq:vsys_psn}. Surviving binaries (i.e. where $0\leq e_{\rm post-SN}<1$) are then evolved in \cosmic\ (simulating the post-SN evolution of each binary), until Hubble time (or the binary merger, whichever is first). This gives us a population of binaries, for which we have time-resolved properties. For each system in our sample, we search these simulated binaries to find those which are consistent with the system's: compact object type (black hole or neutron star); primary (compact object) mass; secondary (companion) mass; orbital period; and systemic velocity (assuming an intrinsic Galactic dispersion velocity). For the X-ray binaries in our sample (i.e. \xtej, \cyg, \aqlx, and \gxthree) we also require the binaries to be mass-transferring (either via RLOF or stellar winds). These are referred to as system `analogues' and represent plausible evolutionary pathways for each system, including information on the allowed natal kicks.}
    \label{Fig:MethodsSchematic}
\end{figure*}

\begin{figure*}
    \centering
    \includegraphics[width=0.8\linewidth]{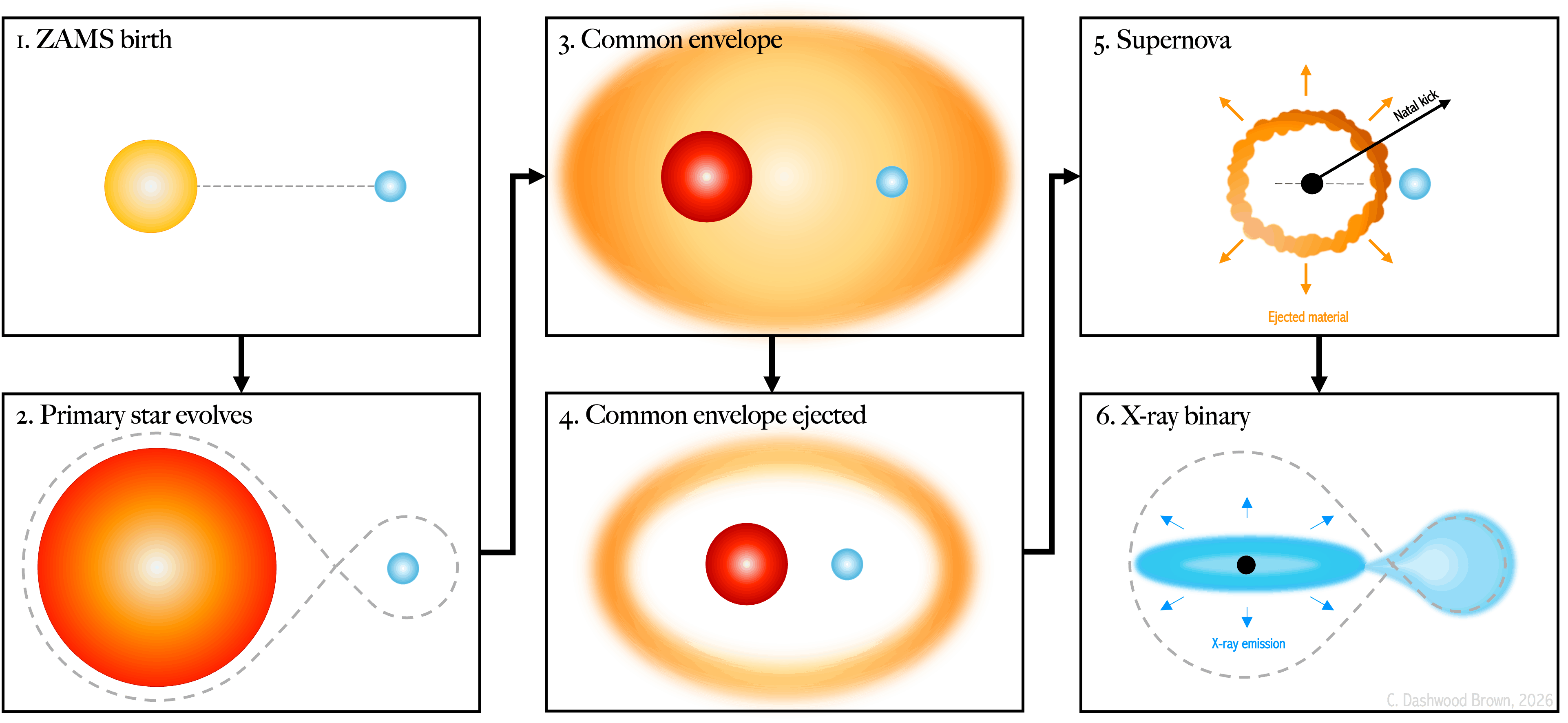}
    \caption{Schematic cartoon illustrating a plausible lifecycle for one of our sources, in this case a low-mass X-ray binary system such as \xtej. 1) A binary forms from a common cloud, (we consider isolated binary formation vs dynamical capture). 2) The primary (larger) star evolves and expands, superseding its Roche lobe. 3) Mass-transfer becomes dynamically unstable and the primary expands to engulf its companion (secondary) star. 4) Orbital separation shrinks, gravitational and orbital energy is transferred to envelope, which is then ejected. 5) The primary undergoes a supernova forming a black hole (or neutron star), and the ejected matter imparts a mass-loss kick to the binary, which may be accompanied by a natal kick imparted to the compact object. 6) The secondary (companion) star supersedes its Roche lobe and enters a period of stable mass transfer onto the compact object, resulting in X-ray emission as infalling matter loses gravitational potential energy.}
    \label{fig:LMXBTypical}
\end{figure*}

\subsection{Overview}\label{SubSect:MethodsOverview}

Our work utilises the population synthesis code \cosmic\ \citep{Breivik2020} which, in turn, is based on the binary population synthesis (Binary Star Evolution, or BSE; \citealt{Hurley2002}) formalism. Binary population synthesis utilises fitting formulae for single-star evolution models \citep{Pols1998, Hurley2000}, and folds these together with approximations for physical binary evolution processes derived from semi-analytic prescriptions. 

This approach facilitates speedy evolutionary simulations of populations from birth through to a Hubble time, which is a major advantage for covering a wide range of parameter space uniformly with a single code base. Conversely, this approach naturally also limits the veracity with which some of the underlying physics can be simulated. Below, we present a high-level overview of our simulation approach. For a definitive overview of the binary population synthesis code used in our framework (except for the generation of natal kicks described in Section \ref{sec:vnk_method}), we refer the reader to \citet{cosmic}. Our implementation of the code is discussed in Section \ref{SubSect:BSE}.

The key stages in our methods are as follows. We generate 15 million zero-age main sequence (ZAMS) simulated binaries, spanning a wide range of initial masses, orbital periods, eccentricities, and metallicities (see Table \ref{tab:priors} for these ranges). Each binary undergoes prescribed evolution within \cosmic\ up until the instant of first supernova. We are only interested in systems that have one, and only one, compact object in order to reproduce XRBs as seen today. Each of these binaries is then subjected to 50,000 natal kicks of varying magnitude and direction applied to the compact object. The resultant orbital separation, eccentricity, and systemic velocity are calculated according to Equations \ref{eq:a_psn}, \ref{eq:e_psn}, \& \ref{eq:vsys_psn} below (see \citealt{Kalogera1996,Brandt1995a}). \cosmic\ is then used to further evolve these systems beyond supernova until one of the following endpoints: (i) merger of the binary; (ii) disruption of the binary; or (iii) a Hubble time (whichever comes first). This yields a population of time-resolved binaries, from which we identify systems that resemble those within our X-ray binary sample (hereafter referred to as `analogues', see Sect \ref{SubSect:Matching}). These analogues give the range of plausible initial masses, orbital periods, eccentricities, metallicities, and natal kick magnitudes and directions, as well as providing further detail on their likely formation pathways (i.e. nature of interactions prior to the supernova (if applicable), degree of mass transfer, etc.). Figure\,\ref{Fig:MethodsSchematic} shows a schematic of our method, and Figure\,\ref{fig:LMXBTypical} summarises the typical evolution for a LMXB.  

Below, we outline some of the key prescriptions adopted in our work, both as part of \cosmic\ as well as additional customised assumptions required. One key aspect that we wish to highlight from the outset is that our purpose here is not to determine the precise evolutionary pathway followed by any given source. 

Instead, our primary focus here is to constrain the influence of {\em natal kicks}, given the landscape of potential pathways that the source could have followed. In a sense, we are effectively {\em marginalising} over all evolutionary parameters, in order to extract the allowed natal kicks for each system. We will present full distributions of allowed kicks as well as corner plots, enabling the reader to ascertain for themselves the relevant region of parameter space of interest, and highlighting cases where extra caution is warranted. 

\begin{table}
    \centering
    \caption{Summary of terminology used in this paper. All masses are in units \Msun; orbital periods in days (unless otherwise specified); separations in \Rsun (unless otherwise specified); velocities in \kmps; and angles in degrees.}
    \begin{tabular}{lcr}
    \hline
     Physical Parameter    &  Symbol\\
     \hline
     ZAMS mass (compact object progenitor)  & \monei \\
     ZAMS mass (companion progenitor)           & \mtwoi \\
     ZAMS orbital period                    & \porbi  \\
     ZAMS orbital separation                    & \sepi  \\
     ZAMS eccentricity                      & \ei\\
     Mass at onset of SN (compact object progenitor)  & \monepre \\
     Mass at onset of SN (companion progenitor)           & \mtwopre \\
     Orbital period at onset of SN                    & \porbpre  \\
     Orbital separation at onset of SN                    & \seppre  \\
     Eccentricity at onset of SN                      & \epre\\
     Post-SN mass (compact object)  & \monepost \\
     Post-SN mass (companion)           & \mtwopost \\
     Post-SN orbital period                    & \porbpost  \\
     Post-SN orbital separation                    & \seppost  \\
     Post-SN eccentricity                      & \epost\\
     Current mass (compact object)  & \moneobs \\
     Current mass (companion)           & \mtwoobs \\
     Current orbital period                    & \porbobs  \\
     Current eccentricity                      & \eobs\\
     Compact object mass                       & $\Mc$ \\
     Mass transfer rate                     & \mdot \\
     Mass ratio (=\,$M_2/M_1$)                & $q$ \\
     Relative velocity                      & $\vorb$ \\
     Natal kick magnitude              & $\vnk$ \\
     Blaauw (mass-loss) kick magnitude      & $\vmlk$  \\
     Polar angle of kick (see Figure\,\ref{Fig:NKAngles})  & $\theta$\ \\
     Azimuthal angle of kick (see Figure\,\ref{Fig:NKAngles}) & $\phi$\ \\
 Angle between $\vorb$ and centre of mass (see Figure\,\ref{Fig:NKAngles}) & $\Omega$ \\
     \hline
    \end{tabular}
    \label{tab:anysymbols}
    ~\\
    ...
\end{table}

\begin{table}
    \centering
    \caption{Assumed Prior Ranges for key Simulation Parameters at ZAMS}
    \begin{tabular}{lcr}
    \hline
     Parameter    &  Range  & Notes\\
     \hline
     \monei        & 1--200\,\Msun & Uniform\\
     \mtwoi        & 0.1--150\,\Msun & Uniform\\
     \porbi        & 0.01--10,000\,days & Uniform in log-space\\
     \ei        & 0--0.75 & Uniform\\
     Z & 0.05--2Z$_\odot$ (where Z$_\odot$=0.014) & Uniform\\
     $\vnk$\ & 0--1000 & Uniform\\
     $\phink$\ & 0--360\degree & Uniform\\
     $\cos(\thetank)$\ & -1--1 & Uniform in solid angle\\
     \manomaly\ & 0--360\degree & Uniform\\
     \hline
    \end{tabular}
    \label{tab:priors}
    ~\\
\end{table}

\subsection{Binary Star Evolution with \cosmic}\label{SubSect:BSE}

\cosmic\ is an open-source binary population synthesis suite that simulates the evolution of binary stars from their zero-age main sequence (ZAMS) properties through to the formation and potential merger of each component's stellar remnant. We used v3.6.1 of \cosmic, released July 2025. Populations can be initialised within \cosmic\ to either mimic empirically derived probability distribution functions of present-day stellar populations \citep[e.g.][]{Sana2012,Moe2017}, or be agnostically assigned uniformly over any given parameter range. Here, we choose the latter assumption (spanning ranges in Table\,\ref{tab:priors}) since our focus here is on individual systems rather than stellar populations. \cosmic\ simulates stellar evolution in discrete time steps that are dynamically chosen to ensure numerical stability and are limited such that key quantities like stellar mass and radius do not change by more than $1\%$ in each time step. The output for each binary can be specified in user-defined time step intervals as well as at key evolutionary stages like evolutionary stage change or at Roche-lobe over flow (RLOF) onset. Given binary properties at any one of these stages, \cosmic\ can simulate the forward evolution from that point onward to explore how different assumptions modify the future evolution. We use this feature to directly investigate how natal kicks alter the formation and evolution of XRBs.

We apply the default binary interaction tuning parameter assumptions, but include a description of key assumptions that alter the formation of compact object XRB binaries. Stellar winds are modelled based on theoretical predictions of radiation-driven mass loss (where mass-loss rates scale with metallicity depending on temperature) and Luminous Blue Variable mass loss \citep{Vink2001,Vink2005}. Wind velocities and accretion rates correspond to those employed by \StarTrack, as described in \citet{Belczynski2008}, and accretion rates are limited to 10x Eddington. Mass transfer processes follow those outlined by \citet{Hurley2002}. 
The critical mass ratio, $q_{\rm crit}$, which determines the onset of unstable mass transfer, is consistent with standard binary population synthesis and includes distinct conditions in the case of giant stars \citep{Hjellming1987,Hurley2002}. Stable mass transfer is limited by the thermal timescale of the donor star ($\min\left(10\frac{\tau_{\dot{M}}}{\tau_{\rm KHT}},1\right)$ for MS \& CHeB stars, where $\tau_{\dot{M}}$ is the mass transfer timescale and $\tau_{\rm KH2}$ is the thermal timescale of the secondary star).

Prescriptions for common envelope (CE) evolution and magnetic braking were initially kept constant in our core simulations, using canonical assumptions widely adopted in population modelling. The CE phase is based on the standard $\alpha\lambda$ prescription, characterised by an ejection efficiency parameter $\alpha_{{\rm CE}}\,=\,1.0$. The binding energy factor, $\lambda$, is calculated following \citet{Pols1995}. We consider a `pessimistic' CE scenario, in which mergers are unavoidable in the case of unstable mass transfer in which the core-envelope boundary is ill-defined. Magnetic braking adheres to the prescription of \citet{Ivanova2003} (this is only applicable to systems with low mass stellar companions). However, we also explored limited sets of additional simulations where these assumptions are systematically changed (see Section \ref{SubSect:DifferentParams}). Our main inferences do not appear to be strongly influenced by such changes, at least qualitatively. 

Supernova explosion mechanisms include standard core collapse, electron-capture supernova (ECSNe) at the low to intermediate mass end (with neutron star mass limited to 2.25\,\Msun in this case), as well as pair- and pulsational-instability supernova (PISN) above 30\,\Msun. Default prescriptions for these are based on \citet{Fryer2012, Hurley2002, Belczynski2016}. When considering fallback mechanisms, we adopt the `delayed' supernova prescription, as outlined in \citet{Fryer2012}. The mass of the proto-neutron star and the degree of fallback are both based on the core mass at the instant of supernova (i.e. systems with a core mass $\geq$11\Msun\ do not experience any mass loss; in this paper, this is only relevant for \cyg).  

Using the above framework allows us to explore the influence of a wide swath of initial binary parameter space. But it should be kept in mind that no single set of simulations can explore all possible evolutionary scenarios and formation channels, given the complexity and unknowns surrounding binary evolution. Key aspects of our later work will focus on a comparison between different classes of compact objects. Therefore, more important than full coverage of parameter space is a {\em representative} coverage, such that any biases inherent to the approach similarly impact black holes vs. neutron stars, for example. This will be discussed in greater depth in Paper II.

\subsection{Application of Natal Kicks}\label{SubSect:NKs}
\label{sec:vnk_method}

In this study, natal kicks are assumed to impact the compact object isotropically, in the sense that they are equally likely to occur in any given solid angle as seen from the source, and do not depend upon stellar characteristics; we note that the direction of natal kicks has been the subject of previous study (e.g. \citealt{Valli2025}), however there is not yet a clear consensus on which direction (if any), relative to the binary configuration and/or the progenitor star, is preferred. Here, we sample kick angles uniformly in solid angle ($\phink$ drawn from from a flat distribution $0\leq\phink\leq360^\circ$, $\thetank$ are generated from $\cos(\theta)$, drawn from a flat distribution $-1\leq\cos\left(\thetank\right)\leq1$).

\begin{figure}
    \centering
    \includegraphics[width=0.9\linewidth]{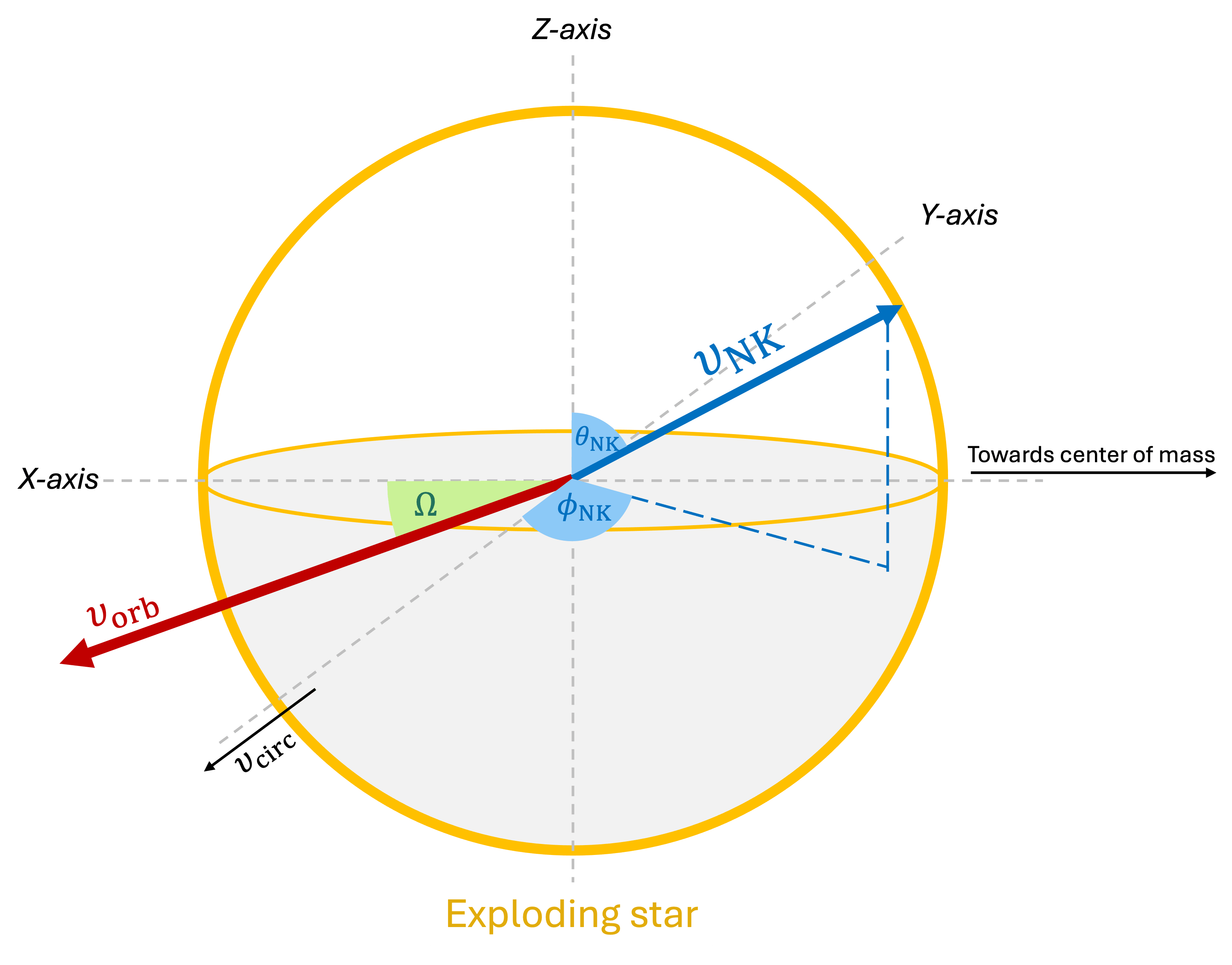}
    \caption{Diagram showing geometry of natal kicks in this study. The X- and Y- axes define the binary orbital plane: X is directed towards the centre of mass and Y is in the direction of orbital motion in the event of a circular orbit. The polar and azimuthal angles are $\thetank$ \& $\phink$, respectively. $\pi-\Omega$ gives the angle between the orbital motion and the centre of mass. Note: our geometry is defined differently than some previous studies.}
    \label{Fig:NKAngles}
\end{figure}

Individual binaries are subjected to kicks with magnitudes spanning the range of 0\,$\leq\vnk\leq$\,1000\,\kmps\ in steps of 1\,\kmps. Directional angles are sampled 50 times, so that each simulated binary is subjected to 50,000 distinct natal kicks intended to maximise coverage of parameter space and encompassing realistic magnitudes and directions, and agnostic to the underlying physical mechanism driving the kick. 

The input kick distribution does not depend on any system parameters (i.e. we do not assume any dependence on remnant mass, any mediation via fallback for black holes, etc). Whilst previous papers have found dependences on mass \citep[e.g.][]{Giacobbo2019,Bray2016,Mandel2020}, our choice of priors for natal kicks means we are insensitive to assumed kick mechanisms and dependencies. This is suitable for the purposes of our study, as we are focused on providing purely observational estimates of natal kicks. In that respect, our work may be considered a {\em test} of these dependencies, by confirming whether observationally inferred natal kicks are consistent with magnitudes which can be explained by current mass-dependent models. Uniform priors for natal kicks are also simpler to account for when fitting distributions to natal kick velocities (see Paper II). However, we acknowledge that this leads to the application of natal kicks with magnitudes that are unfeasible according to current conjecture. For instance, we consider the impact of 1,000\,\kmps natal kicks being applied to black holes - this is widely believed to be physically unreasonable when considering natal kicks as a consequence of conservation of momentum, and the current conjecture that black hole formation is associated with minimal mass loss \citep{Janka2017,Richards2023}. That being said, we note that some hydrodynamic models of supernova have determined black hole natal kicks up to $>$1,000\,\kmps \citep{Burrows2024b}. We hope that, in the future, our work can be considered jointly with various mass-dependent models to find a description of natal kicks that is rooted in both theory and observation.

Equations \ref{eq:a_psn}, \ref{eq:e_psn}, \& \ref{eq:vsys_psn} below describe the effect of a kick with magnitude $\vnk$ and in directional angles $\theta$ and $\phi$ on the post-SN orbital separation, eccentricity, and systemic velocity, respectively. This formalism is implemented in \cosmic, though using slightly different definitions for the natal kick angles - our geometry is defined in Figure\,\ref{Fig:NKAngles}.  

\begin{dmath}\label{eq:a_psn}
    a_{\rm post-SN} = \G \left(\Mc+M_{\rm 2, SN}\right) \left[\frac{2\G \left(\Mc+M_{\rm 2, SN}\right)}{r_{\rm SN}} - \vorb^2 - \vnk^2 - 2\vorb\vnk\left(\salpha\stheta\cphi - \calpha\stheta\sphi\right)\right]^{-1}
\end{dmath}

\begin{dmath}\label{eq:e_psn}    
    e_{\mathrm{post-SN}} =\sqrt{1-\frac{r_{\rm SN}^2\left[\vnk^2\left(\ctheta^2 +\stheta^2\cphi^2\right) + \calpha^2\vorb^2 + 2\vnk\vorb\salpha\stheta\cphi \right]}{\G\left(\Mc+M_{\rm 2, SN}\right)a_{\rm post-SN}}}
\end{dmath}

\begin{dmath}\label{eq:vsys_psn}    
    \vsys^2 = \left(\frac{\Mc}{\Mc+M_{\rm 2, SN}}\right)^2\vnk^2 + \left(\frac{M_{\rm 2, SN}\left(M_{\rm 1, SN}-\Mc\right)}{\left(M_{\rm 1, SN}+M_{\rm 2, SN}\right)\left(\Mc+M_{\rm 2, SN}\right)}\right)^2\vorb^2 - 2\frac{\Mc}{\Mc+M_{\rm 2, SN}}\frac{M_{\rm 2, SN}\left(M_{\rm 1, SN}-\Mc\right)}{\left(M_{\rm 1, SN}+M_{\rm 2, SN}\right)\left(\Mc+M_{\rm 2, SN}\right)}\vnk\vorb\left(\salpha\stheta\cphi-\calpha\stheta\sphi\right)
\end{dmath}

\noindent
The notation here is consistent with Table \ref{tab:anysymbols}: $M_{\rm 1,SN}$ \& $M_{\rm 2,SN}$ are the component masses at the onset of supernova; $M_{\rm CO}$ is the mass of the newly-formed compact object (equivalent to \monepost); $r_{\rm SN}$ \& $\upsilon_{\rm rel}$ are the orbital separation and relative velocity of the components at the onset of supernova; $\vnk$, $\theta$, \& $\phi$ describe the natal kick magnitude and direction; and $\Omega$ describes the angle between the relative velocity component and the centre of mass (i.e. 90\degree\ in the case of a circular orbit). Subscripts $s$ and $c$ denote the sine and cosine of angles respectively (e.g. $\theta_s = \sin \theta$, $\phi_c=\cos \phi$, etc.) and $G$ is the gravitational constant. 

These equations come from application of Kepler's Third Law and angular momentum conservation across the SN event, for a given explosive mass-loss of $M_{\rm 1, SN}-M_{\rm CO}$ \citep{Brandt1995a,Kalogera1996}. Equation\,\ref{eq:vsys_psn}, in particular, shows how any given natal kick is moderated by the total system mass and by the relative orbital velocity of the two binary components $\upsilon_{\rm rel}$ in determining the ultimate systemic velocity. This will be visualised and discussed in later sections.

\subsection{Matching to Observations}\label{SubSect:Matching}

Following the kick, binaries that survive (e.g. those that are not disrupted and those that do not immediately merge, 0\,$\leq$\,$e_{\rm post-SN}<1$) are then further evolved using \cosmic. For each of our systems, we then identify those specific simulations that end up matching the observed binary properties. Specifically, $\mone$, $\mtwo$ and $\porb$ need to {\em simultaneously} match their respective estimates from observations \citetalias{Zhao2023} to within quoted errors or 10\%, whichever is larger. The decision to include a lower limit on uncertainties is to mitigate the impact of differing degrees of precision to which different systems are characterised.

In addition, when matching the X-ray binaries (\xtej, \cyg, \aqlx, \& \gxthree), we require that simulated binaries are accreting at the time of the match. 
In practice, this means that the companion star is in a mass-transferring phase onto the compact object within \cosmic\ ($\dot{M}\geq10^{-12}\,\rm{M_{\odot} yr^{-1}}$): this may be in the form of 
RLOF or stellar winds; whilst these different mechanisms are distinguished in \cosmic, we do not consider the nature of the accretion when matching systems. For the binary pulsars (\jtwo\ \& \btwelve) we do not consider accretion. This means our simulations include binaries in which X-ray emission is due to pulsar activity, {\it and} those which are actively accreting - we allow the potential for analogues to be accreting given that some spider pulsars can have the potential to become transitional millisecond pulsars, sometimes behaving as LMXBs.

Our final match criterion is between the $\vsys$ of our simulations and the $\vpec$ estimate from observations. The assumption here is that the systemic motion resulting from the SN does not change dramatically over time. This is only an approximation; for example, systems may have slowed due to interactions with other celestial bodies, or sped up as they pass through deeper gravitational potential regions of the Galaxy. Accounting for these changes requires knowledge (or assumption of) a Galactic spatial model, introducing additional environment-related uncertainties specific to each target, which are beyond the scope of the present study. For a detailed discussion on the coupling of natal kicks and Galactic potentials we refer the reader to \citet{2025ApJS..276...16W} and \citet{2025AJ....170..192W}.

We do, however, account for the fact that Galactic stellar motions themselves have some intrinsic scatter beyond simple disc rotation. Observations have long established that stars have a non-zero velocity dispersion --- this dispersion is age-dependent and with a magnitude of up to $\sigma_\upsilon$\,=\,40\,\kmps\ \citep{Carlberg1985} in the disc. We account for this dispersion by subtracting $\sigma_\upsilon$ in quadrature from the $\vpec$\ estimates of each system. In a few high-mass XRBs, there is also observational evidence allowing the parent stellar cluster of the system to be identified (e.g., \cyg\ originated in the Cyg\,OB3 cluster and \fouru\ in the NGC\,6231; \citealt{Rao2020, vandermeij2021}). A correction for the systemic motion of the parent cluster is also applied, when known. Otherwise, $\sigma_\upsilon$ is assumed to be sufficient to account for environmental effects. This is likely still an overcorrection (since age effects are not accounted for), resulting in effective peculiar motions consistent with zero when the raw astrometric $\vpec$\,$<$\,40\,\kmps, so our estimates of natal kicks herein should be treated as conservative. 

Matches are only considered as viable if they occur following the first supernova and before any second supernova, or before a Hubble time elapses (whichever comes first), assuming the binary survives.

\subsection{Possibility of introducing additional constraints}

There are, of course, a multitude of other parameters that may be used to match simulated systems to observations, such as: eccentricity; age; characteristics of the luminous star (e.g. temperature etc.); location (e.g. Galactic scale height); local dispersion velocity; etc. We do not employ these here as general match criteria, because knowledge of these constraints can vary significantly for different systems (e.g. eccentricity is only known for 3 of the 6 systems we focus on). Further, for any binary which initiates RLOF, \cosmic\ circularizes the binary under assumption of efficient tides.

Previous studies have used additional constraints, such as; eccentricity \citep{Valli2025, Fortin2022, Wong2012,Willems2005}; spin-orbit misalignment \citep{Valli2025}; Galactic location (typically assuming formation in the Galactic plane) \citep{Willems2005,Fragos2009,Wong2012,Repetto2015}; age \citep{Willems2005,Fragos2009,Wong2012}; and donor properties \citep{Willems2005,Kimball2023}. These studies have used various methods, however the pre-supernova evolution of the binary is not considered jointly with the natal kick.

\section{Resultant Parameter Space}\label{Sect:Results}

Here, we summarise the results for our testbed targets, highlighting their salient properties, plausible evolutionary pathways and inferred natal kicks. For each source, we have created figures profiling their journey from birth to current state as a function of time, and discuss tracks of specific interest. Comprehensive overviews spanning the parameter space of successful simulations are shown as corner plots. These are presented in the flow of the main text for the first few systems, with all remaining figures available in the Appendix (\ref{Sect:AppA} \& \ref{Sect:AppB}). These figures should be referred to in conjunction with the text descriptions below. When quoting values for various parameters, the 68\% confidence intervals ($X_{68\%}$) refer to highest density intervals (HDI).

\subsection{\xtej}\label{SubSect:XTEJ1118}
As a luminous soft X-ray transient that erupted in 2000 \citep{Remillard2000}, \xtej\ has a short (4.1\,hr) orbital period and in quiescence revealed its cool (late K) donor star \citep{Torres2004, Gelino2006, Khargharia2013}, thereby allowing its orbital parameters to be well constrained (Table\,\ref{tab:sample}). A mass function $f(M)$\,$\approx$\,6\,\Msun\ confirmed the black-hole nature of the compact object \citep{McClintock2001, Wagner2001, Khargharia2013}. Optical proper motion measurements with {\em HST}, combined with distance estimates $d$\,$\approx$\,1.7\,kpc, placed the source in the Galactic halo, with a large space velocity likely originating in a supernova kick \citep{Mirabel2001}. 

\begin{figure*}
    \centering
    \includegraphics[width=0.95\textwidth]{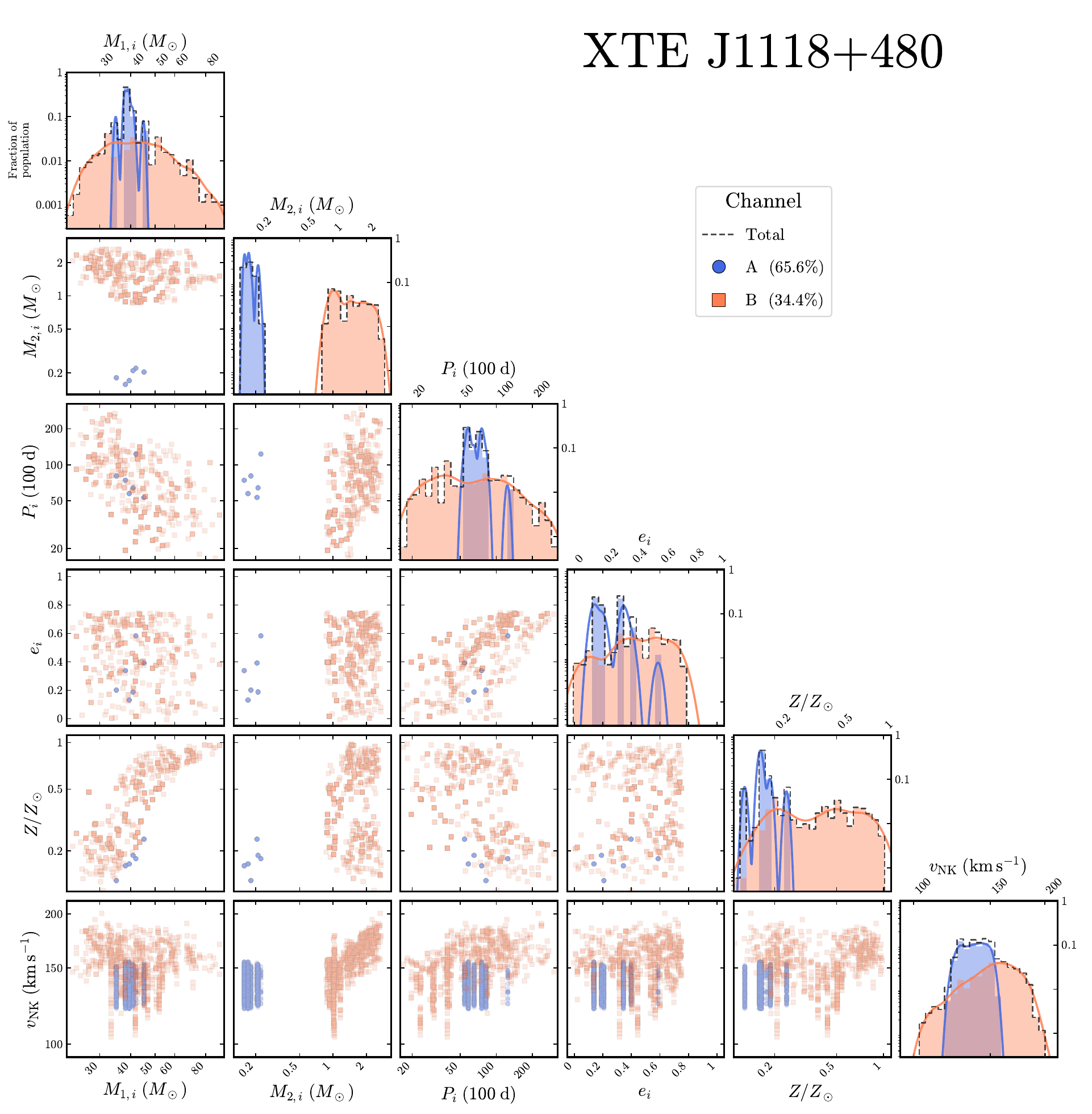}
    \caption{Corner plot highlighting key initial characteristics and acceptable natal kicks for simulated system analogues to \xtej. The ZAMS properties plotted include initial masses (\monei, \mtwoi), in units of Solar masses, orbital period (\porbi, per 100 days), eccentricity (\ei) and metallicity ($Z$ relative to Solar). The final parameter of interest plotted is the natal kick velocity ($\vnk$). Identical parameter space coverage for other systems is shown in later corner plots. For \xtej, two distinct evolutionary channels stand out in our simulations, differing primarily in terms of donor mass --- a low \mtwoi\ channel is plotted in orange and a high \mtwoi\ channel in blue, with respective proportions annotated at the top.}%
    \label{Fig:J1118_cornerplot}
\end{figure*}

\begin{figure*}
    \centering
    \includegraphics[width=0.95\linewidth]{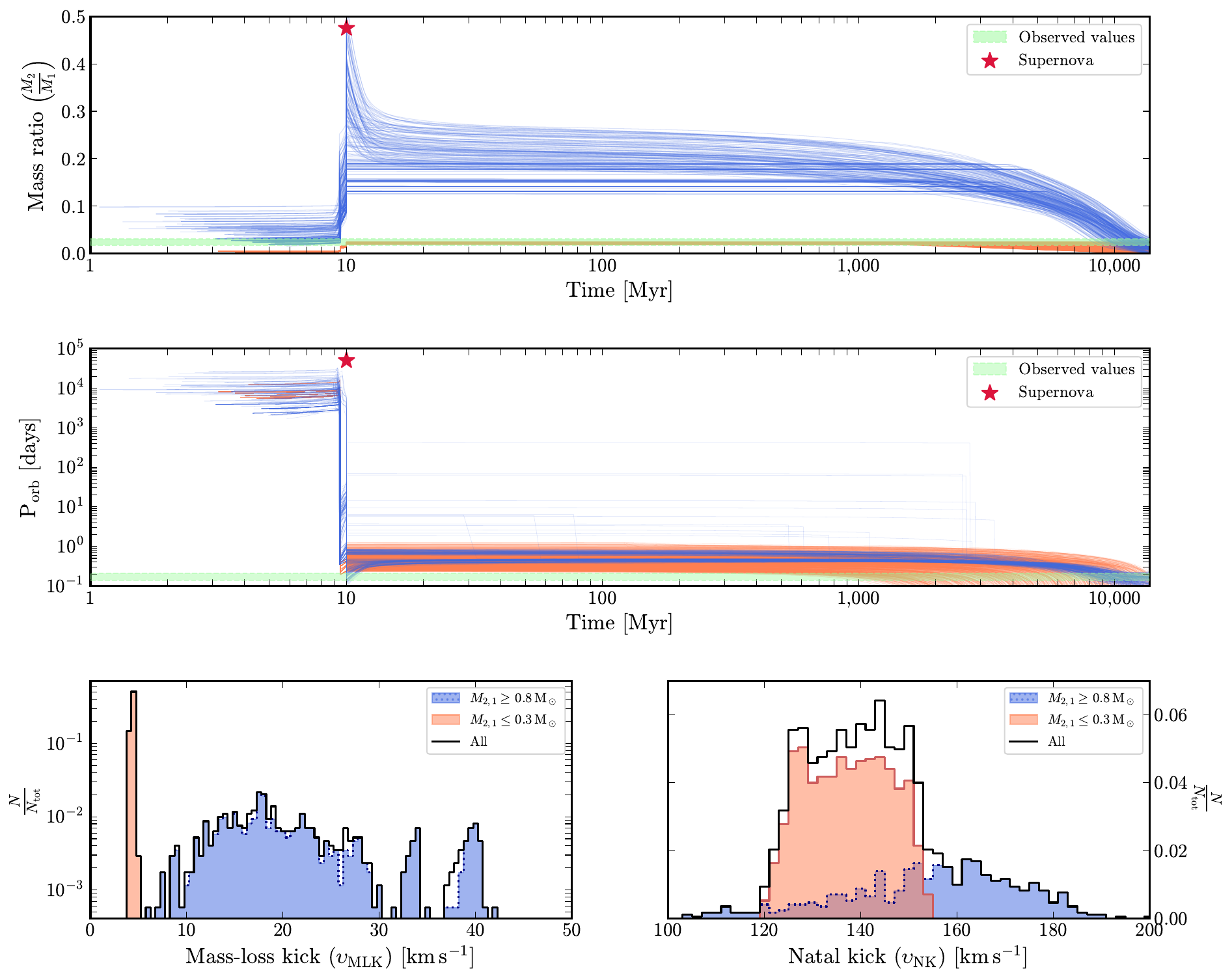}
    \caption{Simulated evolutionary tracks for all potential analogues of \xtej, depicted in terms of the mass ratio $q$ ($M_2/M_1$) ({\em top}) and orbital period $\porb$ ({\em middle}) shown as a function of time (here, time is shifted such that the supernova occurs at 10\,Myr, in order to ease comparison of all simulations), with the observed range of current values indicated in green. The panel on the lower right shows the mass-loss kicks $\upsilon_{\rm MLK}$ for the `high'- and `low'-mass channels (blue and orange, respectively). Finally, on the lower right is shown the distribution of natal kick magnitudes ($\vnk$) split by companion mass. The different evolutionary channels have different natal kick distributions; the `low'-mass channel corresponds to tighter constraints on natal kick magnitude, whereas the 'high'-mass channels allows for a broader range of values.}\label{Fig:J1118_evolution_typical}
\end{figure*}

\begin{figure*}
    \centering
    
\hspace*{-0.3cm}\includegraphics[width=1.02\linewidth]{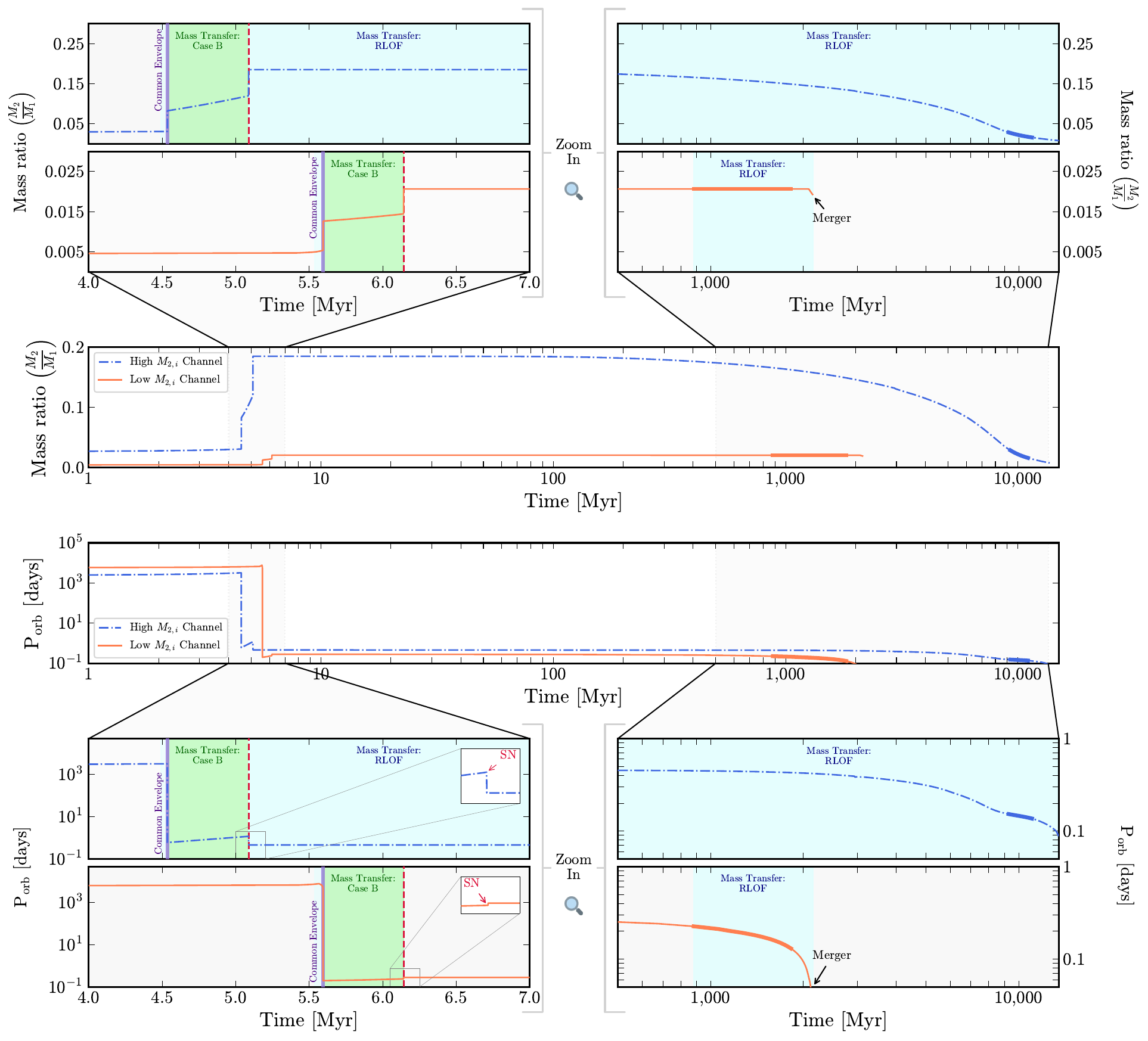}
    \caption{Selected detailed evolutionary tracks from each of the two channels identified for \xtej\ (`high'- (blue dashed line) and `low'- (orange solid line) companion mass). The two long panels show the mass ratio $q$ ({\em middle top}) and orbital period $\porb$ ({\em middle bottom}). The panels at the top and bottom are zoom-ins around early ({\em left}) and late ({\em right}) shaded periods of interest, with the SN, mass transfer and any mergers highlighted. The current range of values, relevant for matching to observations, are denoted by thicker portions of the evolutionary tracks during the late stages.}
    \label{Fig:J1118_evolution_focused}
\end{figure*}

Our simulations (corner plot in Figure\,\ref{Fig:J1118_cornerplot} and evolutionary tracks in Figures\,\ref{Fig:J1118_evolution_typical} \& \ref{Fig:J1118_evolution_focused}) find that \xtej\ follows an evolutionary pathway typical of LMXBs --- the system begins as an unequal-mass, widely separated binary, with \monei\,=\,20--70\,\Msun\ mass and \mtwoi\,=\,0--3\,\Msun\ (with a potential bimodality in the latter discussed below). The compact object progenitor evolves off the main sequence, expanding substantially, and superseding its Roche lobe. The extreme mass ratio results in unstable mass transfer via a common envelope, which rapidly shrinks the orbit. 
We note that in \cosmic, as in most binary population synthesis codes, common envelope phases are treated as instantaneous, with only the initial and final parameters being determined.

The ejection of the envelope leaves a naked helium star that has lost much of its mass, leaving $\simeq$40\,\% of its initial mass. There follows a further period of mass transfer --- in this case, Case B mass transfer, which occurs after the main-sequence phase \citep{Kippenhahn1967} --- until the primary undergoes a core-collapse supernova. With a core mass of $M_{\rm 1,c}$\,=\,7--9\,\Msun, mass loss is mediated by fallback such that the remnant retains 50--70\% of the progenitor mass, i.e., \monepost\,=\,7--8\,\Msun; \,\citet{Fryer2012}. 

Assuming the binary survives the imparted natal kick, what follows depends largely on the immediate post-SN orbit \porbpost\ and companion mass. If the post-SN orbit is sufficiently compact, or the radius of the companion already comparable to its Roche lobe, mass transfer can begin promptly. Alternatively, the binary could remain detached for an extended period before angular-momentum losses due to tidal forces and magnetic braking shrink the orbit and trigger mass transfer. The delay between compact-object formation and the onset of RLOF ranges anywhere between 10$^3$--10$^9$\,years. In many cases ($\approx70$\% of simulations), the binary merges within Hubble time. 

Simulated binaries fall broadly into two groups, distinguished by their initial companion mass; a `high-mass channel', where \mtwoi\,>\,0.8\,\Msun, and a `low-mass channel' where \mtwoi\,<\,0.3\,\Msun\ (Figures \,\ref{Fig:J1118_cornerplot}--\ref{Fig:J1118_evolution_focused}). In particular, Figure\,\ref{Fig:J1118_evolution_focused} illustrates the detailed time evolution of one typical simulation for each of the two channels, with mass-transfer periods and relevant physical processes highlighted. 

The example binary representing the `high-mass' channel begins with \monei\,=\,46.5\,\Msun, \mtwoi\,=\,1.2\,\Msun, \& orbital separation $a_i$\,=\,3200\,\Rsun, whereas the `low-mass' example has \monei\,=\,40.0\,\Msun, \mtwoi\,=\,0.2\,\Msun, \& $a_i$\,=\,4600\,\Rsun. The pre-SN evolution for both systems is similar; the expansion of the more massive star as it evolves leads to the system superseding it Roche lobe after $\sim$4.8\,Myr and $\sim$5.6\,Myr (the difference in time-frames being due to the slightly more massive primary evolving faster in the `high-mass' channel). The extreme mass ratio in both channels results in a common envelope phase, which is ejected leaving a naked helium star $M_1$\,=\,15.2\,\Msun\,or 13.7\,\Msun, and binary separation $a$\,=\,6\,\Rsun\,or 3.5\,\Rsun\, for high- and low- mass channels respectively ($M_2$ is unchanged to within 0.1\%).

These stars continue to evolve and lose mass, the orbital separation increases slightly to compensate for the change in orbital angular momentum (the companion star gains only a small amount of mass during this time period through Bondi-Hoyle wind accretion). The supernova occurs at $\sim$5.3\,Myr or $\sim$6.1\,Myr (again, depending on the mass of the primary star). 

At the instant of supernova, the two example binaries have masses $M_{\rm 1,SN}\approx$11--12\,\Msun, with core masses $M_{\rm 1,c}\approx$9--10\,\Msun. According to the calculations laid out in \citet{Fryer2012}, the supernova results in the formation of a proto-neutron star of mass 1.4\,\Msun, which forms a black hole due to $\sim$70\% of the ejected material ($\sim$7\,\Msun) falling back (this gives the baryonic mass of the remnant; the gravitational mass is reduced by the radiation of neutrinos).

Both the `high-mass' and `low-mass' examples receive natal kicks of 150\,\kmps\, (and both are imparted to within 15\degree\,of the orbital plane). This additional momentum increases the systemic velocity of the binary to 130--150\,\kmps\, --- consistent with the present day peculiar velocity. The kicks introduce some moderate eccentricity to the orbit in both cases ($\sim$0.1--0.4); the orbit of the `high-mass' example is reduced by the natal kick, whereas the orbital separation of the `low-mass' example is relatively unchanged (highlighting the complex and sometimes unintuitive influence of natal kicks on system parameters). 

Following the supernova, the `high-mass' example immediately begins mass transfer via Roche lobe overflow. The binary circularises quickly, slightly reducing the orbital separation (from 7.5\,days to 5.2\,days). Mass-transfer ranges $\dot{M}\sim10^{-12}-10^{-9}\,\rm{M_{\odot}}yr^{-1}$, with the companion star losing half its mass within a few Myr, and the orbital period continually shrinking. The binary resembles current observations around 10,000\,Myr after ZAMS.

The `low-mass' channel example does not interact directly after supernova. Instead, tidal forces circularise and shrink the binary orbit, and Roche lobe overflow begins around 2000\,Myr following the supernova. This period of mass transfer is short-lived, as the binary merges within $<100$\,Myr of RLOF onset, with mass transfer rates peaking at $\dot{M}\approx10^{-9}\,\rm{M_{\odot}}yr^{-1}$.

Note that it is not always the case that the `high-mass channel' results in immediate mass transfer post-SN, though this is more likely given that larger mass and therefore radius of the companion star. Nor is it the case that a natal kick which increases the orbital period precludes a system from immediately undergoing Roche lobe overflow (though again, it makes it less likely). Figure\,\ref{Fig:J1118_evolution_focused} also shows that systems which begin mass transfer immediately after the SN are not guaranteed to resemble \xtej\ any sooner than those which remain detached for some time. 

Figures\,\ref{Fig:J1118_cornerplot} and \ref{Fig:J1118_evolution_typical} demonstrate that different pathways may be associated with systematically different natal kicks though there is formally strong overlap between the respective distribution; for example, the `low-mass channel' favours a tighter range of natal kicks, centred on $\approx$\,140\,\kmps, whereas the `high-mass channel' appears to be less sensitive to natal kicks, allowing for a wider distribution. The relative likelihood of these different channels results in the overall distribution of inferred kicks exhibiting multi-modal peaks and skews. 

We note here that the number of possible progenitors for \xtej\ in the low-mass channel is very small (accounting for less than 1\% of all simulations). However, these binaries are viable when subjected to a large range of natal kicks, meaning they account for a greater proportion of all simulations. When analysing our results, we use the natal kick distribution from all simulations combined, without weighting by progenitor likelihood --- this is because we are trying to mitigate the influence of binary evolution physics on the observational results. However, one could weight possible natal kicks based on the perceived likelihood of a given progenitor/evolutionary channel (determined by whichever criteria one chooses) to further refine the estimate for the actual natal kick which \xtej\ received.

The natural response to these distinct channels is to wonder why it is that systems with intermediary companion masses (i.e. $0.3\leq M_{2,i}\leq0.8$\,\Msun) are apparently not viable for \xtej. These two channels correspond to two distinct post-evolution pathways, in which changes in orbital period are governed by different processes. For the system to be transferring mass via Roche lobe, the orbital separation must be sufficiently small (at least at periastron). Conservative mass transfer has the effect of widening the orbit; shrinking the orbit to the values observed today requires efficient loss of orbital angular momentum, either via magnetic braking (MB) or gravitational wave emission (GW).

Where $M_2<0.3$\,\Msun, the post-supernova orbit may shrink due to a combination of angular momentum loss via magnetic braking and gravitational wave emission --- this orbital shrinking may occur for orbital periods $\lesssim1$\,day (consistent with our simulations), and the system may evolve to match observations. Systems with $M_2\geq0.7$\,\Msun\,can shrink due to magnetic braking alone across a wide range of orbital separations --- this is the region in which magnetic braking is most efficient; that is until the `Kraft break' (1.2--1.5\,\Msun), after which orbital decay efficiency decreases again. Our results show that systems with $M_2\gtrsim1.2$\,\Msun\,begin interacting directly following their formative supernova; they require tighter orbits to interact given that they cannot efficiently reduce their separation. It is worth noting that, given the orbital shrinking occurs over very long timescales ($10^8-10^{10}$\,years), even small variations in shrinkage efficiency can render a system unviable.

Where the secondary mass is between 0.3--0.7\,\Msun, both magnetic braking and gravitational wave emission are too weak to shrink the orbital separation on astrophysical timescales to allow for the onset of mass transfer (if the system is too wide to interact directly after the supernova, i.e. $P_{\rm orb, post-SN}\gtrsim$ a few days). Alternatively, the natal kick may change the orbit such that the system is immediately interacting post-SN, though this appears to be less likely; our simulations find that $>$90\% of systems are too widely separated after the supernova to interact, and of those which are more tightly separated, $\sim85$\% have systemic velocities $\vsys\geq150$\,\kmps, which is inconsistent with observations of \xtej. For the few systems that do end up in this region of parameter space, mass transfer rates are low enough that the orbital evolution remains driven by gravitational wave emission. Tidal effects work to try and synchronise the spin of the donor star with the orbital period, resulting in an angular momentum sink leading to run-away mass transfer and merger; before the companion has been sufficiently stripped down to $\leq0.2$\,\Msun.

Whether this gap in initial companion masses is genuine, or an artifact introduced by \cosmic\ (see Section \ref{SubSubSect:BSEMESA}) is unclear, however, the inclusion of plausible progenitors in this regime would be unlikely to significantly impact the natal kick distribution for \xtej. If all other progenitor parameters are broadly similar to the successfully simulated analogues (i.e. black hole progenitor masses, pre-SN orbital periods, etc.), the allowed natal kick for the `intermediate'-mass channel would be in line with results for the `low'- and `high'-mass channels (i.e. $\sim$100--200\,\kmps).

\xtej\ likely received a natal kick 130--160\,\kmps\ ($\upsilon_{\rm NK,68\%}$, within the range presented by \citet[][80--310\,\kmps]{Fragos2009}). Our tighter constraint partly stems from the improved \gaia\ astrometry and other observational data that have since become available, leading to more stringent constraints on peculiar velocity (i.e. updated distance estimate from $d=1.85\pm0.36$\,kpc to $d=1.72\pm0.1$\,kpc). Additionally, our methodology includes simulating the system's evolution prior to the supernova, which has the effect of further limiting plausible evolutionary channels.

\subsection{\cyg}\label{SubSect:CygX1}

Cygnus\,X--1 (or Cyg\,X--1 for short) is generally renowned as the first Galactic black hole system, and has a bright donor star (HDE\,226868; \citealt{Webster1972}). Unlike \xtej, the donor in this case is a massive, early-type star, making \cyg\ the prototypical black hole HMXB. The system is not eclipsing, and accordingly there remains ongoing controversy as to the precise inclination and compact object mass. Early work by \citet{Orosz2011} estimated a BH mass of $14.8\pm 1.0$\,\Msun. \citet{MillerJones2021} derived a mass of $21.2\pm 2.2$\,\Msun\,with an updated distance estimate from {\em Gaia} and VLBI astrometry; these are the values we adopt in this study. It must be noted, however, that recent evaluations of high-resolution optical spectroscopy of the donor give slightly lower masses (\moneobs\,=\,12--18\,\Msun\,\& \mtwoobs$\approx$29\,\Msun), albeit still covering a wide range \citep{Ramachandran25}. Estimates of the system age, together with the direction and magnitude of its centre-of-mass motion, point to Cygnus\,OB3 being the parent stellar cluster from which the source is now receding \citep{Rao2020}. 

\begin{figure*}
    \centering
    \includegraphics[width=00.95\textwidth]{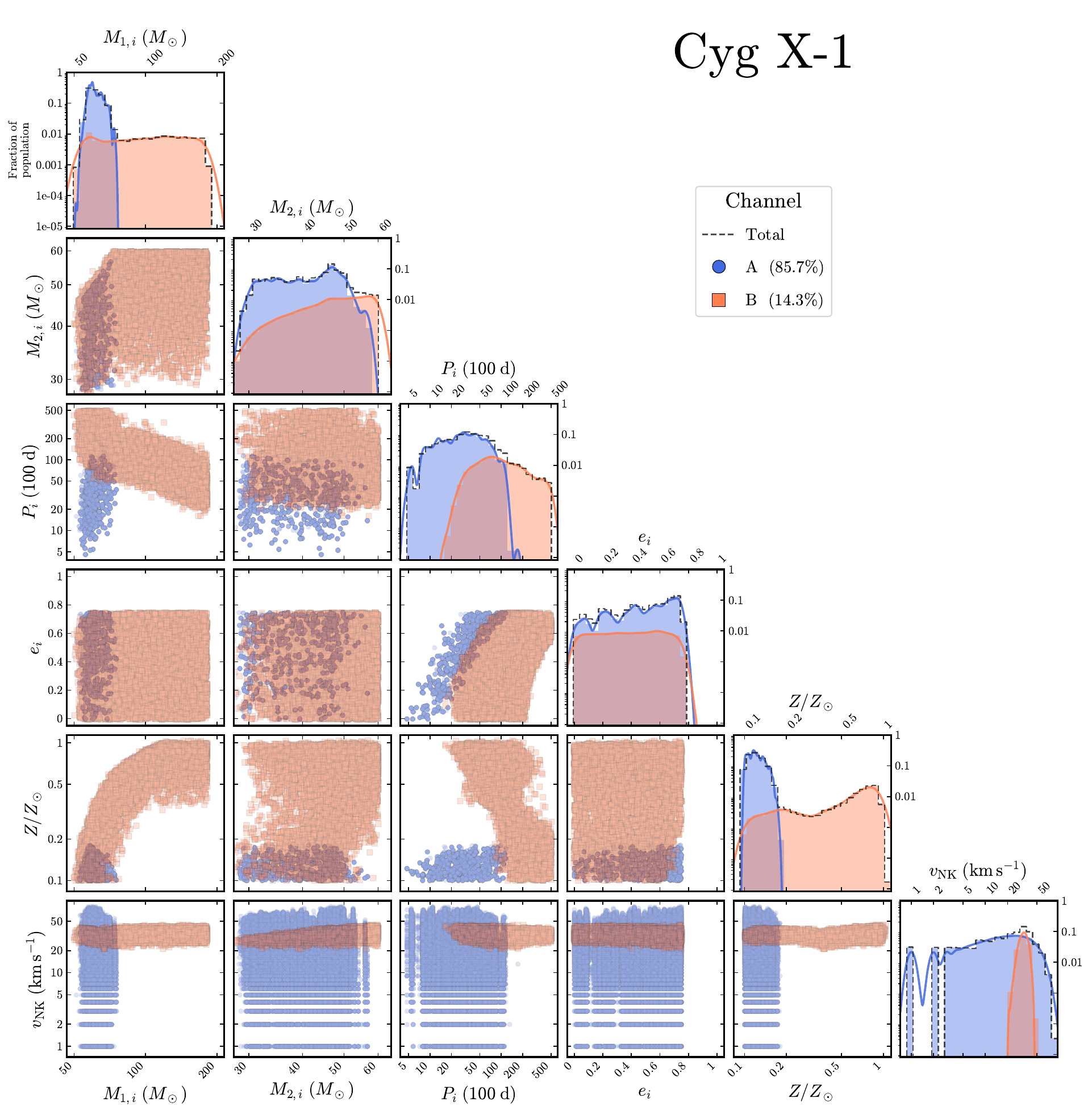}
    \caption{Corner plot highlighting key initial characteristics and acceptable natal kicks for simulated system analogues to \cyg. In this case, distinct evolutionary channels are indicated by colour and marker style; blue circles refer to simulations that undergo a CE phase before supernova, whilst orange squares are those simulations that do not. Refer to Figure\,\ref{Fig:J1118_cornerplot} for further details. Whilst the non-CE channel allows for a broader range of initial binary characteristics, the CE channel dominates in number of successful analogues.}
    \label{Fig:CygX1_cornerplot}
\end{figure*}

\begin{figure*}
    \centering
    \includegraphics[width=0.95\linewidth]{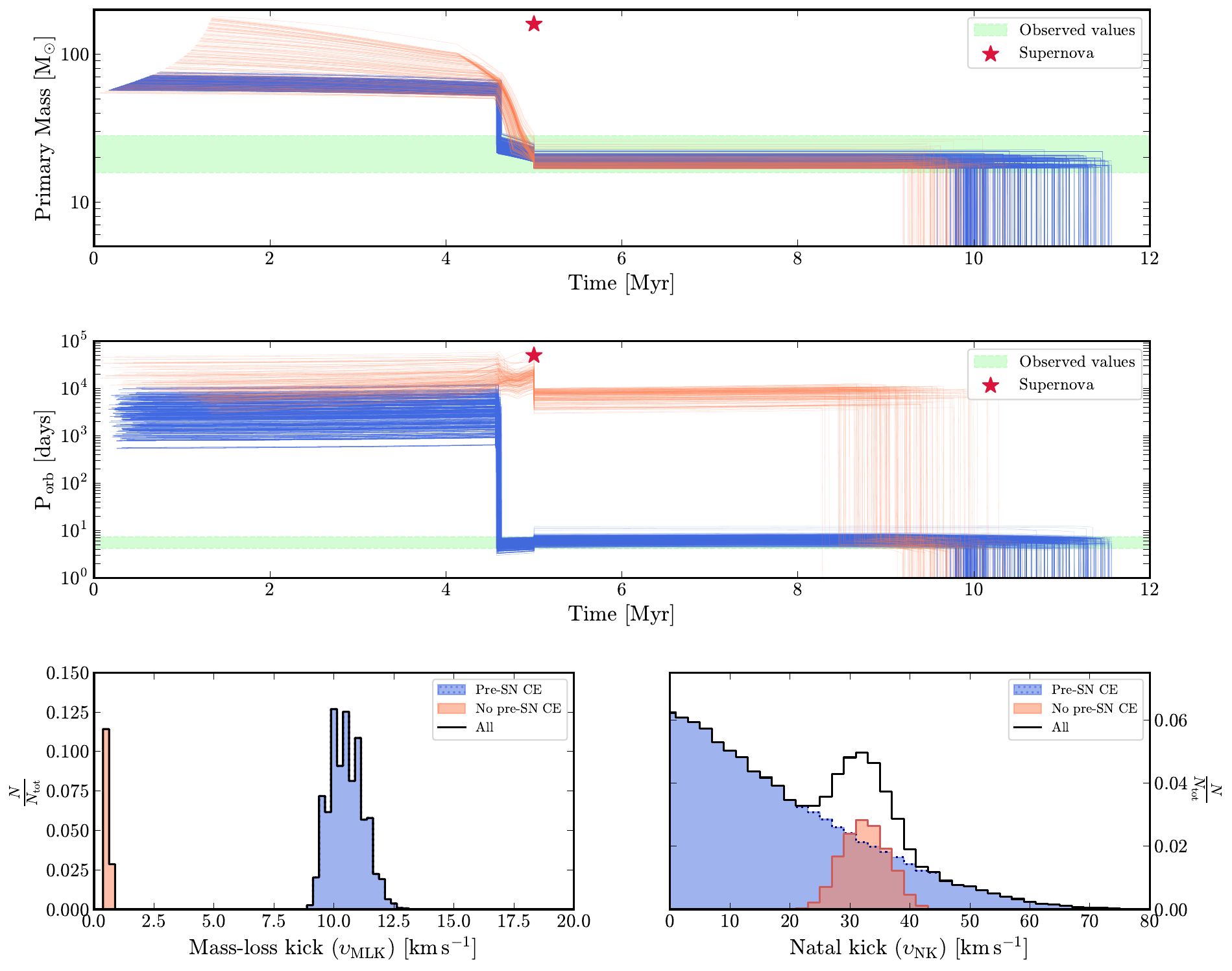}
    \caption{Summary of the likely evolution of \cyg. Primary mass (top) and orbital period (middle) as a function of time for all potential analogues (time is normalised such that the supernova occurs at 5\,Myr, in order to ease comparison of all simulations), with the observed range of current values indicated in green. The two channels distinguished here are based on whether the system undergoes a CE phase prior to supernova (blue), or not (orange). Other panels as in Figure\,\ref{Fig:J1118_evolution_typical}. Again, the two evolutionary pathways have distinct natal kick distributions, with a much narrower range of allowed kicks for the non-CE channel - this is due to higher likelihood of disrupting the system (given its orbital period is much larger at the instant of supernova) requiring a more finely-tuned kick to keep the system bound and impart a substantial eccentricity to allow for future interactions.}
    \label{Fig:CygX1_evolution_typical}
\end{figure*}

\begin{figure*}
    \centering
    \includegraphics[width=0.95\linewidth]{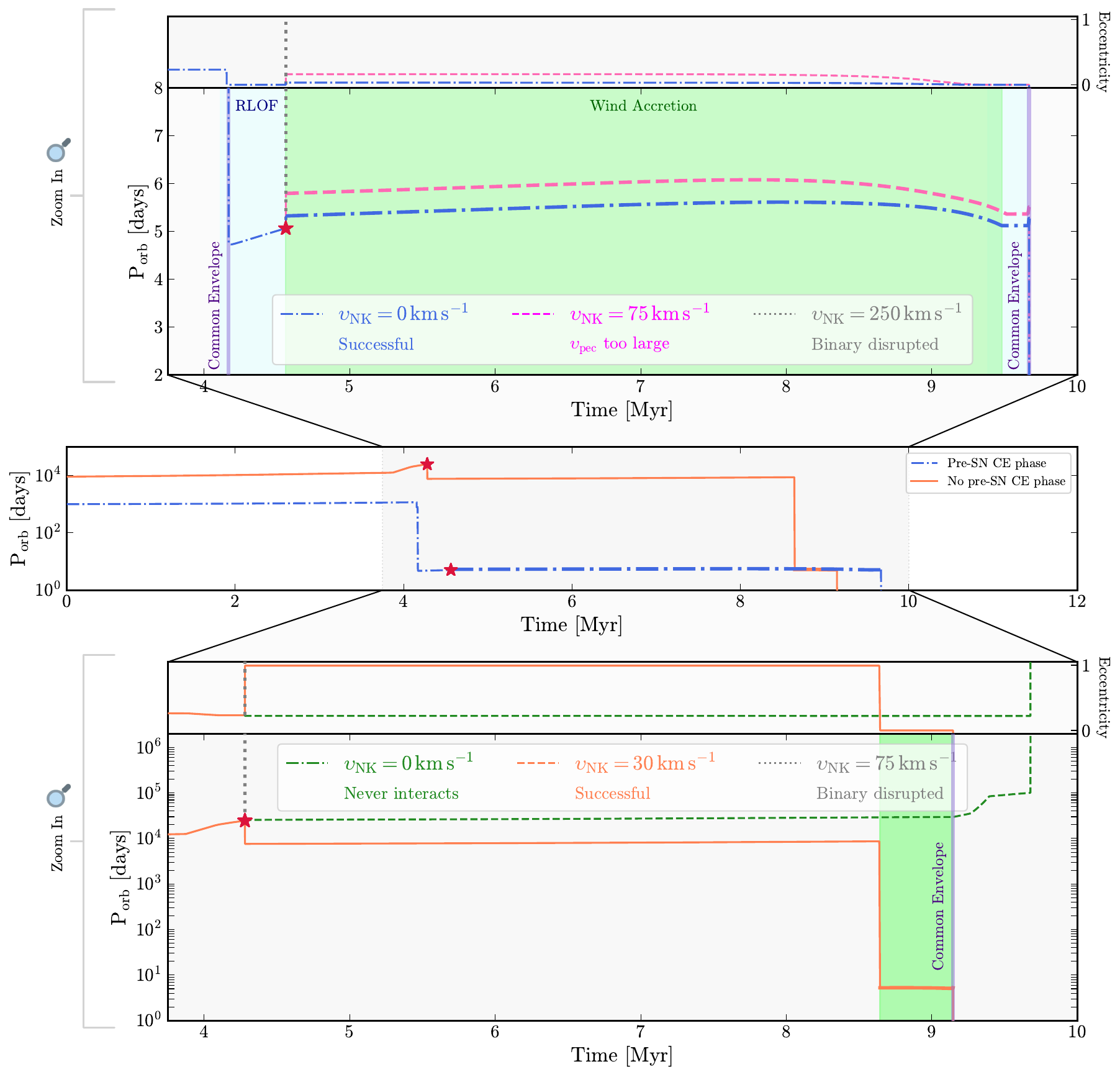}
    \caption{Detailed evolution of orbital period $\porb$\ as a function of time for two putative \cyg\ analogues --- one of which underwent a CE phase before supernova (blue) and one that bypassed this (orange). The full history is shown in the middle panel, including one simulation from each channel. The upper and lower panels are zoomed in versions of this central panel, each focusing on a different channel. {\em Upper panel:} Zoom-in on the CE channel, showing orbital period and eccentricity as a function of time. The blue dash-dot line shows the simulated system; the effect of infeasibly high natal kicks are shown in pink (dashed) and grey (dotted). {\em Lower:} Zoom-in on a system that bypassed the CE phase. Different phases of evolution, such as common envelope, wind-driven accretion, and Roche lobe overflow are indicated by shaded regions. Again, three separate outcomes are shown, depending on natal kick strength. Thicker portions of the evolutionary tracks indicate times where the orbital period is consistent with current observations. This figure demonstrates the importance of natal kicks on both the systemic velocity of a system and its orbital configuration.}
    \label{Fig:CygX1_evolution_focused}
\end{figure*}

The most frequent evolutionary pathway for \cyg\ as simulated by \cosmic\ sees the binary beginning its life with an initial mass ratio $q_{\rm init}$ of 0.4--0.8, where $50\leq{\rm M}_{1,i}\leq150$\,\Msun\ and $30\leq{\rm M}_{2,i}\leq60$\,\Msun, and an orbital separation 1000\,$\leq$\,$R_{\rm i}$\,$\leq$\,30,000\,\Rsun\ (corresponding to orbital periods 100\,$\leq$\,$P_{\rm orb,i}$\,$\leq$\,50,000\,days) (Figure\,\ref{Fig:CygX1_cornerplot}). The more massive star expands as it evolves, exceeding its Roche lobe, and the system is engulfed in a common envelope, resulting in rapid mass loss/transfer and a shrinking of the orbital period from a few thousand days to less than 10 days. The primary continues to transfer mass through stable Roche lobe overflow and eventually undergoes a core-collapse supernova, leaving behind a black hole with \moneobs\,$\approx$\,20\,\Msun\ (see Figure\,\ref{Fig:CygX1_evolution_typical}). 

According to the core collapse prescriptions laid out in \citet{Fryer2012}, systems with Carbon-Oxygen core masses $M_{\rm c}\geq11$\,\Msun\ will experience complete fallback, resulting in zero effective ejecta from the supernova, and thus no natal kick above the Blaauw recoil. The conversion from baryonic to gravitational remnant mass still removes mass from the binary, typically parameterised as a fractional binding-energy loss of $\sim 10$ per cent of the remnant baryonic mass for black holes. At the same time, it is also generally considered physically implausible for a system to radiate more than $\sim$\,0.5\,\Msun\ as neutrinos. Therefore, for \cyg, we run additional simulations in which the mass lost via conversion is limited to 0.5\,\Msun. The inferred natal kicks for \cyg\ show minimal variation between the two scenarios, and the results we present here include both sets of analogues for completeness. 

The system continues to interact via stellar winds; the donor star evolves, eventually expanding, and the system is expected to merge before the secondary undergoes a supernova (within 5--10\,Myr). This differs from the predictions of other studies, which propose \cyg\ will form a double black hole binary and become a source of gravitational waves. This may be, in part, due to the fact that codes using \textsc{MESA} typically predict lower merger rates across simulations compared to BSE type codes.

There exists an alternative channel, in which the binary avoids a common envelope phase prior to the first supernova, instead interacting via stable mass-transfer or not interacting at all, remaining widely separated up until supernova. Such wide orbits are more readily disrupted by natal kicks, however a finely tuned kick (favouring $20\leq\vnk\leq45$\,\kmps and $150\leq\phink\leq210$\degree) may result in a highly eccentric orbit, leading to the interaction of the black hole and the large companion star at periastron, resulting in a shrunken orbit and the X-ray binary configuration we observe today. This channel is considerably less likely (accounting for fewer than 1 in 10 simulations), and although a wider range of initial masses and larger orbital periods are allowed in this regime, it requires natal kicks to be within a narrow range, and directed almost entirely opposite the orbital motion of the progenitor binary. Additionally, binaries evolving along this channel match observations at later times (typically $\gtrsim8$\,Myr compared to 4-5\,Myr for the CE-channel).

Figure\,\ref{Fig:CygX1_evolution_focused} shows the evolutionary path followed by a putative \cyg\ progenitor impacted by a variety of natal kicks. Natal kicks of $\leq$\,35\,\kmps are favoured to reproduce observations (when considering the overall distribution of successful natal kicks, and not weighting based on progenitor likelihood, kick direction, etc). Stronger natal kicks may result in higher systemic velocities than observed, or inconsistent orbital periods (one should note that higher natal kicks do not always lead to wider separated orbits, as shown in Figure\,\ref{Fig:CygX1_evolution_focused}). For $\vnk\geq200$\,\kmps, the binary is disrupted, forming an isolated black hole and a high mass run-away star. 

The distribution of natal kicks for \cyg\ spans $\vnk$\,=\,0--70\,\kmps, with two peaks at $\leq10$\,\kmps\ and at $\sim30$\,\kmps\ (corresponding to the two channels discussed above). This is in agreement with previous work (e.g., \citealt{Wong2012} reported a range of $0\leq\vnk\leq60$\,\kmps). The dominant channel includes a CE phase, outnumbering the non-CE channel in terms of simulated analogues by about 9:1. 

\subsection{\jtwo}\label{SubSect:J2339}

Discovered as a Fermi $\gamma$-ray source \citep{Abdo2010}, \jtwo\ is a redback millisecond pulsar in a 4.6\,hr orbit and consists of a rapidly (2.9\, ms) spinning neutron star and a low-mass ($\approx 0.3\,\mathrm{M_\odot}$) companion \citep{Romani2011, Kong2012, Strader2019}. 

The evolution of \jtwo\ is largely governed by the interchange between two different mass transfer mechanisms: Roche lobe overflow and strong stellar winds. A wide orbit binary (period $\sim1,000-10,000$\,days) with a primary mass $10\leq$\,\monei$\,\leq30$\,\Msun\, and a much smaller companion (see Figure\,\ref{Fig:J2339_cornerplot}) experiences a common envelope (initiated by the expansion of the primary as it evolves) which shrinks both the orbital period and the mass of the compact progenitor. This is followed by a period of stable Roche lobe overflow, culminating in the supernova of the $\approx$3--5\,\Msun\,progenitor. Following the supernova, the companion/donor star loses mass due to stellar winds, after which the system goes into another common envelope phase. This results in a very short period binary with a sub-solar mass companion; colloquially known as a `redback'. The predicted natal kicks for \jtwo\ range 0--120\,\kmps, peaking between 20--60\,\kmps (Figure\,\ref{Fig:J2339_khist}). 

\subsection{\btwelve}\label{SubSect:B1259}
\btwelve\ is one of the gamma-ray binaries containing a neutron star \citep{Johnston92} and a massive companion \citep{Negueruela2011}. The orbit is wide (orbital period $\sim$3.5\,yrs) and highly eccentric \citep{Shannon14}. It is relatively nearby (2.3\,kpc \citealt{Negueruela2011}) and has an early type (O9.5Ve) donor, but the mass of this component which is only poorly constrained. \citet{MillerJones2018} estimate a donor mass of 23\Msun, which we use in our analogue matching.

The neutron star likely started as a circa 20\,\Msun star co-existing with another star of a similar mass with an orbital period of a few hundred days (see Figure\,\ref{Fig:B1259_cornerplot}). The expansion of the primary results in a period of mass transfer via Roche lobe overflow. The compact object progenitor decreases to $\sim$5\,\Msun, the companion star increases in size, and this is accompanied by a slight increase in orbital period. At the instant of supernova, \monepre\,$\approx$\,4\,\Msun, \mtwopre\,$\approx$\,20--30\,\Msun, and \porbpre\,$\approx$\,2000--5000\,days. The compact progenitor loses approximately 2--3\,\Msun, and the orbital period remains largely similar (within the same order of magnitude). As the system evolves and the companion experiences mass loss via strong stellar winds, the orbital period evolves to match that which is observed today. That the system may have evolved from a tighter period binary directly after supernova to the wide period system observed today negates the need for a strong natal kick to substantially increase the orbital separation (though a substantial eccentricity, like that observed for \btwelve\ ($\sim$0.87) is more likely if kicks are on the higher end). The supernova of the companion star will, in all likelihood, disrupt the binary, leaving behind an isolated neutron star and isolated black hole. 

Its $\vpec$ is uncharacteristically low when compared to the broader population of pulsars. Our simulations imply a low-to-moderate natal kick velocity, $12\leq\vnk\leq50$\,\kmps, and the system may be consistent with no natal kick at all (see Figure\,\ref{Fig:B1259_khist}).

\citet{Fortin2022} calculated a moderate natal kick (ranging between 30--100\,\kmps). Figure\,\ref{Fig:B1259_fortincomparison} shows the values for progenitor mass, orbital period, and natal kick values inferred from this study compared to those inferred by \citet{Fortin2022}, demonstrating some differences but significant overlap of values. When considering the spin-orbit misalignment in the system, \citet{Valli2025} found that, for \btwelve, natal kicks cannot be confined to the orbital plane, and are instead likely to be closely aligned with the progenitor rotation axis, and moderate in magnitude (averaging 100\,\kmps, and inconsistent with the commonly cited Maxwellian distribution presented by \citet{Hobbs2005}). Our results do not show such significant preferences in natal kick direction (with natal kicks slightly favouring the orbital plane: $\theta_{\rm 68\%}\approx$ 30--150, though it is possible that, should we also apply constraints based on spin-orbit misalignment, our results would be more in line with that study.

\subsection{\aqlx}\label{SubSect:AqlX1}
\aqlx\ is one of the first discovered transient NS-LMXBs \citep{Kunte1973, Kaluzienski1977}. The NS nature is confirmed through archetypal features such as Type-I X-ray bursts \citep[e.g.,][]{Koyama1981, Galloway2008}, burst oscillations \citep{Zhang1998}, and millisecond X-ray pulsations \citep{Casella08}. The donor was identified as a K-type star \citep{Chevalier1999, MataSanchez2017}, in a 19-hr orbit with the NS \citep{Chevalier1991, Welsh00}.
\citet{MataSanchez2017} managed to single out the counterpart from the interloper star and obtained new constraints on the orbital parameters, which we used in this study (Table\,\ref{tab:sample}).

There appear to be a large number of different channels of ZAMS binaries, evidenced by clustering in some regions of parameter space (see Figure\,\ref{Fig:AqlX1_cornerplot}). These likely correspond to distinct evolutionary channels, as is the case with \xtej\ and \cyg. 

We find that plausible natal kicks span 0--90\,\kmps, peaking around $\upsilon_{\rm NK,mean}=20$\,\kmps (Figure\,\ref{Fig:AqlX1_khist}). This results in a more complex distribution of inferred natal kicks, however the magnitudes of natal kicks for each channel span a similar range: 0--100\,\kmps.

\subsection{\gxthree}\label{SubSect:GX301}
\gxthree\ is one of the classic supergiant HMXBs containing a neutron star and a massive B-type donor \citep{Kaper2006}. The orbit is highly eccentric with a period of 41.5 days, and accretion onto the NS is fed by slow but dense stellar winds from the hypergiant donor \citep{Kaper2006}. The distance is around 3.6$\pm$0.2\,kpc \citep{Kaper2006}. However, with it not being eclipsing, there remains controversy over the mass of this star, and those we have used in Table\,\ref{tab:sample} are not dynamical.

We find this system must have experienced a natal kick $\geq250\,$\kmps\ (see Figure\,\ref{Fig:GX301_cornerplot} for ZAMS binary characteristics and Figure\,\ref{Fig:GX301_khist} for natal kick distribution). This is the source with the highest predicted $\vnk$ of systems explored here. At face value, this seems surprising, given the systemic velocity is substantially lower than that of \xtej; in fact, it is the prime example of the non-linear dependence of $\vsys$ and $\vnk$, as is further discussed in Section \ref{SubSect:VNKEq}. 

\citet{SimazBunzel2023} conduct a thorough study of the evolutionary history of \gxthree, providing constraints on the progenitor properties and natal kick strength. They find strong natal kicks are favoured ($\geq450$\,\kmps), and the non-linear dependences of systemic velocity on natal kick velocity is also demonstrated (their Figure\,4). However, the evolutionary pathway they find differs to the results of our simulations --- most notably in their initial (ZAMS) orbital separation. Our simulations favour an initial separation of 500--4000\,\Rsun (orbital period of 100--1000\,days), whereas \citet{SimazBunzel2023} find a much shorter initial orbital period, with \porbi$\approx$2--5 days, and initial component masses also being smaller than those in our simulations. In our simulations, such short-period, high-mass systems always merge prior to the first supernova. It is possible that fine tuning the various prescriptions in the binary evolution code may make the evolutionary channel presented by \citet{SimazBunzel2023} more plausible, however neither our core simulations, nor the additional simulations we run varying some of the physical assumptions (see Section \ref{SubSect:DifferentParams}) can replicate these results.

\subsection{\fouru}\label{SubSect:4U170037}

\fouru\ is among the bright X-ray sources detected by the {\it Uhuru} mission \citep{Jones1973}. Accretion onto the object is from the wind of the massive O-type supergiant, HD 153919 \citep{Clark2002, Hainich2020}. Despite intensive studies, the nature of its compact object still remains elusive. \citet{Murakami1984} claimed the detection of 64-second X-ray pulsation with the {\em Tenma} satellite, but this was not confirmed by follow-up studies \citep[e.g.,][]{Gottwald1986}. A recent X-ray study by \citet{Bala2020} revealed a possible cyclotron feature indicative of a NS, 
but no periodicity was detected. \citet{Clark2002} estimated a dynamical mass of $2.44\pm 0.27\,\mathrm{M_\odot}$, indicating an unusually massive NS or a mass-gap BH. We adopt the more recent estimate by \citet{Falanga2015}, who report a lower estimate of \moneobs\,=\,$1.97\pm 0.19\,\mathrm{M_\odot}$. The system has a compact (\porbobs\,=\,3.41\,days) but eccentric ($e\approx 0.2$) orbit \citep{Clark2002}. We adopt the values from \citet{Falanga2015} in our Table\,\ref{tab:sample}. We note that it is also one of the few systems known to show a bow-shock likely associated with its systemic motion through the interstellar medium \citep{Privsegen2019}. 

Previous studies of the natal kicks and evolution of \fouru\ have determined natal kicks ranging 20- $\geq400$\,\kmps. \citet{Fortin2022} found $23\leq\vnk\leq192$\,\kmps\,for \fouru, based on inferred progenitor properties of $5.9\leq M_{\rm 1, SN}\leq8.9$\,\Msun\,\&\,$2.2\leq P_{\rm orb, SN}\leq3.8$\,days. Regrettably, across all our simulations, we cannot identify systems which exist in that parameter space. 

Systems with the progenitor (\& donor) mass predicted by \citet{Fortin2022} all exist in much wider orbits (of order 10s-1000s of days; only $\sim10^{-3}$\,\% with \porbpre\,$\leq100$\,days and none shorter than 25\,days). Systems within the predicted region of progenitor mass and orbital period can have companion masses of no more than 30\,\Msun\,(with more than 90\% being lower than 15\,\Msun). Finally, we find systems with companion mass greater than 30\,\Msun\,and orbital period shorter than 10\,days exceedingly rare (only a couple hundred in a population of $>$\,20 million). All of these systems have a compact progenitor mass greater than 12\,\Msun, and all form black holes rather than neutron stars.

Whilst the uncertainties in binary evolution mean we cannot definitively rule out the existence of systems matching the progenitor properties presented by \citet{Fortin2022} binary population synthesis simulations indicate these are unlikely evolutionary pathways for \fouru. We summarize that systems with such extreme mass ratios cannot survive the common envelope phase which they are destined to undergo if they exist in orbits tighter than $\sim$1\,AU. Therefore, our study appears to rule out the suggested evolution proposed in \citet{Fortin2022} and, consequently, their values of $\vnk$. 

In another study \citet{Shah2026} propose that \fouru\ could have evolved from a binary in which \monepre\,$\approx$\,8.5\,\Msun\, \mtwopre\,$\approx$\,40\,\Msun, and \porbpre\,$\approx$\,4.4\,days, with these parameters identified through Monte Carlo simulations emulating the impact of the supernova \& natal kick on binary properties (masses, orbital periods, and system velocity). Having identified preferred pre-SN parameters, they use \mesa\ to reconstruct the system's evolutionary history, and identify the ZAMS binary parameters as \monei\,=\,40\,\Msun, \mtwoi\,=\,28\,\Msun, and \porbi\,=\,3\,days. A strong natal kick, $\vnk=237^{+156}_{-105}$\,\kmps\, applied during the supernova reproduces the present-day observations. When attempting to reproduce these results with \cosmic, we find that binaries with the aforementioned ZAMS parameters are unable to avoid merger, which typically occurs within a few Myr, well before the primary is sufficiently evolved to undergo a supernova.

Our failure to reproduce the results of \citet{Shah2026} may be the result of differences between rapid population synthesis codes like \cosmic, and \mesa-based codes (for further discussion of the comparison between the two approaches, see \ref{SubSubSect:BSEMESA}). Alternatively, it may be that \fouru\ can form from the progenitor identified by \citeauthor{Shah2026}, but under specific assumptions of binary evolution outside the scope of our simulations (e.g. a value of $\alpha$ we have not considered). Future studies will seek to address this discrepcncy, but for the meantime, \fouru\ exists as an example of the limitations of our methodology, and a challenge for models more widely.

\begin{figure}
    \centering
    \includegraphics[width=0.99\linewidth]{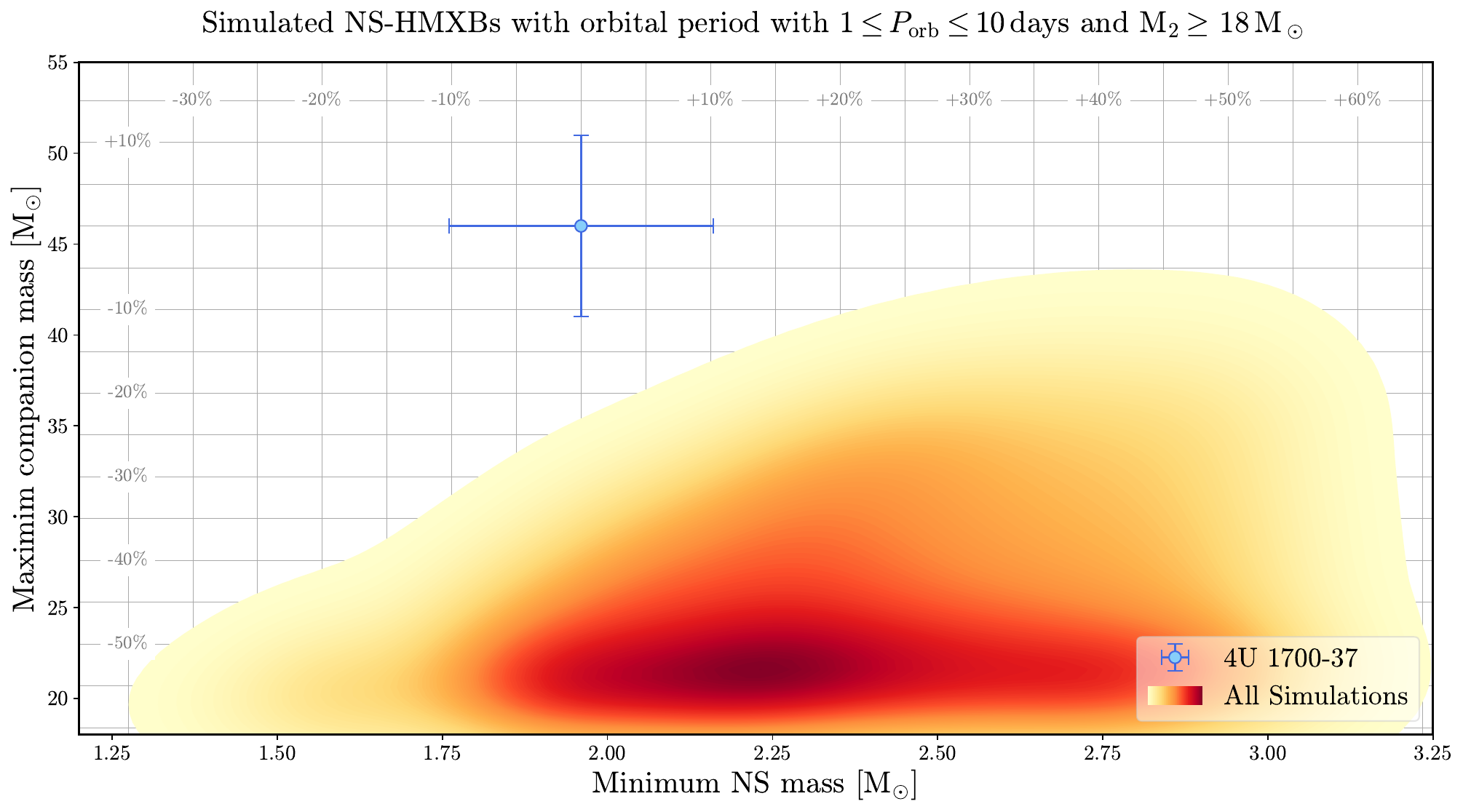}
    \caption{Mass limits for simulated short-period ($1\leq\porb\leq10\,$day) NS-HMXBs, coloured according to number density. Canonical values for \fouru\ are indicated in blue.  The extreme mass ratio of \fouru\ is impossible to reproduce with our simulations; such large companions ($\geq40$\,\Msun) are only feasible when the compact object exceeds 2.5\,\Msun. This suggests that our population synthesis with \cosmic\ has not spanned enough parameter space, or that the system requires very specific evolutionary models, or that the inferred mass(es) of the binary components of \fouru\ require further scrutiny.}
    \label{Fig:AllHMXBs}
\end{figure}

It is also worth noting that the masses of both components of \fouru\ remain controversial; we consider the masses presented by \citet{Falanga2015} --- \moneobs\,=\,$1.96\pm0.19$\,\Msun\,and \mtwoobs\,=\,$46\pm5$\,\Msun\ for the neutron star and the luminous companion, respectively. These were determined from X-ray eclipse timing and optical radial velocity studies and MC simulations, using the radius of the optical component as one of the input parameters. Prior to that, \citet{Abubekerov2004} conducted a similar study, this time using the radius of the optical component and a mass-luminosity relation for $M_2$, determining $M_1\approx1.4$\,\Msun\ and $M_2\approx27$\,\Msun. Though these studies have subtle differences in their methods, they both rely on the findings of \citet{Clark2002}, specifically the radius of the optical companion (21.9\,R$_\odot$) and its luminosity ($\log L$\,=\,5.82). These values themselves were determined from modelling of the stellar spectra using a non-LTE radiative transfer code presented by \citet{Hillier1998}; however, this code is based on perfectly spherical geometry (i.e. not accounting for distortion due to the Roche potential, or oblation caused by rapid rotation) and is sensitive to the assumed atmospheric abundances, which are difficult to determine accurately in highly evolved stars that have undergone significant accretion phases. Furthermore, \citet{Clark2002} use the distance obtained by \citet[][$d$\,=\,1.9\,kpc]{Ankay2001}, which differs significantly from more recent distance estimates (cf., $d$\,=\,1.5\,kpc based on {\em Gaia} astrometry; \citealt{Zhao2023}). 

This may explain why this system is so difficult to reproduce, and why simulated companion masses err on the lower end (with $M_2\geq40$\,\Msun\,being rare in our simulations). In Figure\,\ref{Fig:AllHMXBs} we show the minimum neutron star mass and maximum companion mass for all short-period ($\porb\leq10$\,days) NS-HMXBs in our simulations (and further highlight those with systemic velocities consistent with \fouru). Such an extreme mass ratio in a short period HMXB appears to not be viable, with the systemic velocity proving even more troublesome to explain. 

If all other SN parameters are the same, a lower mass companion would have the effect of decreasing constant $B$ and increasing $A$, meaning the relative contributions of the natal kick is determining the systemic velocity increases, resulting in perhaps more stringent constraints on $\vnk$. Natal kicks would result in smaller orbital separations and eccentricities. However, robust constraints on natal kicks applied to \fouru\,if it indeed has a lower mass companion than the most recent literature suggests would involve conducting further full-scale simulations, beyond the scope of this paper.

\section{Discussion}\label{Sect:Discussion}

We have successfully reproduced the evolution of 6 compact objects in interacting binaries, utilising the BSE-based population synthesis code \cosmic, and determined the plausible natal kick distributions (magnitudes and directions) for each of these systems. In this section, we discuss these results; the implications thereof; the robustness of our conclusions; and the limitations of our methodology.

\begin{table}
    \centering
    \caption{Summary of natal kicks results for each target. $\vsys$ is the observed peculiar velocity (see \citetalias{Zhao2023}) without Galactic dispersion subtracted; $\upsilon_{\rm MLK, max}$ is the maximum velocity imparted to the system due to mass loss alone; $\upsilon_{\rm NK, min}$ is the minimum natal kick magnitude; and $\vnkconf$ gives the 68\% HDI for the natal kick magnitudes in our simulations.}
    \begin{tabular}{lcccc}
    \hline
     System    &  $\vsys$  & $\upsilon_{\rm MLK, max}$ & $\upsilon_{\rm NK, min}$ & $\vnkconf$\\
         &  (\kmps)  & (\kmps) & (\kmps) & (\kmps)\\
     \hline
     \aqlx\    &  42\,$\pm$\,12  & 21 & 0   & 10\,--\,46 \\ 
     \btwelve\ &  25\,$\pm$\,8   & 13 & 0   & 15\,--\,58 \\ 
     \cyg\     &  22\,$\pm$\,3   & 14 & 0   & 10\,--\,43 \\ 
     \gxthree\ &  58\,$\pm$\,4   & 27 & 282 & 336\,-\,407 \\ 
     \jtwo    &  64\,$\pm$\,13  & 69 & 0   & 28\,--\,89 \\ 
     \xtej\    &  143\,$\pm$\,11 & 42 & 104 & 127\,--\,152 \\ 
     \fouru\  &  71\,$\pm$\,12  & & & \,--\,\\
     \hline
    \end{tabular}
    \label{tab:detailed_sys_results}
    \\
\end{table}

\begin{figure}
    \centering
    \includegraphics[width=\linewidth]{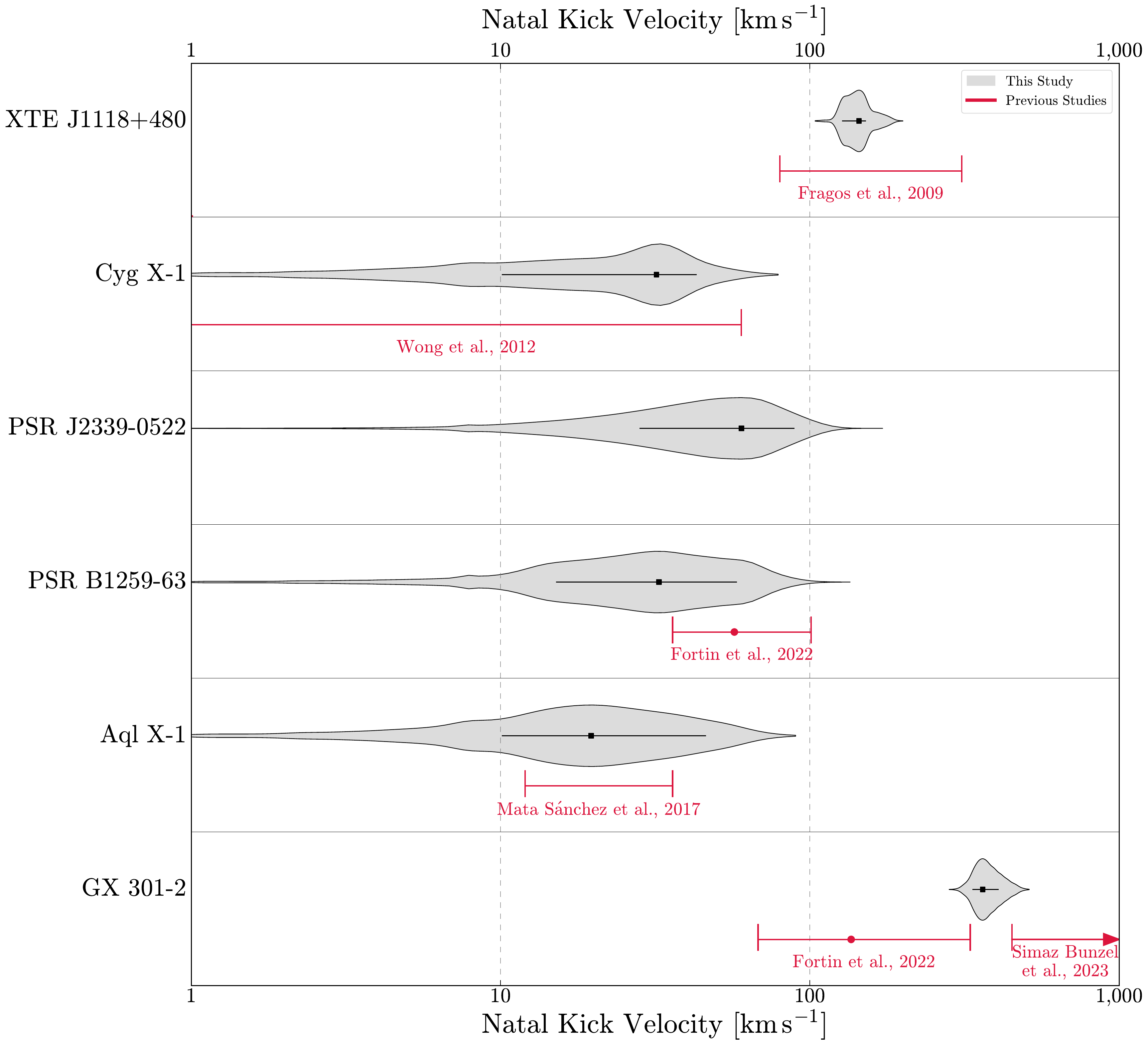}
    \caption{Violin plots showing distributions of plausible natal kicks for all six systems in this study (grey shaded), compared to values determined by previous studies (red lines), where applicable. Violin plots show distribution of all viable natal kick magnitudes; internal black horizontal line indicates 68\% HDI, and MAP values are indicated as a small black square. \citet{Fragos2009}; \citet{Wong2012}; and \citet{MataSanchez2017} quote ranges for the natal kicks of \xtej, \cyg, and \aqlx\ respectively. \citet{Fortin2022} present mean (red circle) and asymmetric errors for \btwelve\ and \gxthree. \citet{SimazBunzel2023} present a lower limit for the natal kick applied to \gxthree. Comparisons between our results and those from the literature are discussed in \ref{SubSect:PrevResults} - they generally arise from; updated observational constraints; inclusion of pre-SN evolution in our study; and/or inclusion of an intrinsic Galactic dispersion velocity, which decreases the value of $\vsys$ which we match to in our simulations.}
    \label{Fig:KickViolins}
\end{figure}

\subsection{Comparison against published results}\label{SubSect:PrevResults}

A comparison of the inferred natal kick estimates across our samples with those presented by previous studies is summarised in Figure\,\ref{Fig:KickViolins}. The average values of the kicks that we infer overlap with published ranges in 4 of our 6 sources, with no prior inference available for \jtwo\ and a significant difference in the estimates for \gxthree. Individual sources are discussed below.

Improved constraints on \xtej\ compared to those presented by \citet{Fragos2009} stem from our more comprehensive modelling of both pre- and post- supernova evolution, as well as the tighter observational confidence interval on the present-day kinematics derived from \gaia\ astrometry.

For \btwelve, we obtain a wider $\vnk$ confidence interval compared to that of \citet{Fortin2022}, predominantly at the lower end of estimates. Our simulations suggest that the system could have formed {\em without} a significant natal kick, whereas \citet{Fortin2022} find natal kicks in excess of 30\,\kmps. These differences are due to the different ways of determining pre-supernova orbital parameters, with our progenitor properties coming from binary evolution. Our compact progenitor mass and orbital period ($M_{\rm 1,SN, 68\%}$\,=\,3.5--4.5 and $P_{\rm orb,SN, 68\%}$\,=\,1820--4980\,days) differ slightly from those inferred by \citet{Fortin2022} ($M_{\rm 1,SN}$\,=\,$6.2^{+2.6}_{-3.0}$\,\Msun, $P_{\rm orb,SN}$\,=\,1592$^{+1081}_{-1264}$\,days), identified using Markov Chain Monte Carlo simulations assuming uniform priors for progenitor attributes, resulting in different kick distributions (see Figure,\,\ref{Fig:B1259_fortincomparison}).

Our natal kick constraints for \gxthree\ are broader than those presented by \citet{SimazBunzel2023}, owing again to different progenitor characteristics from different methods. Additionally, they require a much stronger $\vnk$\,$\sim$\,450\,\kmps, as compared to our lower limit of $\sim$\,250\,\kmps. This is, in part, due to matching for different values of $\vpec$ --- our study uses the value calculated in \citetalias{Zhao2023} (58\,\kmps) with an assumed Galactic dispersion velocity of 40\,\kmps subtracted in quadrature, and requires our simulations to be consistent with a systemic velocity of $42\pm6$\,\kmps --- whereas \citet{SimazBunzel2023} match to the nominal value of 56\,\kmps\ \citep{Fortin2022}. Another reason for the discrepancy is the different progenitor properties found in each study.

\subsection{Black hole natal kicks $>100$\,\kmps}
Our results reaffirm that \xtej\ must have experienced a strong natal kick \citep{Fragos2009,Belczynski2016}. This system was amongst the first to be identified as a high-velocity system based on optical astrometric studies \citep{Mirabel2001}. 

The inference of high $\vnk$ values, exceeding 100\,km\,s$^{-1}$ in this case, is at odds with theoretical predictions of black hole natal kicks (expected to be of order a few \kmps\ as a consequence of fallback mechanisms; \citealt{Fryer2012, Janka2013,Burrows2024a,Janka2024}). Not only does fallback reduce the mass lost due to the supernova and, consequently, the mass-loss kick, but the low-mass companion in \xtej\ also limits the ejected mass. Therefore, the momentum imparted due to this mass loss is limited to $<45$\,\kmps, with $\sim$70\% of simulations finding $\vmlk\leq10$\,\kmps. The observed $\vpec$ distribution of \xtej\ peaks at $143\pm11$\,km\,s$^{-1}$, invoking the need for an additional source of momentum. 

That hydrodynamical simulations of supernovae struggle to consistently reproduce scenarios in which black holes are subject to high natal kicks \citep[e.g.][]{Burrows2024a,Burrows2024b,Janka2024} suggests further research is needed into the differences (or, perhaps, lack thereof) in supernovae and related natal kick mechanisms between neutron stars and black holes. 

\subsection{Non-linear relationship between $\vnk$ and $\vsys$}\label{SubSect:VNKEq}

Many studies have implicitly assumed observed peculiar velocity or inferred nascent peculiar velocity by integrating orbits back in time as being commensurate with natal kick velocity. Whilst fast moving systems can be indicative of strong natal kicks, the influence of orbital characteristics, both before and after the supernova, alter the imprint of a given $\vnk$ on $\vsys$, meaning the latter cannot be used as a direct proxy for the former.

In order to highlight the importance of this issue, we here present a worked example of a HMXB, where the complex and non-linear impact of a given $\vnk$ on $\vsys$ stands out starkly. We choose \gxthree, whose resulting $\vnk$ spans $\approx$\,300--500\,\kmps (Figure\,\ref{Fig:GX301_khist}). This range is extreme as compared to the system's moderate $\vpec$\,$\approx$\,58\,\kmps\ (reduced to $\approx$\,40\,\kmps when corrected for Galactic stellar dispersion).

Examining Eq.\,\ref{eq:vsys_psn}, the magnitude of the velocity imparted to the centre-of-mass of a system ($\vsys$) subjected to a $\vnk$ is determined by:

\begin{align*}
    \vsys^2 & =   \left(A\vec{\vnk}\right)^2 + \left(B\vec{\vorb}\right)^2 - 2AB\left(\vec{\vnk}\cdot\vec{\vorb}\right),
\end{align*}

\noindent
where $A = \frac{M_{\rm 1,post-SN}}{M_{\rm 1,post-SN}+M_2}$ and $B = \frac{M_{\rm 2, SN}\left(M_{\rm 1, SN}-M_{\rm 1,post-SN}\right)}{\left(M_{\rm 1, SN}+M_{\rm 2, SN}\right)\left(M_{\rm 1,post-SN}+M_{\rm 2, SN}\right)}$. 
If the orbit is circular at the time of supernova (as is the case for all \gxthree\ analogues), $\vec{\vnk}\cdot\vec{\vorb} = \vnk\vorb\cos\phi\sin\theta$ where $\theta$ and $\phi$ describe the polar and azimuthal angles of the natal kick (Figure\,\ref{Fig:NKAngles}).

Our simulations find that the neutron star in \gxthree\ had a progenitor mass of $M_{\rm 1,SN, 68\%}$\,=\,5.6--6.0\,\Msun\ (which ejected $\sim$\,4\,\Msun\ during the supernova) in an orbit of $P_{\rm orb,SN, 68\%}$\,=\,18--38\,days.

The contribution of the ejected mass to $\vsys$ enters through the $B$ term and $\vorb$\ in the equation above. In our simulations for \gxthree, $B$ averages to $\approx$\,0.085, and the relative velocity at the instant of supernova is $\vorb$\,$\sim$\,250\,\kmps, resulting in a mass-loss-kick $\vmlk$\,$\sim$\,15--30\,\kmps\ (see Figure\,\ref{Fig:GX301_khist}). This is insufficient to explain the current systemic velocity; an additional natal kick is therefore required. \gxthree's large $M_2$ results in a small $A \approx 0.04-0.06$, and therefore $\vsys$ has a weak dependence on $\vnk$.

Using typical initial system parameters determined from our population synthesis runs: \monepre $\approx5.8$\,\Msun; \porbpre$\approx$30\,days; $\vorb\approx250$\,\kmps; one can substitute: $A = \frac{1.85}{1.85+36}\approx0.05$, $B=\frac{36\left(5.8-1.85\right)}{\left(5.8+36\right)\left(1.85+36\right)}\approx0.09$ (Note: here we adopt $M_2=36$\,\Msun; the nominal value is $43\pm10$\,\Msun, but our simulations favour companion masses on the lower end of this range). If one assumes the kick is directed in a way that maximises the systemic velocity (i.e. $\cos(\phi)\sin(\theta)=-1$), this leads to:

\begin{align*}
    \vsys^2 & = \left(A\vnk\right)^2 + \left(B\vorb\right)^2 + 2AB\vnk\vorb\\
    & =  \left(0.05 \vnk\right)^2 + \left(0.09 \times 250\right)^2 + 2\times0.05\times0.09\times\vorb\vnk \\
    & \approx 0.0025 \vnk^2 + 2.25\vnk + 506
\end{align*}

\noindent
Substituting in the observed $\vsys$ allows one to determine the minimum acceptable value for $\vnk$ after correcting for Galactic dispersion. For $\vsys$ to be greater than 35\,\kmps, $\vnk$ must be $\gtrsim250$\,\kmps. One can prove that increasing this to $\vnk\,=\,500\,$\kmps\,results in only a moderate increase in systemic velocity ($\sim$45\,\kmps). This system highlights the nuance surrounding the relationship between $\vnk$ \& $\vsys$, and serves as a reminder that one cannot use the latter as a direct proxy for the former.

\begin{figure*}
    \centering
    \includegraphics[width=0.9\textwidth]{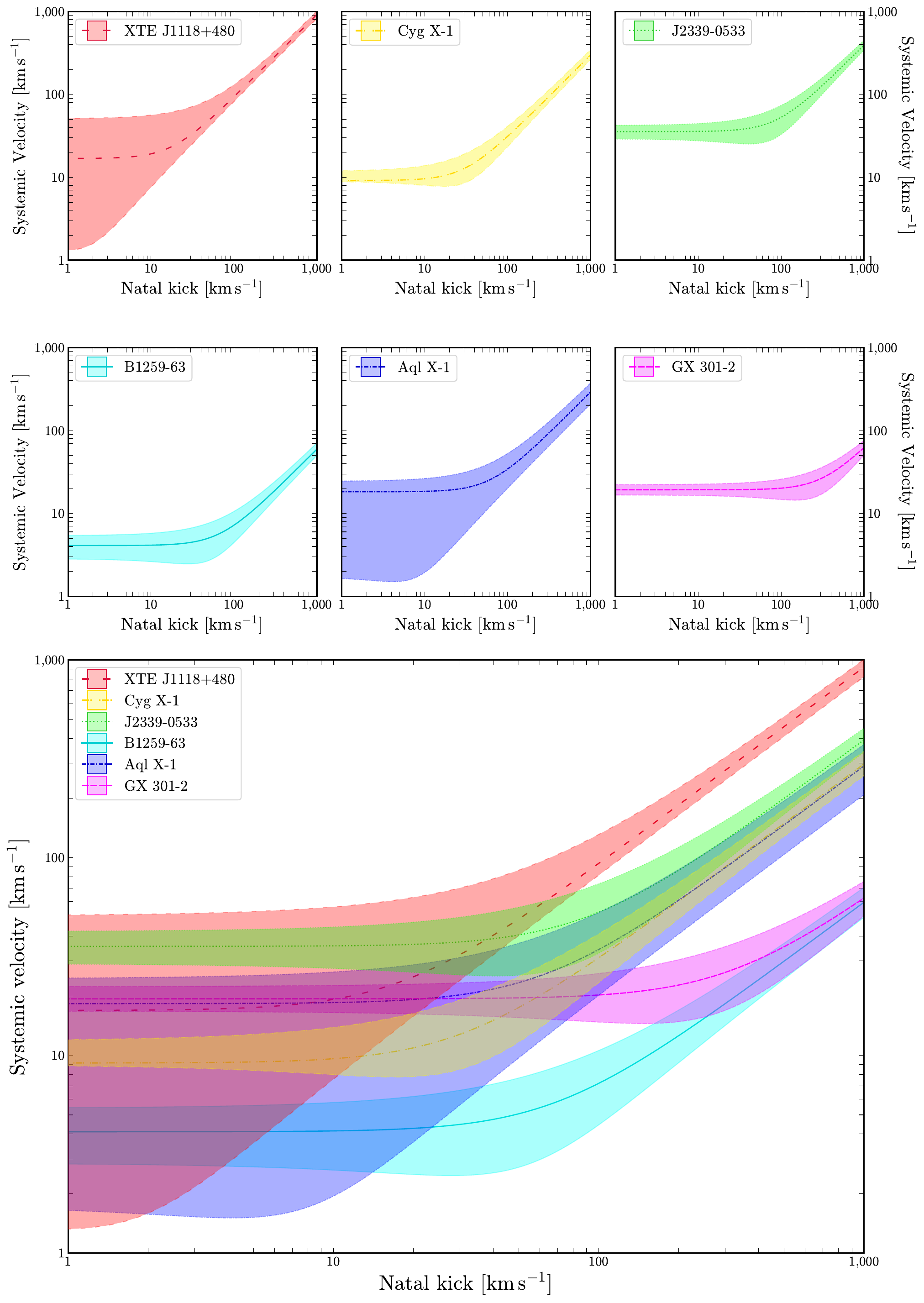}
    \caption{Systemic velocity as a function of natal kick magnitude on systemic velocity for each of the systems in this study (from Equation \ref{eq:vsys_psn}), showing the mean (line) and 68\% confidence intervals (shaded region). These figures are made using the binary parameters determined from \cosmic, and assuming the natal kick is directed to maximise and minimise the resultant systemic velocity (for the upper and lower limits), respectively. Each system is distinguished by colour and linestyle, annotated on the top left.}
    \label{Fig:VNKVSYS}
\end{figure*}

Figure\,\ref{Fig:VNKVSYS} illustrates how the relation between $\vnk$ \& $\vsys$ varies across different systems. These figures use the system progenitor parameters determined from \cosmic; the different shapes of the relation are primarily governed by compact object progenitor mass, ejected mass, and companion mass (at time of supernova).

The shapes of these relations are broadly similar across sources, consistent with the quadratic form of Eq.\,\ref{eq:vsys_psn}. At low values of $\vnk$ (low being relative for each system), the systemic velocity is dominated by the mass-loss, and the relationship is flat (with the natal kick contributing very little). As $\vnk$ ramps up, there is an upwards inflection, which can be interpreted as the point at which the $\vnk^2$ term dominates, and is dependent (amongst other variables) on the mass ratio $q$. For example, for \xtej, $\vnk$ dominates systemic motion beyond $\sim10$\,\kmps\,(as the mass-loss kick is limited to $\lesssim20$\,\kmps). In contrast, for \gxthree, $\vnk$ contributes very little until it exceeds a few hundred \kmps. At `high' values of $\vnk$, the relationship becomes quasi-linear, though $\vsys$ remains impacted by $\vmlk$ (which varies significantly depending on numerous system parameters), and is therefore not a direct proxy for $\vnk$.

\subsection{Influence of physical assumptions}\label{SubSect:DifferentParams}

\begin{figure*}
\centering
    \begin{subfigure}[b]{0.49\textwidth}
        \centering
        \includegraphics[width=\textwidth]{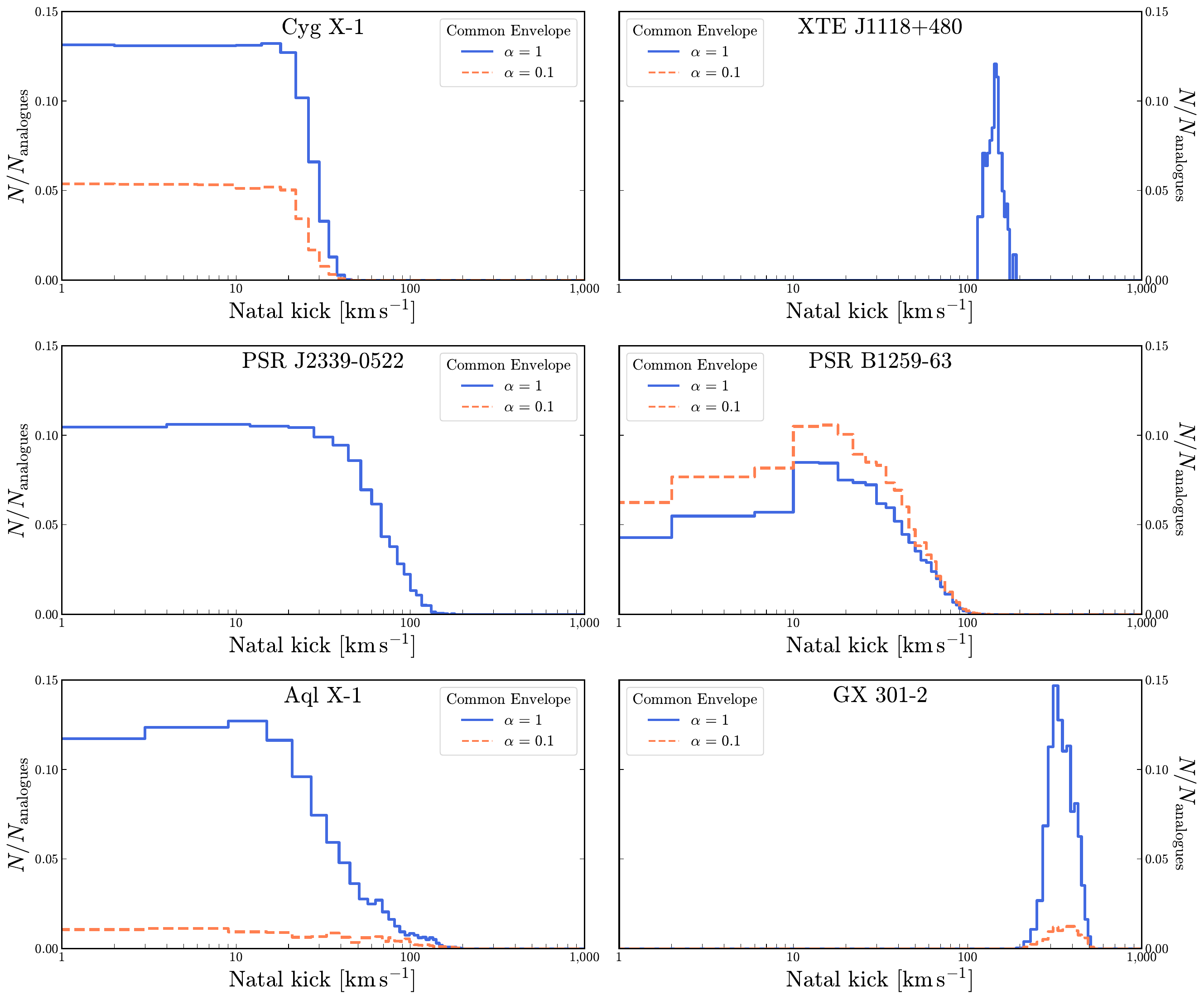}
        \caption{Common Envelope Efficiency $\alpha$}%
    \label{Fig:DifPhysics_Alpha}
    \end{subfigure}
    \hfill
    \begin{subfigure}[b]{0.49\textwidth}  
        \centering 
        \includegraphics[width=\textwidth]{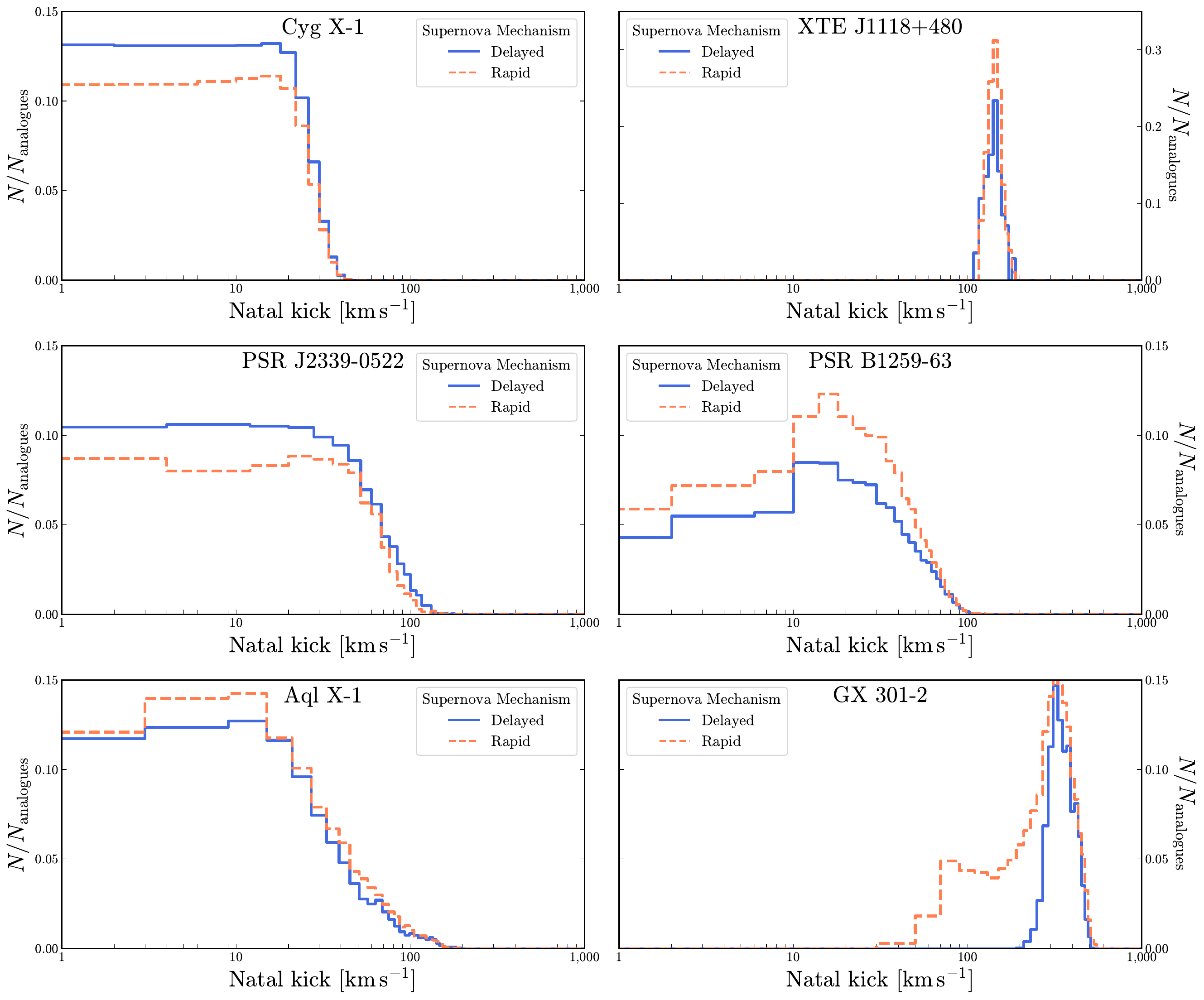}
        \caption{Supernova mechanism}%
    \label{Fig:DifPhysics_SN}
    \end{subfigure}
    \vskip\baselineskip
    \begin{subfigure}[b]{0.49\textwidth}   
        \centering 
        \includegraphics[width=\textwidth]{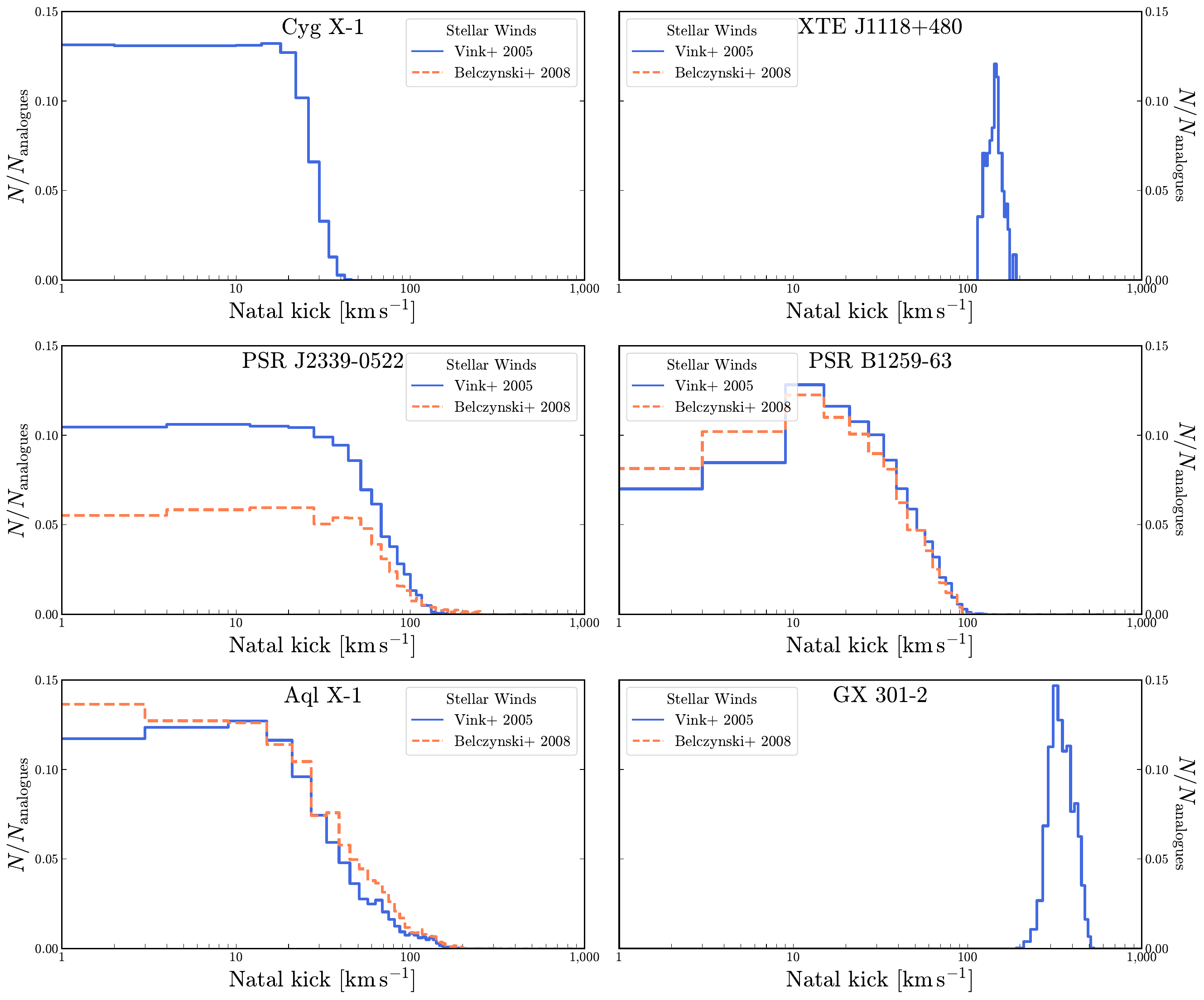}
        \caption{Stellar winds}%
    \label{Fig:DifPhysics_Winds}
    \end{subfigure}
    \hfill
    \begin{subfigure}[b]{0.49\textwidth}   
        \centering 
        \includegraphics[width=\textwidth]{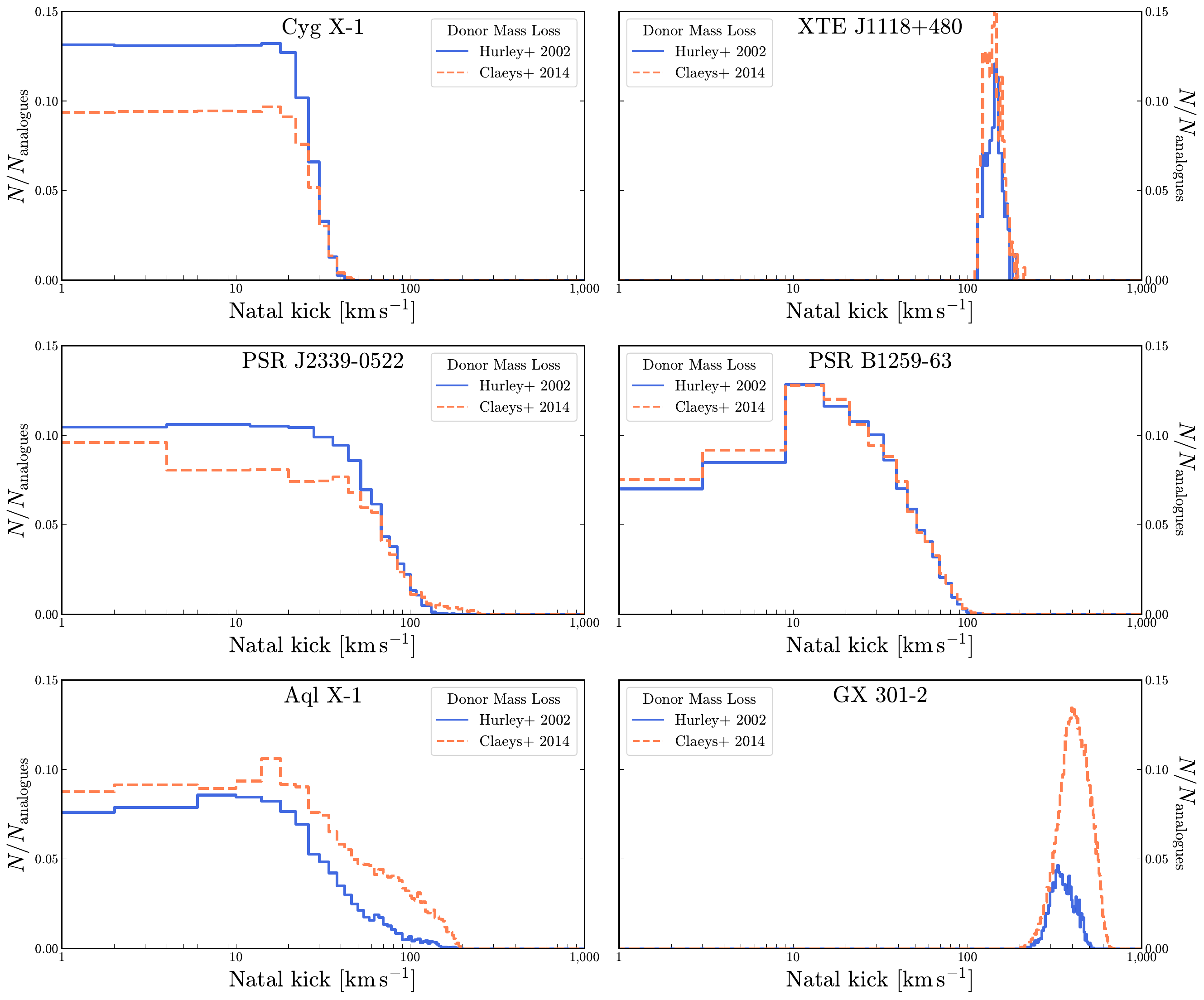}
        \caption{Stable RLOF mass loss}%
    \label{Fig:DifPhysics_MassTransfer}
    \end{subfigure}
    \caption{Influence of different physical assumptions on natal kick distributions for all 6 systems: $\alpha$ (upper left); supernova mechanism (upper right); stellar winds (lower left); mass transfer (lower right). Blue solid line indicates the parameters adopted for the core simulations in this paper, orange dashed line indicates an alternative scenario (where orange lines are absent, alternative scenario is non-viable according to our simulations). Employing different binary evolution prescriptions/parameters results in different kick distributions in some cases, though the effect does not appear to be systematic (i.e. consistently higher/lower natal kick magnitudes). In some cases, simulations are unable to reproduce systems (i.e. \xtej\ appears unable to form with $\alpha=0.1$).} 
    \label{Fig:DifferentPhysics_Separate}
\end{figure*}

\begin{figure*}
    \centering
    \includegraphics[width=0.9\textwidth]{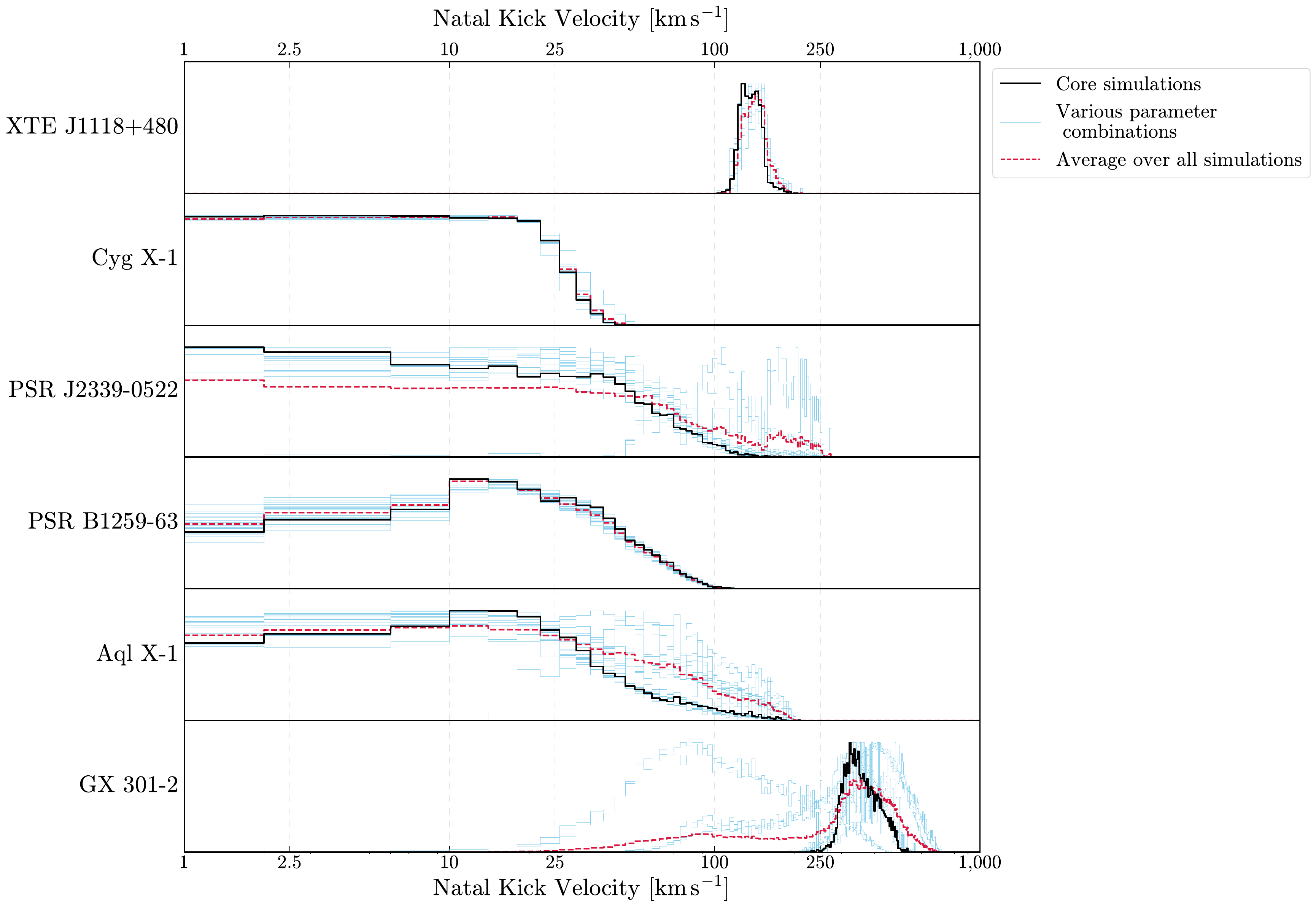}
    \caption{Influence of different physical assumptions on natal kick distribution of all 6 systems. Thick black solid line indicates the parameters adopted for the core simulations in our work, the light blue lines show the distributions for independent checks using various combinations of metaparameters described in Section \ref{SubSect:DifferentParams}, and the dashed red line denotes the `average' distribution across all scenarios. Varying assumptions about binary \& stellar evolution from our core simulations have a significant impact on the inferred natal kick distributions - however, the extent to which the results change varies between systems. For example, the magnitudes of plausible natal kicks for both \xtej\ and \cyg\ are broadly unchanged, with the caveat that simulations are unable to reproduce any analogues for these systems under certain conditions (see main text). Changes in the metaparameters do not show obvious systemic trends, e.g. there is no consistent increase/decrease in natal kick magnitude, nor any consistent change in the overall shape of the distributions.}
    \label{Fig:DifferentPhysics_All}
\end{figure*}

Our workflow samples a wide swathe of binary parameter space, varying $M_1$, $M_2$, $e$, $\porb$, as well as the imparted natal kick magnitude and direction for each simulated system. Beyond these, are several {\em metaparameters} that systemically control evolutionary phases and whose influence remains debated. These include (i) the common envelope ejection efficiency; (ii) the strength of stellar winds and their dependence on metallicity; (iii) supernova explosion duration and ejecta fallback, and (iv) mass transfer prescriptions. Here, we consider the impact of some of the canonical choices that we have adopted for our presented simulations. For this purpose, 32 sets of focused simulations are conducted and compared to our core results. Uncertainties in binary evolution are numerous, and covering all possible physical scenarios is not computationally feasible. Our aim here is to determine, at least qualitatively, how sensitive our output natal kick distributions are to some of the input physical assumptions --- the resultant natal kick distributions under different assumptions are shown in Figures \ref{Fig:DifferentPhysics_Separate} and \ref{Fig:DifferentPhysics_All}, and we refer the reader to these in conjunction with the discussion below.

\subsubsection{Common envelope efficiency, $\alpha$}

A common envelope phase may occur if mass transfer within a binary is unstable, resulting in both stellar cores being engulfed in a shared stellar envelope. Within this envelope, both stellar cores lose orbital and angular momentum to their surroundings, rapidly decreasing their orbital separation. This stage may end in the total merger of the two components, or the successful ejection of the envelope, leaving a more tightly separated binary. Common envelopes are typically described with the $\alpha\lambda$ model:

\begin{equation}
    \alpha\left(\frac{M_{\rm 1,c}M_{\rm 2,c}'}{2a_i}-\frac{M_{\rm 1,c}M_{\rm 2,c}'}{2a_f}\right)=\frac{G}{\lambda}\left(\frac{M_1M_{\rm 1,env}}{R_1}+\frac{M_2M_{\rm 2,env}'}{R_2}\right)
\end{equation}

\noindent
where $a_i$ \& $a_f$ are the orbital semi-major axes directly before and after the common envelope phase; $M_1$ \& $M_2$ are the masses of the two components; $M_{\rm 1,c}$ is the core mass of the primary star; $M_{\rm 2,c}'$ is the effective core mass of the secondary, which is equal to $M_2$ if the secondary star is main sequence, or the core mass ($M_{\rm 2,c}$) otherwise; $M_{\rm 1,env}$ is the envelope mass of the primary star; $M_{\rm 2,env}'$ is the effective envelope mass of the secondary, which is equal to $M_{\rm 2,env}'=M_2-M_{\rm 2,c}'$; and $R_1$ \& $R_2$ are the radii of the two components.

The constant $\alpha$ is the common-envelope effective efficiency parameter (describing the transfer of orbital energy to the common envelope, and consequently, the likelihood that the system will eject the envelope), and $\lambda$ parametrises the envelope binding energy \citep{Tout1997,Hurley2002} (typically, $\lambda=0.5$, which is the value adopted in our simulations). $\alpha=1$ indicates that all of the orbital energy has gone into ejecting the envelope, whereas $\alpha<1$ is interpreted as meaning that only some of the energy has been transferred, and $\alpha>1$ requires an additional source of energy (such as the recombination of cooling atoms). Commonly adopted values of $\alpha$ vary (indeed, it is likely that $\alpha$ is not constant \citealt{Regos1995}); $\alpha$\,=\,1 is fairly typical in the literature \citep[e.g.][]{Nelemans2001}, although it may be as low as 0.2 \citep{Zorotovic2010}, or up to several times larger than 1 in particular systems \citep{Fragos2019}.

The influence of changing values for $\alpha$ varies across the different systems in this study (see Figure\,\ref{Fig:DifPhysics_Alpha}). This is unsurprising, as not all systems experience a common envelope phase, in which case this parameter is somewhat irrelevant (e.g. \btwelve). Some systems appear unable (or at least very unlikely) to form with a low value of $\alpha$ --- \xtej, for example, cannot be reliably reproduced under the assumption of a smaller common envelope efficiency. This is expected; systems where $M_1\gg M_2$ at onset of common envelope require high values of $\alpha$ to eject the envelope \citep{DeMarco2011}. Conversely, \btwelve\ appears to be slightly more viable with a lower value of $\alpha$, with more analogues generated from these simulations. The distribution of natal kicks appears broadly unchanged, irrespective of $\alpha$ - we can summise that, at least of the systems in our sample, the common envelope efficiency impacts the likelihood of a given binary evolving to resemble the observed system, however, if the binary survives this common envelope phase, the pre-SN parameters must be consistent across different common envelope descriptions, meaning natal kicks have the same influence on the binary system, and therefore the same kicks are viable. However, we stop short of claiming any definitive systemic dependence, or lack thereof, on natal kick with $\alpha$.

\subsubsection{Supernova mechanism}

\citet{Fryer2012} outline the prescriptions for remnant mass according to different supernova mechanisms. Our core simulations follow the `delayed' prescription (their equations 18--20); however we conduct additional simulations using the `rapid' mechanism (their equations 15--17). The key difference between the two is that the `delayed' supernova mechanism populates the lower `mass-gap' (i.e. compact objects may form with masses 2--5\,\Msun).

Interestingly, varying the supernova mechanism between `delayed' and `rapid' appears to have very little influence on the inferred natal kicks for all but one system (see Figure\,\ref{Fig:DifPhysics_SN}). The system in question, \gxthree, exhibits two very different distributions under the two assumptions. Whilst the probability peak occurs around similar $\vnk$ in both cases ($\approx400\,$\kmps), a rapid supernova engine can allow the system to form with a much smaller natal kick ($\leq50$\,\kmps) whereas the minimum natal kick for a delayed supernova must exceed 200\,\kmps. This is primarily due to higher typical mass-loss under the delayed supernova mechanism, which is the dominant factor in determining the resultant systemic velocity (as discussed in Section \ref{SubSect:GX301}).

This comparison suggests that different engines could broaden the inferred $\vnk$ confidence intervals, without substantially affecting the mean estimators. The link between supernova mechanisms in both determining mass-loss and the consequent natal kicks is complex. Our natal kicks are agnostic to the physical mechanisms which drive them; however, it has been suggested that the {\em intrinsic} distribution of natal kicks imparted to a compact object is contingent on supernova mechanisms \citep[e.g.][]{Fryer2012,Burrows2024a}, meaning a fair comparison of the two would require normalising the distributions against the probability of the simulated binary receiving a given natal kick. This is beyond the scope of this study (and, indeed, is contrary to our desire to remove theoretical expectations from observational studies of natal kicks, nevertheless we think it necessary to point this out. 

We have compared \& discussed the two supernova prescriptions presented by \citet{Fryer2012}, though it is necessary to point out that there are a number of other studies focusing on remnant mass determination, many of which can be incorporated into \cosmic and other population synthesis codes \citep[e.g.][]{Belczynski2002, Belczynski2008, Eldridge2004, Giacobbo2019}. Whilst a complete review of the influence of supernova prescriptions on (observationally-inferred) natal kick magnitudes is beyond the scope of this work, the intrinsic link between supernova processes and natal kick drivers means this is an area worthy of further investigation.

\subsubsection{Stellar winds}

Our core simulations employ the description of stellar winds laid out by \citet{Vink2001} and \citet{Vink2005}.

\begin{dmath}
    \log\dot{M}_{\rm don} =k_1\log\left(Z/Z_{\odot}\right) + k_2\log\left(L_{\rm don}/10^5\right) + k_3\log\left(M_{\rm don}/30\right) + k_4\log\left(\frac{v_{\infty}/v_{\rm esc}}{2}\right) + k_5\log\left(T_{\rm eff}/k_{6}\right) \\
    + k_7\log\left(T_{\rm eff}/k_{6}\right)^2 + k_8
\end{dmath}

\noindent
where $Z$ is stellar metallicity; $L_{\rm don}$ and $M_{\rm don}$ are the luminosity and mass of the donor star, respectively; $v_\infty$ and $v_{\rm esc}$ are the terminal and escape velocities of the material from the donor star (the ratio $v_{\infty}/v_{\rm esc}$ is $\sim$1.3--2.6); and $k_1$, $k_2$, $k_3$, $k_4$, $k_5$, $k_6$, $k_7$, and $k_8$, are constants which depend on stellar spectral type.

We conduct additional simulations where stellar winds are modelled as in \StarTrack, i.e. where the orbit-averaged wind accretion rate is given by:

\begin{equation}
    \dot{M}_{\rm acc} = -\frac{F_{\rm wind}}{\sqrt{1-e^2}}\left(\frac{GM_{\rm acc}}{V_{\rm wind}^2}\right)\frac{\alpha_{\rm wind}}{2a^2}\frac{\dot{M}_{\rm don}}{\left(1+V^2\right)^{3/2}}
\end{equation}

\noindent
where $M_{\rm acc}$ \& $M_{\rm don}$ are the masses of the accretor and donor respectively; $a$ and $e$ are the orbital semi-major axis and eccentricity respectively; $V_{\rm wind}$ is calculated from $V_{\rm wind}^2=2\beta_{\rm wind}\frac{GM_{\rm don}}{R_{\rm don}}$ (where $\beta_{\rm wind}$ is a constant, dependent on the donor star spectral type, and $R_{\rm don}$ is the radius of the donor star); $V^2=V_{\rm acc,orb}^2 / V_{\rm wind}^2$ with $V_{\rm acc, orb}$ being the orbital velocity of the accretor; $\alpha_{\rm wind}$ is accretion efficiency in the Bondi-Hoyle model (set at 1.5); and $\dot{M}_{\rm don,wind}=-4\pi r^2\rho V_{\rm wind}$ is the wind mass-loss rate from the donor (where $r$ is the distance between the two stars and $\rho$ is the density of the stellar wind matter). $F_{\rm wind}$ exists to ensure the mass accretion rate never exceeds the donor mass-loss rate (i.e $\dot{M}_{\rm acc,wind} \leq 0.8\dot{M}_{\rm don,wind}$), which may occur when averaging over highly eccentric orbits.

The impact of different stellar wind prescriptions varies across different systems (see Figure\,\ref{Fig:DifPhysics_Winds}). Of the three high-mass systems in our sample, only one (\btwelve) can be reliably reproduced when stellar winds are described according to \citet{Vink2001} and \citet{Vink2005}, as opposed to the description employed in \StarTrack. This is likely due to greater mass stripping and consequently smaller compact object progenitors in the latter scenario. There is not an obvious trend in one given direction when considering the influence of mass-loss physics on natal kick inferences. There is a subtle tendency to favour natal kicks on the lower end of the distribution for \btwelve\ and \aqlx, the \textit{slightly} larger natal kicks are allowed for \jtwo.

\subsubsection{Stable RLOF}

In our simulations, stable mass loss rates via Roche lobe overflow are calculated from:

\begin{equation}
    \dot{M_{\rm don}}=f\cdot3\times10^{-6}\left(\log\left[\frac{R_{\rm don}}{R_L}\right]\right)^3\left(\min\left[M_{\rm don},5\right]\right)^2
\end{equation}

\noindent
where $\dot{M}_{\rm don}$ is the mass loss from the donor star; $M_{\rm don}$ is the donor star mass; and $R_{\rm don}$ and $R_L$ are the donor star and Roche-lobe radii. In our core simulations, $f$ is equal to 1 (as in \citet{Hurley2002}), however, we also test the modified prescriptions from \citet{Claeys2014}, in which $f$ is a dimensionless factor, dependent on the stability of mass transfer, according to: 

\[ f = \begin{cases} 
      1000 & q'<1 \\
      \max\left[1,\frac{1000}{q'}\cdot\exp\left(-\frac{\left(\ln q'\right)^2}{0.045}\right)\right] & q'>1\\
   \end{cases}
\]
where $q'=M_{\rm accretor}/M_{\rm donor}$. The qualitative difference in these prescriptions is enhanced mass transfer rates when using the \citet{Claeys2014} prescriptions. In both cases, mass transferred is limited by Kelvin-Helmholtz time-scales (i.e. $\dot{M}\leq\dot{M}_{\rm KH,don}$).

\begin{equation}
    \dot{M}_{\rm KH,don}=1\times 10^{-7}\frac{M_{\rm don}}{M_{\rm don}M_{\rm don}'}R_{\rm don}L_{\rm don}
\end{equation}

\noindent
in which $M_{\rm don}'$ is either the mass of the donor star (for main sequence stars) or the mass of the donor star's envelope, and $L_{\rm don}$ is the luminosity of the donor star.

Changing the description of donor mass-loss has a limited impact on the distribution of natal kicks for most systems (see Figure\,\ref{Fig:DifPhysics_MassTransfer}), however both \aqlx\ and \gxthree\ are allowed larger natal kicks when mass transfer is governed according to \citet{Claeys2014}. For \gxthree, the upper limit for the natal kick increases from $\upsilon_{\rm NK, max}=530$\,\kmps to $\upsilon_{\rm NK, max}\sim700$\,\kmps, whilst $\upsilon_{\rm NK,min}$ remains consistent. This is primarily due to larger companion masses in the case of \citet[][recalling that the errors on observed companion mass for the system are broad, \mtwoobs\,=\,43\,$\pm$\,10\,\Msun]{Claeys2014}, resulting in a diminished contribution from $\vnk$ to the final system velocity (see \ref{SubSect:VNKEq}).

\subsection{Limitations of our Framework}\label{SubSect:Limitations}

\subsubsection{Uniform Population Priors}\label{SubSubSect:GalacticPopulation}
The priors adopted for the populations in this study (outlined in Table \ref{tab:priors}) are intended to cover a broad range of initial parameters and, consequently, are not representative of the Galactic population. Specifically, we do not adopt any initial mass function, nor physically motivated orbital periods, eccentricities or metallicity distributions. This means that the identified progenitor properties and evolutionary channels are dependent on the likelihood of the initial binaries existing in the first place. For example, where we find multiple possible channels for any individual binary, (e.g. \xtej\ \& \cyg), it is entirely possible that one channel is implausible due to a low probability of forming the initial binary (e.g. for \xtej, the `low'- and `high'- mass channels are both identified in our simulations, however normalising our initial parameter space to a more physically realistic Galactic scenario may eradicate one of these). 

Our choice of priors is motivated by our focus on examining the influence of natal kicks, and our attempt to present kick distributions that are agnostic to assumptions regarding binary populations. The primary impact of this is that whilst the individual natal kicks deemed tenable for a given analogue are robust, the shape of the overall {\it distribution} of natal kicks for a given system may change. The possibility remains for future studies to quantify the `likelihood' of evolution channels (and related natal kick distributions) using a more physically motivated initial population.

\subsubsection{Matched Observables and their Uncertainties}\label{SubSubSection:OtherObs}
The observed parameters that we match in this study are compact object mass (\moneobs), companion mass (\mtwoobs), and orbital period (\porbobs), in addition to the requirement of having an active ongoing mass transfer phase where relevant. An increasing number of systems now have additional measured properties, including eccentricity, spin-orbit misalignment, and system age. Incorporating these additional constraints could potentially tighten allowed evolutionary pathways and change commensurate natal kicks (see, e.g., \citealt{Valli2025} who fold in eccentricity measurements for a sub-sample of accreting neutron-star binaries).

However, there remains much non-uniformity across the population when it comes to these additional constraints. Eccentricity is known for only three of the six systems considered here, for instance. The uncertainties involved in determining some of these characteristics can also span a wide range; e.g., reported age estimates for \xtej\ are as young as 11\,Myr \citep{Psaltis2007} or, 2--5\,Gyr \citep{Fragos2009}. 

We have thus chosen not to match to these when identifying analogous simulated systems in our workflow, preferring instead consistency and interpretability of evolutionary solutions across the population. The spread of natal kicks calculated here is thus likely to be more conservative than the available data necessitates, and future studies could improve upon these by bringing in additional constraints on individual systems.

Similarly, relevant uncertainties in our pipeline are likely to be overly conservative. For the observed masses and orbital periods, these are set to the quoted errors from observational samples \citetalias{Zhao2023}, or set at 10\% of the observed value (whichever is larger). This decision, again, reflects our desire to reduce the influence of observational limitations which can sometimes lead to inconsistent observational inferences across different systems. It also partly mitigates the impact of physical processes that are imperfectly modelled in the BSE framework.

\subsubsection{Failure to reproduce some systems}\label{SubSubSect:Failure}
Our methodology cannot reliably determine the evolutionary pathway and related natal kicks for at least one system, \fouru. As outlined in Section \ref{SubSect:4U170037}, the observed values for this system are somewhat controversial and subject to many uncertainties, which may explain our inability to identify analogues in our simulated populations. Whilst this is curious (and, perhaps, troubling) we believe that in this instance there is sufficient uncertainty in observed parameters that this doesn't negate the efficacy of our method overall. Further studies targeting \fouru\ may shed light on the system's evolution, perhaps requiring binary evolution prescriptions beyond those explored here.

\subsubsection{The \cosmic\ semi-analytic approach vs. the MESA approach to structural and evolutionary solutions }\label{SubSubSect:BSEMESA}
Our core simulation package, \cosmic\ \citep{cosmic}, is a rapid binary population synthesis, designed to conduct large scale simulations of whole populations, using analytical prescriptions of stellar evolution and binary interactions built off of BSE \citep{Hurley2000, Hurley2002}. It does not account for donor or accretor responses to mass loss or gain, and instead smoothly updates each stellar component mass along the fitted evolution of stars under thermal equilibrium. By contrast, one-dimensional stellar evolution codes, such as MESA, model single stars in high fidelity, including full stellar structure and composition evolution calculations, and then tie these to binary evolution \citep{Paxton2011, Paxton2013, Paxton2015, Paxton2018, Paxton2019}.

Codes like BSE can struggle to adequately describe phases of stable mass transfer in which the thermal and mass-transfer time-scales of the mass donor are comparable. In such cases, the evolution of the binary system depends sensitively on the detailed response of the donor to the mass-loss it experienced, i.e. the degree to which it is driven out of thermal equilibrium. Since BSE models all stars as being {\em in} thermal equilibrium, the code can produce unphysical outcomes in such cases. For example, BSE-based population synthesis codes cannot reproduce the famous period gap in cataclysmic variables (CVs). This gap is associated with a sharp reduction in the angular momentum loss (and mass-transfer) rate experienced by CVs when their donors become fully convective \citep{Knigge2006,Knigge2011}. Prior to this point, the donors are slightly out of thermal equilibrium, and therefore bloated. When the mass-transfer rate is sharply reduced, the donor relaxes to its thermal equilibrium radius, causing it to become completely detached. Mass-transfer therefore stops completely as the CV enters the gap. In BSE-like codes, the donor is {\em always} in thermal equilibrium -- by assumption -- so it {\em cannot} relax and lose contact with the Roche lobe.
 
We believe a similar issue is responsible for the bimodal progenitor distribution suggested by \cosmic\ for \xtej. In Section 4.1, we discussed the apparent dearth in viable progenitors with companion masses $0.3\leq M_{\rm 2,i}\leq0.8$\,\Msun. In \cosmic, simulated binaries in this regime (with the relevant primary masses and orbital periods) merge prematurely (before mass transfer is able to sufficiently strip the companion). However, this is likely to be a consequence of the code being unable to precisely resolve the response of the stellar companion rather than a physically realistic outcome. Indeed, the Roche-lobe overflow factor during the stable mass-transfer phase just prior to merger can reach values as high as 3-4, whereas the donor radius ($R_2$) and Roche-lobe radius in stably mass-transferring systems will typically match to within a factor $\frac{H}{R_2}\sim10^{-4}$, where $H$ is the scale height in the donor's atmosphere.

To determine the natal kicks relevant for our systems requires extensive coverage of possible progenitor parameter space, and therefore rapid binary population synthesis is more appropriate for our goals --- however, this does introduce uncertainty as the interaction between the binary components are not treated as rigorously on a case-by-case basis, impacting the resultant evolutionary channels and, consequently, potentially affecting the natal kick inferences. Whilst MESA based codes (or similar) such as \textsc{posydon} \citep{posydon} are arguably more suitable for detailed studies of \textit{individual} systems, \cosmic\ is suitable for the large-scale studies we are conducting, particularly when comparing different systems/classes. \cosmic\ also allows for more flexibility when running binaries with different assumptions about binary evolution metaparameters such as descriptions of common envelope, stellar winds, accretion efficiency, etc..

\section{Summary}

In this work, we have presented our method for determining the natal kicks applied to compact objects in interacting binaries. This method employs a broad range of observed properties, including the current peculiar velocity, in tandem with binary population synthesis, accounting for stellar and binary evolution before and after the supernova. This combined approach is an improvement upon previous studies, by covering a broad swathe of parameter space; justifying the identified progenitor properties for each system with binary evolution physics rather than reverse-engineering them from MCMC simulations; and accounting for post-SN evolution of orbital period. Our method may be applied to any binary containing a compact object, provided there are observational constraints on the system's properties. The salient points of our work are as follows.

\begin{enumerate}
    \item We present six testbed systems for which we successfully trace plausible evolutionary histories and determine the natal kick(s) necessary to explain current observed physical properties as well as kinematics. These have been purposely chosen to span varied source types (low-mass, high-mass, pulsar binaries, black hole and neutron star systems) to test our pipeline. The resultant natal kick values are summarised in Table \ref{tab:detailed_sys_results}, and are broadly consistent with values determined from previous studies (where available).
    \item We independently conclude that the black hole LMXB \xtej\ must have formed with a natal kick $\vnk\gtrsim100$\,\kmps\ (Section\,\ref{SubSect:XTEJ1118}). Together with other black hole systems in the literature, such findings are beginning to challenge the view that natal kicks applied to black holes are lower than those applied to neutron stars.
    \item We identify two possible evolution channels for \cyg, distinguished by presence of a common envelope phase, and the associated natal kicks for the different pathways (Section\,\ref{SubSect:CygX1}). This highlights the crucial role of binary evolution in constraining natal kicks.
    \item We provide plausible natal kick magnitudes for \jtwo\ ($\vnkconf$\,=\,20--71\,\kmps), \btwelve\ ($\vnkconf$\,=\,13--53\,\kmps), and \aqlx\ ($\vnkconf$=\,6--36\,\kmps). Our results for \btwelve\ are consistent with previous work by \citet{Fortin2022}, although we find a broader distribution (particularly at the lower end). 
    \item We propose that the neutron-star high-mass X-ray binary \gxthree\ formed with a strong natal kick ($\vnkconf$\,=\,341--417\,\kmps), in agreement with previous work by \citet{SimazBunzel2023}.
    \item The impact of a natal kick on a given system is complex and non-linear, as is discussed in Section \ref{SubSect:VNKEq}; we therefore recommend that future studies use combined constraints from systemic velocity and orbital properties (including accounting for post-SN evolution) to determine the magnitude of the natal kick applied to any given system.
    \item Limitations and caveats of our framework are discussed in Section\,\ref{SubSect:Limitations}. These are primarily related to uncertainties in binary evolution physics and the limited fidelity with which stellar components can be evolved using rapid population synthesis codes. Observational limitations also play a role here, with system properties such as eccentricity and metallicity not yet uniformly available for large samples of compact object binaries.
    \item We also highlight systems such as \fouru\ that remain difficult to reproduce (Section\,\ref{SubSect:4U170037}). This inability may be a consequence of the limitations of our methodology, or suggestive of the need to revise measurements of observed properties. Whilst such difficulties are not unique to our work, it is important to keep this limitation of parameter space coverage in mind.
\end{enumerate}
\vspace*{.5cm}

\noindent
This study demonstrates that present-day systemic/peculiar velocities of compact objects cannot be used as proxies for natal kick velocities. This is especially important when considering the intrinsic distribution of natal kicks (which cannot necessarily be represented by the distribution of measured velocities) and when comparing natal kicks applied to different classes of objects (i.e. compact object type, high- vs low- mass companion). Our results demonstrate that black holes may be subject to high natal kicks, (the distribution of the natal kicks applied to black holes is discussed in Paper II), and that population synthesis studies would benefit from employing kick distributions motivated by more thorough analysis of observations.

In our companion paper, Dashwood Brown et al., in prep., Paper II, we present comprehensive results on a much larger sample of X-ray and pulsar binaries, and conduct a statistical study examining the difference between the natal kick distributions for neutron stars vs. black holes.

\section*{Acknowledgments}
\label{LastPage}

CDB, PG, YZ and CK thank UKRI Science \& Technology Facilities Council for support. PG also acknowledges funding from a Royal Society Senior Leverhulme Trust Fellowship (SRF$\backslash$R1$\backslash$241074). PN is supported by the Thai Government’s Development and
Promotion of Science and Technology Talents Project (DPST). PC is supported by a Leverhulme Trust Emeritus Fellowship. PG and KB acknowledge fruitful discussions with participants at the KITP workshop ``Stellar-Mass Black Holes at the Nexus of Optical, X-ray, and Gravitational Wave Surveys'' coordinated by Kareem El-Badry, Jessica Lu, Selma de Mink and Tejaswi Venumadhav, and supported by grant NSF PHY-2309135 to the Kavli Institute for Theoretical Physics (KITP). KB further acknowledges support from the Falco-DeBenedetti Career Development Professorship, \emph{Chandra} Award Numbers GO223031A, GO324037X, and GO425032X, NASA ATP Grant 80NSSC24K0768, and NASA LISA Preparatory Science Program Grant 80NSSC24K0361.

We thank the referee for their helpful comments and for providing some useful references as well as food-for-thought.  

Part of the simulations/calculations were completed using a 36-core Intel Xeon machine with 512 GB RAM at the AI Center of Excellence, Shiv Nadar Institution of Eminence, NCR Delhi. 

We wish to thank all those involved in the development and maintenance of \cosmic.

\section*{Data Availability}
No new data are presented in this work. For details on the observational parameter boundary conditions for individual compact object binaries, we refer the reader to \citetalias{Zhao2023} and references therein. \textsc{cosmic} is publicly available on GitHub --- refer to \citet{cosmic}.

\bibliographystyle{mnras}
\bibliography{references} 

\include{appendix/ExtraFigures}

\bsp	

\label{lastpage}

\end{document}

%% file: appendix/ExtraFigures.tex
\appendix

\onecolumn 

\section{ZAMS characteristics for testbed systems}\label{Sect:AppA}

\begin{figure}
    \centering
    \includegraphics[width=0.95\textwidth]{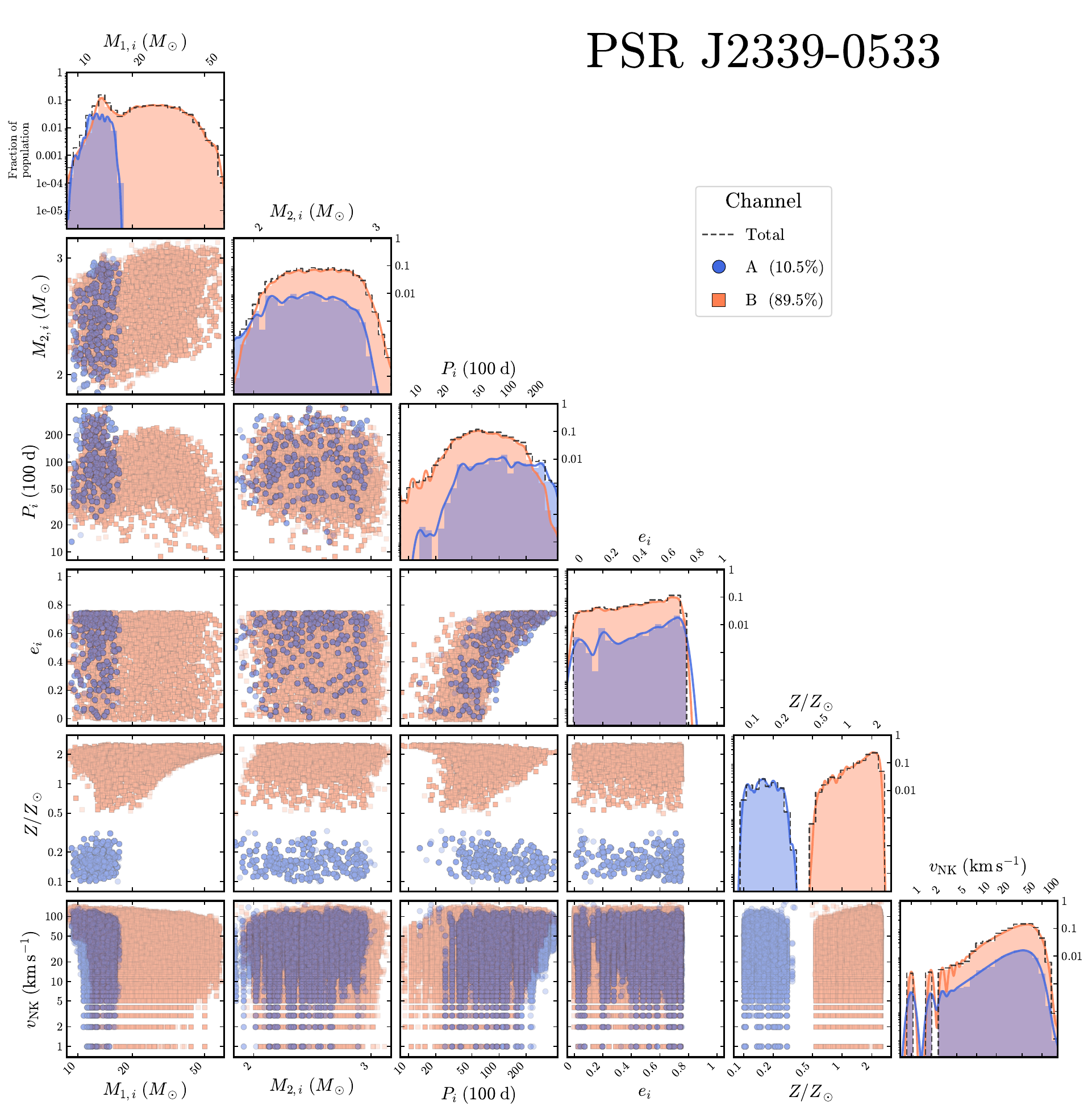}
    \caption{Corner plot highlighting key initial characteristics and acceptable natal kicks for simulated system analogues to \jtwo. In this case, distinct evolutionary families (based on metallicity) are indicated by colour and marker style. Refer to Figure\,\ref{Fig:J1118_cornerplot} for further details.}
    \label{Fig:J2339_cornerplot}
\end{figure}

\begin{figure}
    \centering
    \includegraphics[width=0.95\textwidth]{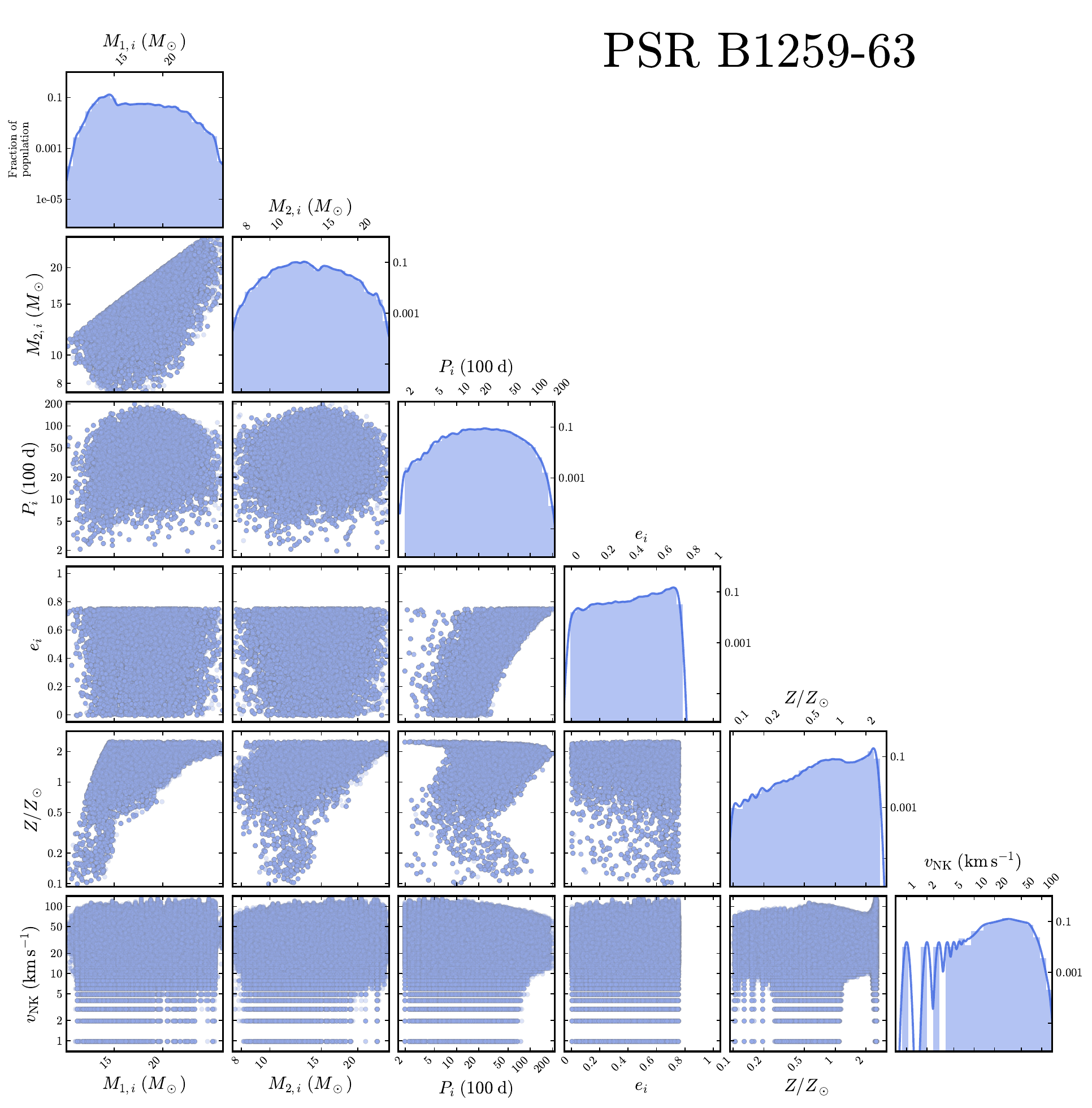}
    \caption{Corner plot highlighting key initial characteristics and acceptable natal kicks for simulated system analogues to \btwelve. In this case, distinct evolutionary families are indicated by colour and marker style. Refer to Figure\,\ref{Fig:J1118_cornerplot} for further details.}
    \label{Fig:B1259_cornerplot}
\end{figure}

\begin{figure}
    \centering
    \includegraphics[width=0.95\textwidth]{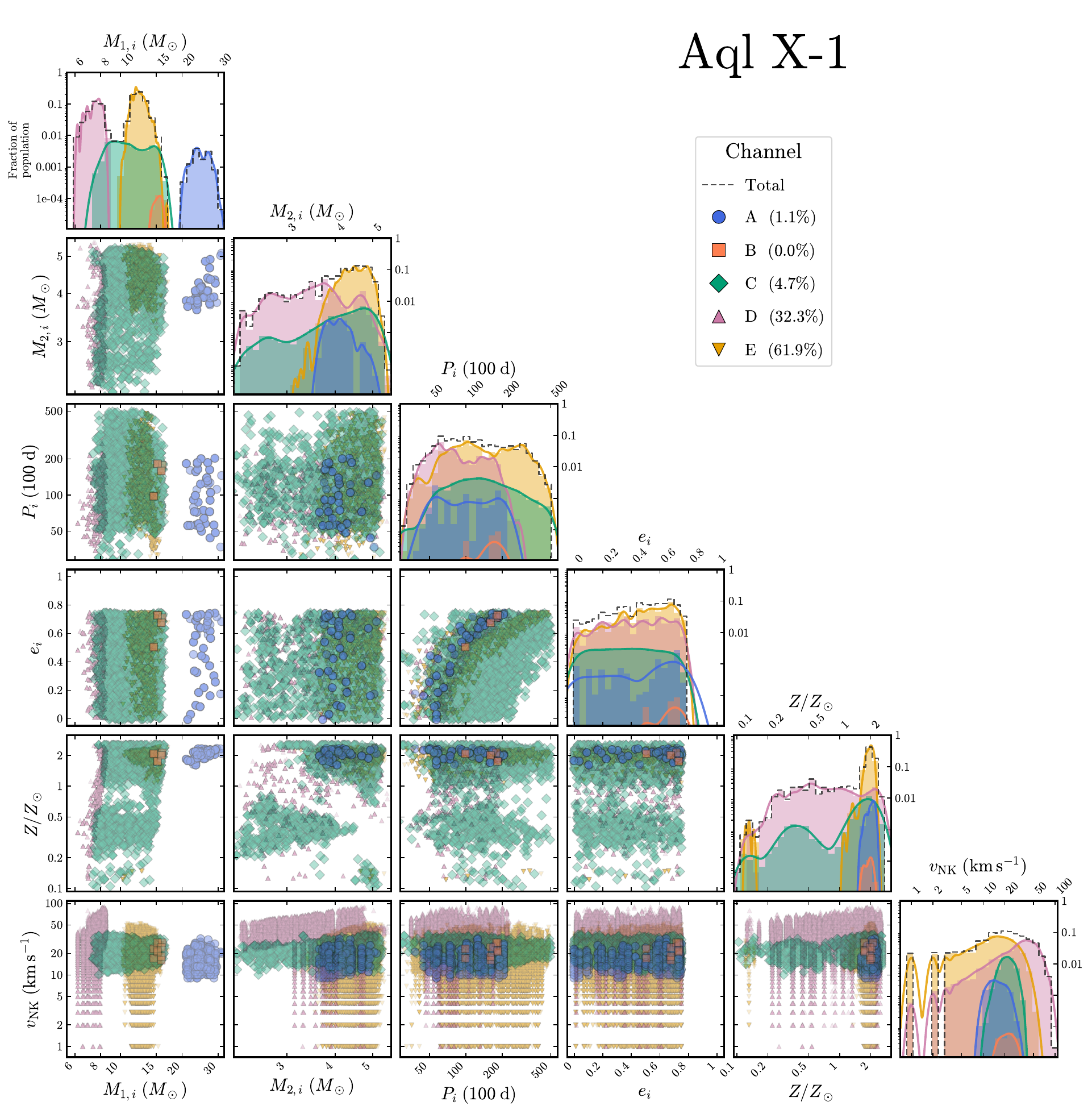}
    \caption{Corner plot highlighting key initial characteristics and acceptable natal kicks for simulated system analogues to \aqlx. In this case, distinct evolutionary families are indicated by colour and marker style. Refer to Figure\,\ref{Fig:J1118_cornerplot} for further details.}
    \label{Fig:AqlX1_cornerplot}
\end{figure}

\begin{figure}
    \centering
    \includegraphics[width=0.95\textwidth]{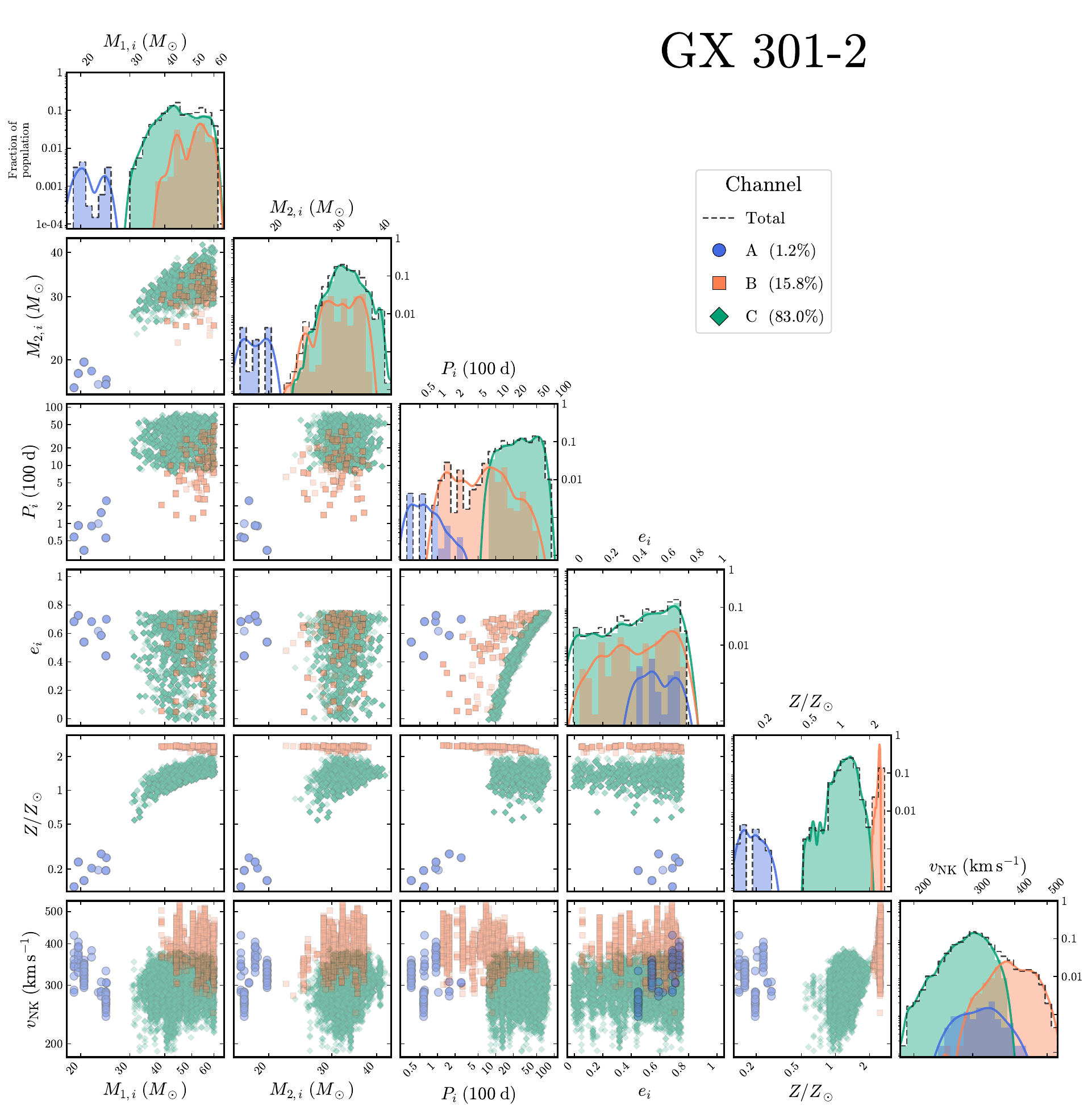}
    \caption{Corner plot highlighting key initial characteristics and acceptable natal kicks for simulated system analogues to \gxthree. In this case, distinct evolutionary families are indicated by colour and marker style. Refer to Figure\,\ref{Fig:J1118_cornerplot} for further details.}
    \label{Fig:GX301_cornerplot}
\end{figure}

Here we present corner plots showing the binary parameters at ZAMS for successfully simulated analogues to \jtwo\ (Figure\,\ref{Fig:J2339_cornerplot}), \btwelve\ (Figure\,\ref{Fig:B1259_cornerplot}), \aqlx\ (Figure\,\ref{Fig:AqlX1_cornerplot}) and \gxthree\ (Figure\,\ref{Fig:GX301_cornerplot}), respectively. These are equivalent to Figures \ref{Fig:J1118_cornerplot} and \ref{Fig:CygX1_cornerplot} in the main text. 

\section{Additional Kick characteristics for testbed systems}\label{Sect:AppB}

Here we present the distributions for mass-loss and natal kicks for simulated analogues \jtwo, \btwelve, \aqlx\ and \gxthree\ (equivalent to bottom pannels in Figures \ref{Fig:J1118_evolution_typical} and \ref{Fig:CygX1_evolution_typical}).

\begin{figure}
    \centering
    \includegraphics[width=0.95\textwidth]{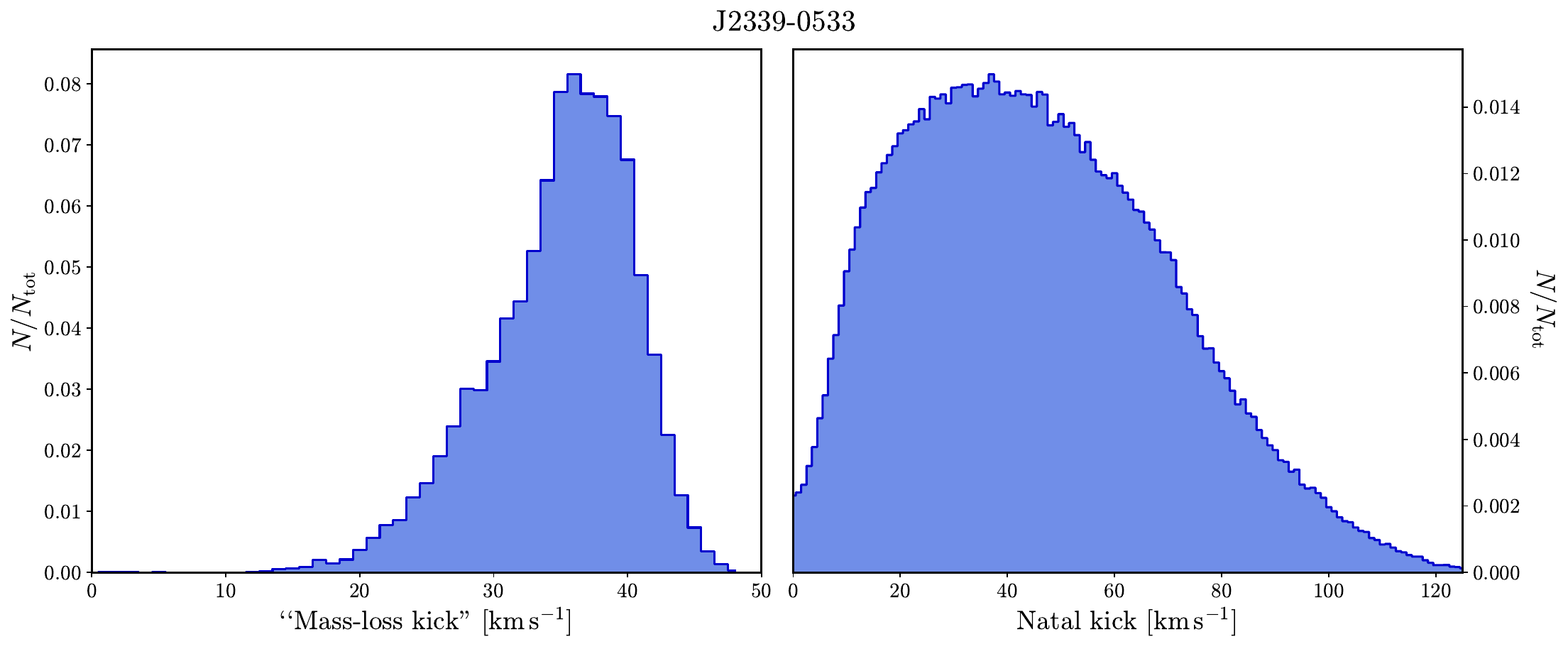}
    \caption{Distribution of mass-loss kicks ($\vmlk$) magnitude (left) and natal kick ($\vnk$) magnitude (right).}
    \label{Fig:J2339_khist}
\end{figure}

\begin{figure}
    \centering
    \includegraphics[width=0.95\textwidth]{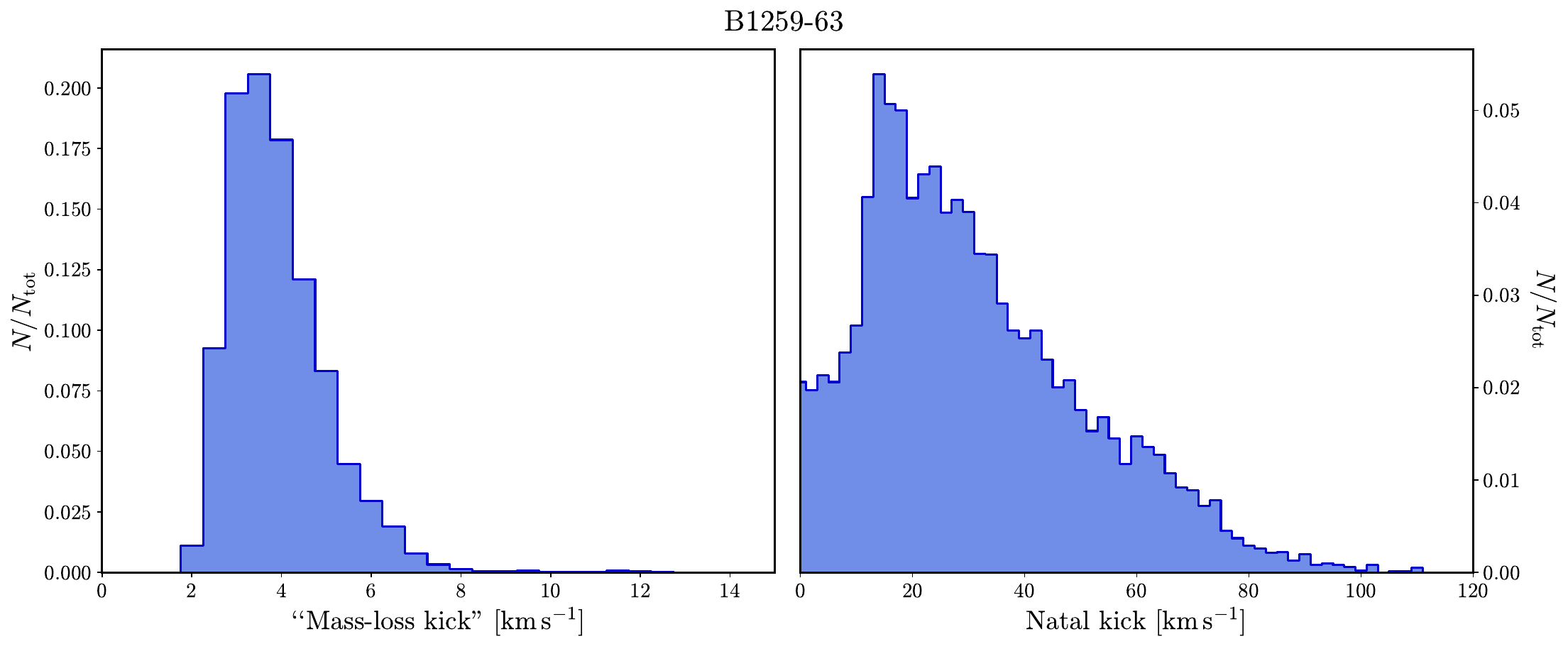}
    \caption{Distribution of mass-loss kicks ($\vmlk$) magnitude (left) and natal kick ($\vnk$) magnitude (right).}
    \label{Fig:B1259_khist}
\end{figure}

\begin{figure}
    \centering
    \includegraphics[width=0.95\textwidth]{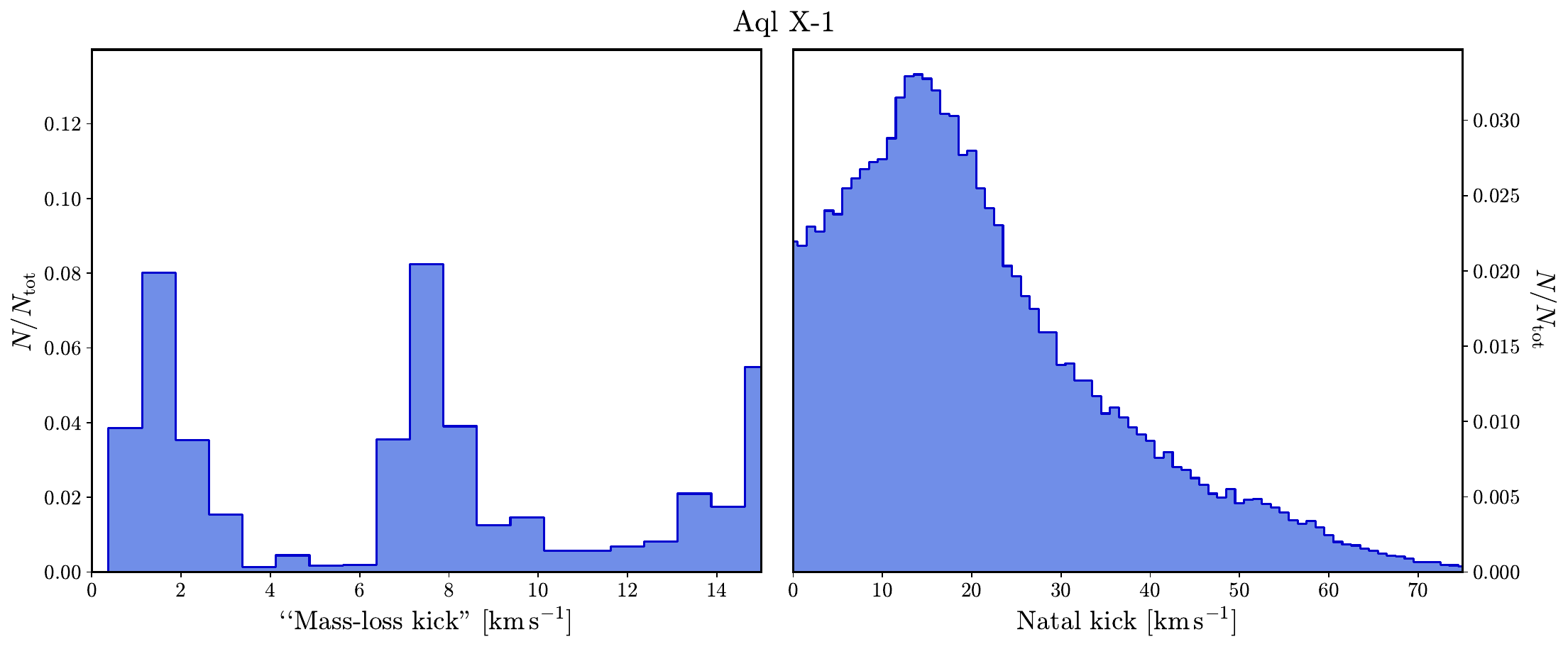}
    \caption{Distribution of mass-loss kicks ($\vmlk$) magnitude (left) and natal kick ($\vnk$) magnitude (right).}
    \label{Fig:AqlX1_khist}
\end{figure}

\begin{figure}
    \centering
    \includegraphics[width=0.95\textwidth]{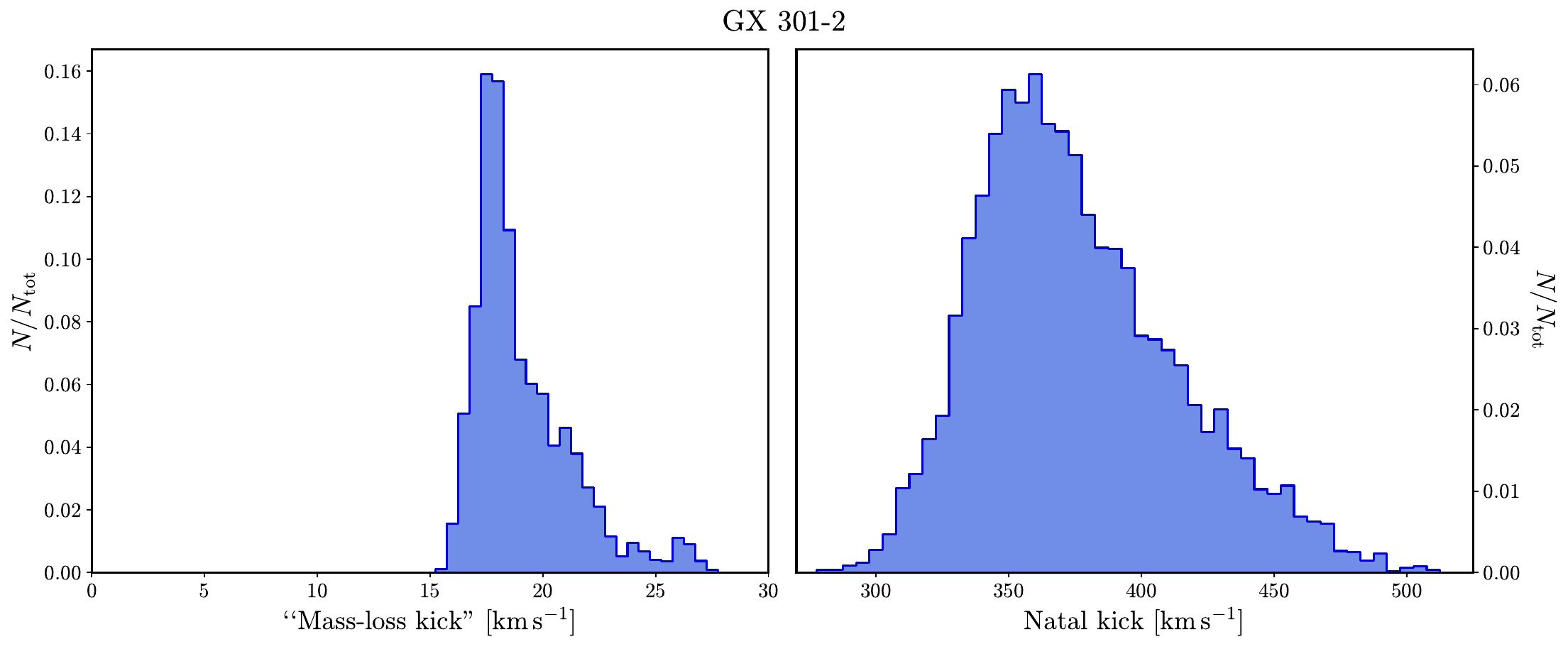}
    \caption{Distribution of mass-loss kicks ($\vmlk$) magnitude (left) and natal kick ($\vnk$) magnitude (right).}
    \label{Fig:GX301_khist}
\end{figure}


\section{Comparison with previous literature }\label{Sect:AppC}

Here we present some additional figures demonstrating the difference in our results with that of \citet{Fortin2022} for \btwelve. 

\begin{figure}
    \centering
    \includegraphics[width=0.95\linewidth]{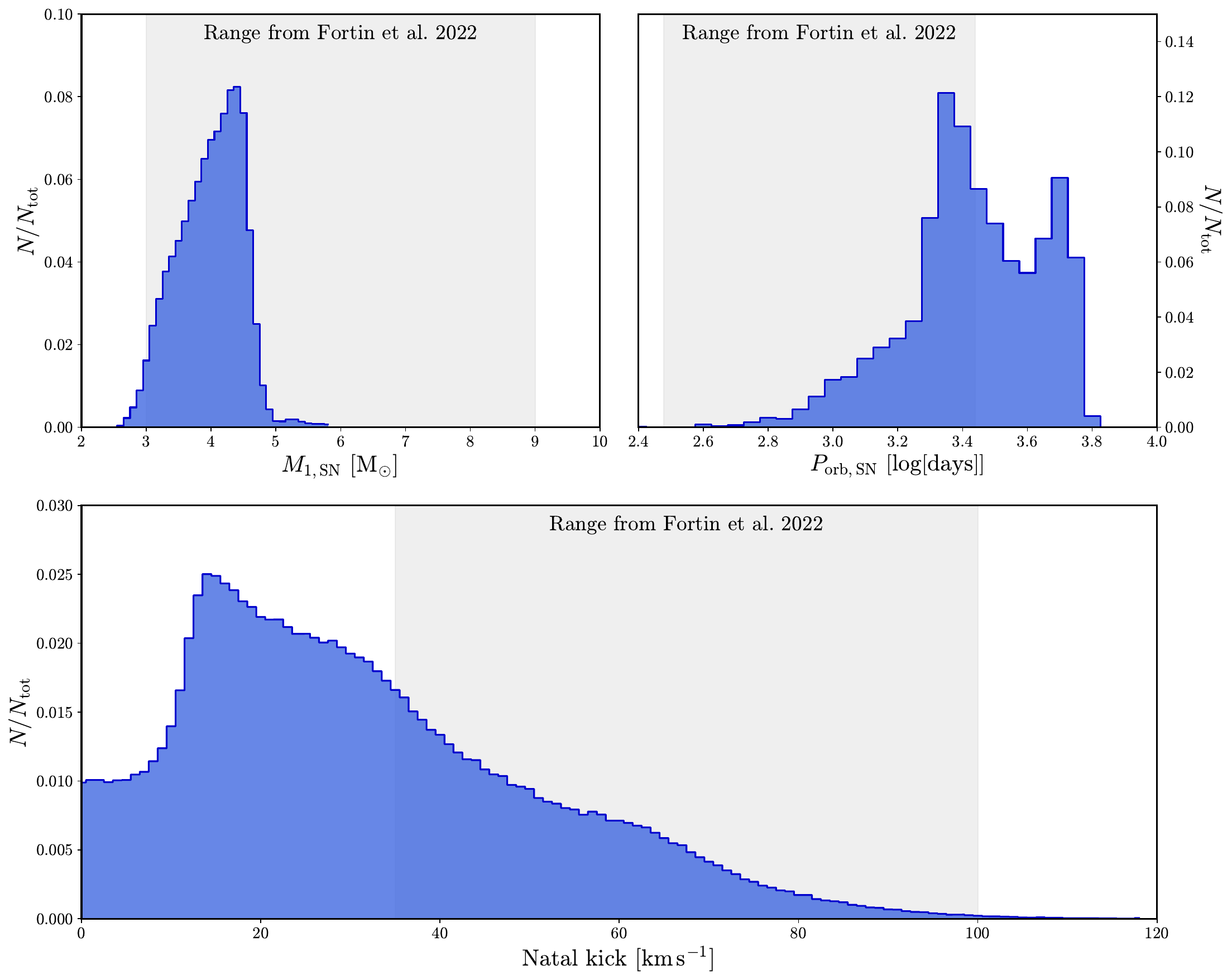}
    \caption{Distribution of compact object progenitor mass, orbital period, and natal kick for \btwelve (blue), alongside equivalent results from \citet{Fortin2022}. Small variations in initial progenitor mass and orbital period result in a slightly different natal kick range - notably, our results allow this system to form without a natal kick.}
    \label{Fig:B1259_fortincomparison}
\end{figure}

\section{Dependence of $\vsys$ on $\vnk$}\label{Sect:AppD}

\begin{figure}
    \centering
    \includegraphics[width=0.9\textwidth]{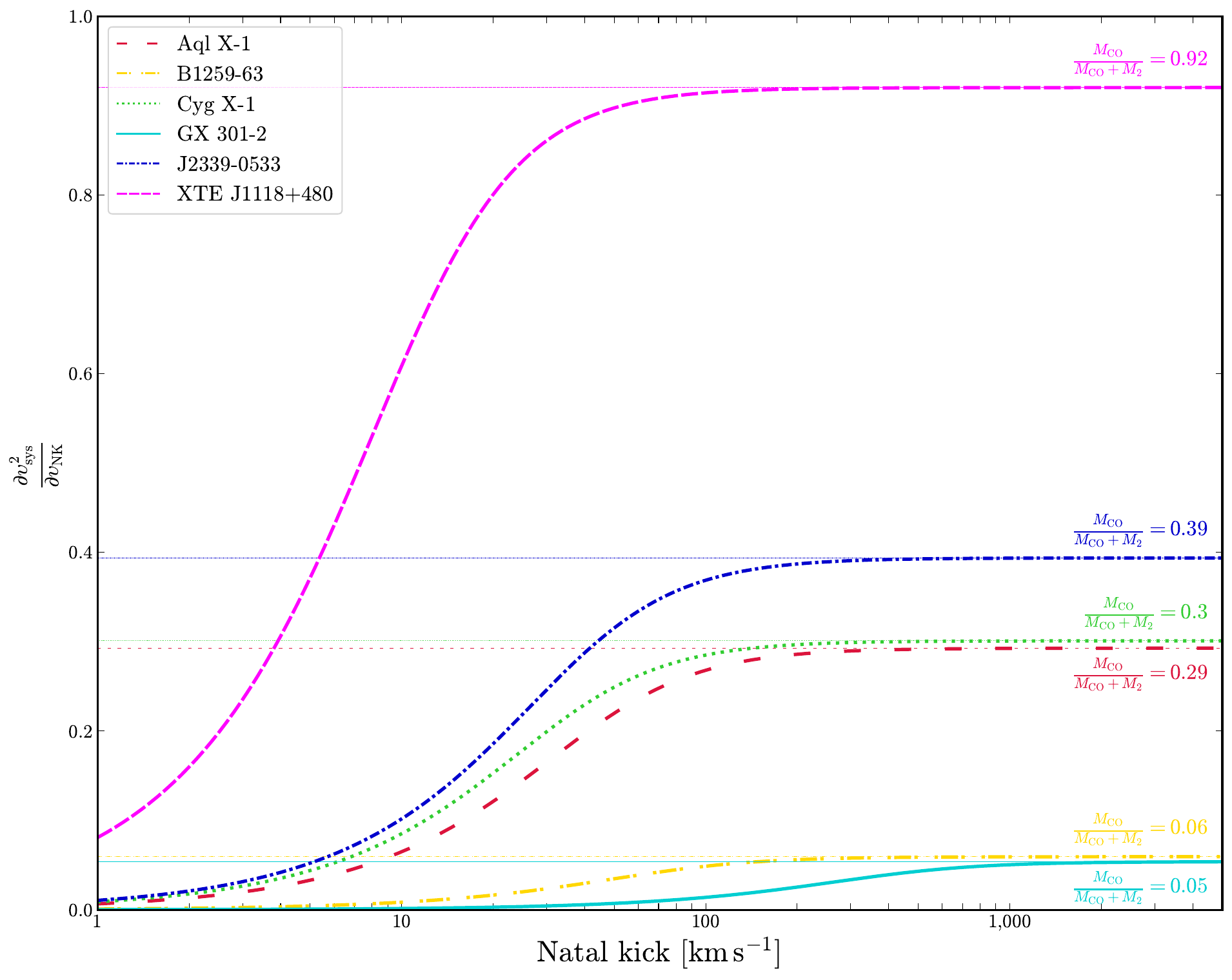}
    \caption{The influence of natal kick magnitude on systemic velocity for each of the systems in this study (from the differential of Equation \ref{eq:vsys_psn}), using binary parameters determined with \cosmic. As natal kick magnitude increases, its impact asymptotes at $\frac{\Mc}{\Mc+M_2}$; the point at which this happens is determined by the post-SN mass ratio. Each system is distinguished by colour and linestyle.}
    \label{Fig:VNKDifferential}
\end{figure}